\documentclass[aps,twocolumn,floats,prd,nofootinbib,superscriptaddress,10pt,longbibliography]{revtex4-1}
\usepackage[utf8]{inputenc}
\usepackage{amsmath}
\usepackage{graphicx}
\usepackage{dcolumn}
\usepackage{bm}
\usepackage{hyperref}
\usepackage{lipsum}
\usepackage{xcolor}
\usepackage{calc}
\usepackage{accents}
\usepackage{comment}
\usepackage{float}
\usepackage{diagbox}
\usepackage{soul}
\usepackage{ulem}
\usepackage{multirow}

\newcommand{\code}[1]{\texttt{#1}}
\makeatletter
\newcommand\footnoteref[1]{\protected@xdef\@thefnmark{\ref{#1}}\@footnotemark}
\makeatother

\begin{document}

\preprint{APS/123-QED}

\title{Can acoustic and axion-like early dark energy still resolve the Hubble tension?}
\author{Th\'eo Simon}
\email{Electronic address:  theo.simon@umontpellier.fr}
\affiliation{Laboratoire Univers \& Particules de Montpellier (LUPM), CNRS \& Universit\'e de Montpellier (UMR-5299), Place Eug\`ene Bataillon, F-34095 Montpellier Cedex 05, France}

\begin{abstract}

In this paper, we reassess the ability of the acoustic early dark energy (ADE) and axion-like early dark energy (EDE) models to resolve the Hubble tension in light of the new Pantheon+ and S$H_0$ES data on the one hand, and the BOSS LRG and eBOSS QSO data, analyzed under the effective field theory of large-scale structures (ETFofLSS) on the other hand.
We find that the Pantheon+ data, which favor a larger $\Omega_m$ value than the Pantheon data, have a strong constraining power on the ADE model, while the EFTofLSS analysis of the BOSS and eBOSS data only slightly increases the constraints.
We establish that the ADE model is now  strongly disfavored as a solution to the Hubble tension, with a remaining tension of $3.6\sigma$ (according to the $Q_{\rm DMAP}$ metric).
In addition, we find that the axion-like EDE model performs better when confronted to the same datasets, with a residual tension of $2.5\sigma$.
This work shows that the Pantheon+ data can have a decisive impact on models which aim to resolve the Hubble tension.

\end{abstract}

\maketitle

\section{Introduction}

The $\Lambda$ cold dark matter ($\Lambda$CDM) model provides a remarkable description of a wide variety of data from the early universe -- such as cosmic microwave background (CMB) or big bang nucleosynthesis (BBN) --, as well as observations of large scale structure (LSS) from the late universe -- including the
baryon acoustic oscillation (BAO) and the uncalibrated luminosity distance to supernovae of type Ia (SNIa). However, as the accuracy of cosmological observations has improved, the concordance cosmological model starts showing several experimental discrepancies.
Among them, the Hubble tension refers to the inconsistency between local measurements of the current expansion rate of the Universe, quantified by the Hubble constant $H_0$, and the values inferred from early universe data assuming the $\Lambda$CDM model.
More precisely, this tension is essentially driven by the \textit{Planck} Collaboration's observation of the CMB, which predicts a value of $H_0 = 67.27 \pm 0.60$ km/s/Mpc~\cite{Planck:2018vyg} within the $\Lambda$CDM model, and the value measured by the S$H_0$ES Collaboration using the Cepheid-calibrated cosmic distance ladder, whose latest measurement yields $H_0 = 73.04\pm1.04$ km/s/Mpc \cite{Riess:2021jrx,Riess:2022mme}. 
The disagreement between these observations results in an $\sim 5\sigma$ tension.
Experimental efforts are under way to establish whether this discrepancy can be caused by systematic effects (see, \textit{e.g.}, \cite{Dainotti:2021pqg,Dainotti:2022bzg,Mortsell:2021nzg,Mortsell:2021tcx,Follin:2017ljs,Brout:2020msh}), but no definitive explanation has yet been found.
This tension could therefore be indicative of new physics, most likely located in the prerecombination era, which involves a reduction in the sound horizon before recombination \cite{Bernal:2016gxb,Aylor:2018drw,Knox:2019rjx,Camarena:2021jlr,Efstathiou:2021ocp,Schoneberg:2021qvd}. Early dark energy (EDE) models are capable of producing such an effect by increasing the total energy density of the Universe before recombination with the addition of a scalar field~\cite{Karwal:2016vyq,Poulin:2018cxd,Smith:2019ihp,Niedermann:2019olb,Niedermann:2020dwg,Ye:2020btb,Agrawal:2019lmo,Berghaus:2019cls,Braglia:2020bym,Braglia:2020auw,Gonzalez:2020fdy,Rezazadeh:2022lsf,Herold:2021ksg,Gomez-Valent:2022hkb,Poulin:2021bjr,Hill:2020osr,LaPosta:2021pgm,Reeves:2022aoi,Herold:2022iib,Eskilt:2023nxm,Murgia:2020ryi,Goldstein:2023gnw} (for review of EDE models see Ref.~\cite{Poulin:2023lkg}, and for a review of models that could resolve the Hubble tension see Refs.~\cite{DiValentino:2021izs,Schoneberg:2021qvd}). In the following, we consider the specific case of acoustic dark energy (ADE) developed in Refs.~\cite{Lin:2019qug,Lin:2020jcb}.\\

In this paper, we reassess the constraints on the ADE and axion-like EDE models (paying particular attention to the former) and their ability to resolve the Hubble tension, by successively evaluating the impact of the effective field theory (EFT) full-shape analysis applied to the BOSS LRG~\cite{BOSS:2016wmc} and eBOSS QSO~\cite{eBOSS:2020yzd} data, and the impact of the Pantheon+ data~\cite{Brout:2022vxf}.
On the one hand, we make use of developments of the one-loop prediction of the galaxy power spectrum in redshift space from the effective field theory of large-scale structures (EFTofLSS)~\footnote{The first formulation of the EFTofLSS was carried out in Eulerian space in Refs.~\cite{Carrasco:2012cv,Baumann:2010tm} and in Lagrangian space in \cite{Porto:2013qua}. Once this theoretical framework was established, many efforts were made to improve this theory and make it predictive, such as the understanding of renormalization \cite{Pajer:2013jj, Abolhasani:2015mra}, the IR-resummation of the long displacement fields \cite{Senatore:2014vja, Baldauf:2015xfa, Senatore:2014via, Senatore:2017pbn, Lewandowski:2018ywf, Blas:2016sfa}, and the computation of the two-loop matter power spectrum \cite{Carrasco:2013sva, Carrasco:2013mua}. Then, this theory was developed in the framework of biased tracers (such as galaxies and quasars) in Refs. \cite{Senatore:2014eva, Mirbabayi:2014zca, Angulo:2015eqa, Fujita:2016dne, Perko:2016puo, Nadler:2017qto}.} applied to the BOSS~\cite{DAmico:2019fhj} and eBOSS~\cite{Simon:2022csv} data in order to constrain the ADE and axion-like EDE models.
This novel theoretical framework has made it possible to determine the $\Lambda$CDM parameters at precision higher than that from conventional BAO and redshift space distortions analyses, and even comparable to that of CMB experiments. 
In addition, the EFTofLSS provides an important consistency test for the $\Lambda$CDM model and its underlying assumptions, while allowing one to derive competitive constraints on models beyond $\Lambda$CDM (see, \textit{e.g.}, Refs.~\cite{DAmico:2019fhj,Ivanov:2019pdj,Colas:2019ret,DAmico:2020kxu,DAmico:2020tty,Simon:2022ftd,Simon:2022lde,Chen:2021wdi,Zhang:2021yna,Philcox:2021kcw,Kumar:2022vee,Nunes:2022bhn,Lague:2021frh,Philcox:2020xbv,Smith:2022iax,Simon:2022csv,Moretti:2023drg,Rubira:2022xhb,Schoneberg:2023rnx,Holm:2023laa}).
The study of the EFTofLSS impact on the ADE constraints is similar to what has already been done for the axion-like EDE model in Ref.~\cite{Simon:2022adh} (see also Refs.~\cite{DAmico:2020ods,Smith:2020rxx,Ivanov:2020ril}), which showed that this model leaves signatures in the galaxy power spectrum on large scales that can be probed by the BOSS data.
On the other hand, we update the constraints on the ADE and axion-like EDE models by considering the Pantheon+ data from Ref.~\cite{Brout:2022vxf}.
It has already been shown in Ref.~\cite{Simon:2022adh} that the combination of the Pantheon+ data with a S$H_0$ES prior provides better constraints on the axion-like EDE model than the equivalent analysis including Pantheon data.
This can be interpreted as a consequence of the fact that the Pantheon+ data prefer a value of $\Omega_m=0.334 \pm 0.018$ which is higher than that of the Pantheon data. Together with the measured value of $H_0=100\cdot h$ km/s/Mpc by S$H_0$ES, it leads to an increased value of $\omega_{\rm cdm} = \Omega_{\rm cdm}\cdot h^2$ (see Ref.~\cite{Blanchard:2022xkk}), which cannot be fully compensated by the presence of EDE, therefore degrading slightly the fit to CMB data.\\

In Sec.~\ref{sec:model_data}, we provide a review of the ADE and axion-like EDE models, as well as a description of the analysis method and the datasets to which these models will be subjected.
In Sec.~\ref{sec:cosmological_results}, we present the constraints of the ADE model and compare them to the axion-like EDE case, while in Sec.~\ref{sec:discussion} we consider some additional variations of the model under study.

\section{The model and the data}
\label{sec:model_data}

\subsection{Review of the ADE model}

\begin{figure*}
    \centering
    \includegraphics[width=1.5\columnwidth]{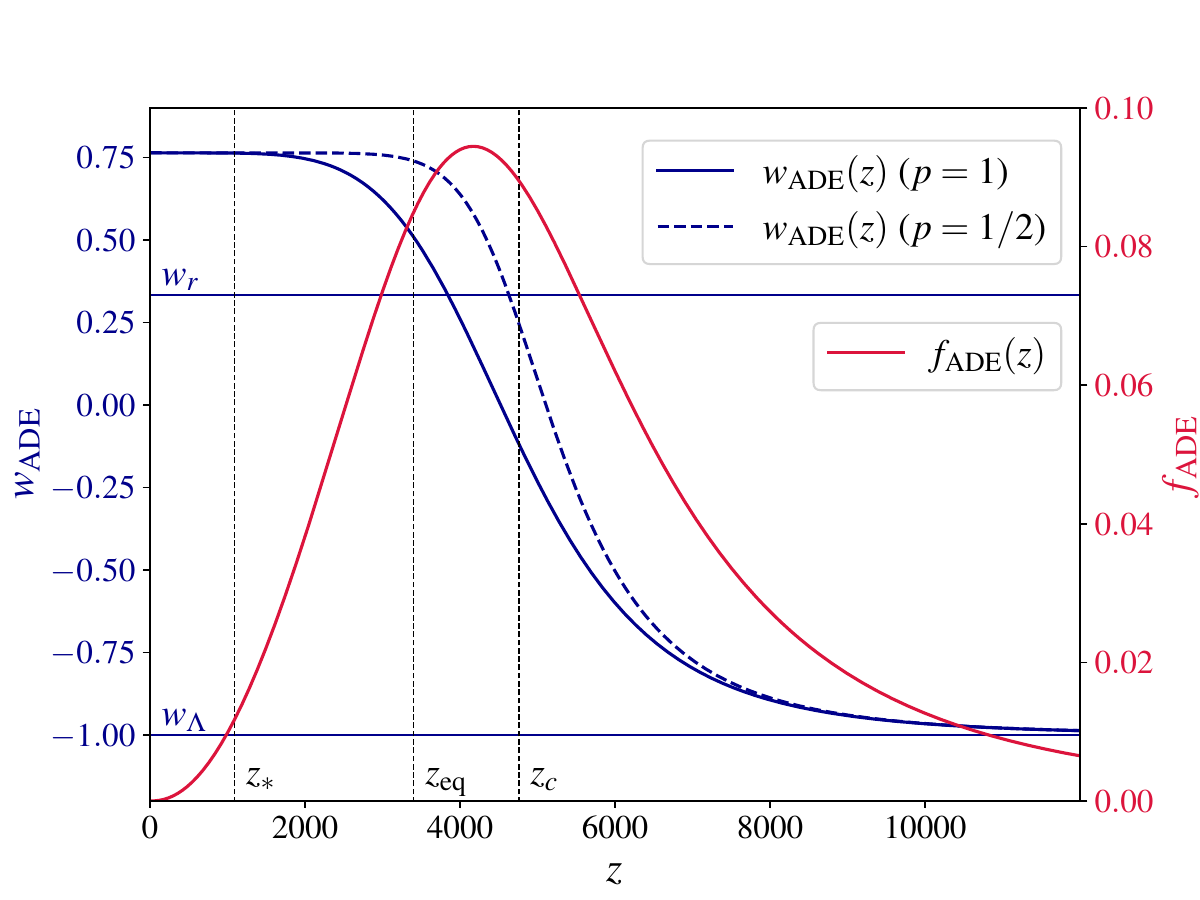}
    \caption{Evolution of the ADE equation-of-state parameter $w_{\rm ADE}$, as well as the  ADE fractional energy density $f_{\rm ADE}(z_c)$ as a function of the cosmological redshift $z$.
    To perform this plot, we use the best-fit values of the BAO/$f\sigma_8$ + Pan + $M_b$ analysis (see Tab.~\ref{tab:cosmoparam}). For the ADE equation-of-state parameter, we set $p=1$, which corresponds to our baseline setup, and $p=1/2$, which corresponds to the setup of Ref.~\cite{Lin:2019qug}.
    The horizontal lines correspond to the radiation and dark energy equation-of-state parameters, $w_r$ and $w_{\Lambda}$, respectively, while the dashed vertical lines correspond to the redshift of recombination $z_*$, the redshift of the matter-radiation equality $z_{\rm eq}$, and the ADE critical redshift $z_c$.}
    \label{fig:w_a}
\end{figure*}

In this paper, we focus on the acoustic dark energy (ADE) model proposed in Ref.~\cite{Lin:2019qug} (see Ref.~\cite{Poulin:2023lkg} for a general introduction).
In this model, the ADE equation-of-state parameter, $w_{\rm ADE}(a)= P_{\rm ADE}(a) / \rho_{\rm ADE}(a)$, is modeled as
\begin{equation}
\label{eq:w_ADE_pheno}
w_{\rm ADE}(a) = \frac{1+w_f}{\left[1+(a_c/a)^{3(1+w_f)/p}\right]^p} -1\, .
\end{equation}
In Fig.~\ref{fig:w_a}, we plot the evolution of $w_{\rm ADE}$ as a function of the cosmological redshift $z$.
This figure clearly illustrates that in this model the critical redshift $z_c = (a_0-a_c)/a_c$ sets a transition in the ADE equation-of-state from $w_{\rm ADE} \to -1$, when $z \gg z_c$, to $w_{\rm ADE} \to w_f$, when $z \ll z_c$.
Therefore, this parametrization allows the ADE component to behave in a similar way to dark energy before the critical redshift (exactly like the axion-like EDE model), while it allows the late-time value of the ADE equation-of-state to be set thanks to the parameter $w_f$. 
As shown in Fig.~\ref{fig:w_a}, the rapidity of this transition is controlled by the parameter $p$, which is set at $p=1$ for our baseline model, corresponding to the modeling of the time-averaged background equation-of-state of the axion-like EDE model~\cite{Poulin:2018dzj}.
Similar to the axion-like EDE case where $w_{\rm EDE}(z\ll z_c) = 1/2$ (see below), the ADE dilutes faster than the radiation (\textit{i.e.}, $w_f>w_r$) below the critical redshift, in order to suppress the contribution of this component to the total budget of the Universe at the moment of the CMB.\\

Let us note that the parametrization of Eq.~\eqref{eq:w_ADE_pheno} can be achieved in the K-essence class of dark energy models. In particular, the dark component is here a perfect fluid represented by a minimally coupled scalar field $\phi$ with a general kinetic term~\cite{Armendariz-Picon:2000ulo}. For the specific case of a constant sound speed $c_s^2$, the Lagrangian density is written as~\cite{Gordon:2004ez}:
\begin{equation}
    P(X,\phi)=\bigg(\frac{X}{A}\bigg)^{\frac{1-c_s^2}{2c_s^2}}X-V(\phi)\,,
\end{equation}
where $X=-\nabla^2\phi/2$ and $A$ is a constant density scale \cite{Lin:2019qug}. In this category of models, $w_{\rm ADE}\to c_s^2$ if the kinetic term dominates, whereas $w_{\rm ADE}\to-1$ if the potential $V(\phi)$ dominates.
The main advantage of the ADE model over the axion-like EDE model is that the former provides a general class of exact solutions, while the latter requires a specific set of initial conditions to achieve a similar phenomenology~\cite{Lin:2019qug}.\\

Since the ADE equation-of-state parameter changes over time, the conservation equation gives
\begin{equation}
\rho_{\rm ADE}(a) = \rho_{\rm ADE,0} e^{3\int_a^1 [1+w_{\rm ADE}(a')]da'/a'}\, ,
\end{equation}
which allows us to define the ADE fractional energy density as
\begin{equation}
    f_{\rm ADE}(a) = \frac{\rho_{\rm ADE}(a)}{\rho_{\rm tot}(a)}\, .
\end{equation}
In Fig.~\ref{fig:w_a}, we also plot the evolution of $f_{\rm ADE}$ as a function of the cosmological redshift $z$.
We notice that this parameter is maximal around the ADE equation-of-state transition, set by the critical redshift $z_c$, namely, when $f_{\rm ADE}(z \sim z_c)$. Then, this parameter becomes subdominant at the time of recombination, with $f_{\rm ADE}(z_*) \sim 1\%$~\cite{Gomez-Valent:2021cbe}.\\

Finally, the ADE model we are considering is described by the three following parameters
\begin{equation}
    \{z_c, \, f_{\rm ADE}(z_c), \, w_f \}\, .
\end{equation}
Ref.~\cite{Lin:2019qug} also considers the variation of a fourth parameter that determines the behavior of the ADE perturbations, namely, their rest frame sound speed $c_s^2(k,a)$. 
Unlike the standard axion-like EDE model (see below), we assume for this model the scale independence of this parameter, \textit{i.e.}, $c_s^2(k,a)=c_s^2(a)$, which is equivalent to assuming a perfect fluid with a linear dispersion relation.
In addition, because of the sharp transition of the $w_{\rm ADE}$ parameter, the impact of the ADE component on the perturbed universe is localized in time, which implies that we can approximate this parameter as a constant.
Thus, Ref.~\cite{Lin:2019qug} varies this parameter to its critical redshift value, namely, $c_s^2 = c_s^2(a = a_c)$, in addition to the three other parameters listed above. In our baseline model, we consider that $c_s^2 = w_f$, insofar as it has been shown to be a good approximation near the best-fit~\cite{Lin:2019qug}.
However, in Sec.~\ref{sec:discussion}, we consider two model variations of our baseline model: (i) the $c_s^2$ADE model, where we free these two parameters independently, and (ii) the cADE model, where we set $c_s^2 = w_f =1$.
Let us note that there exists a second difference between Refs.~\cite{Lin:2019qug,Lin:2020jcb} and our baseline analysis, since these references set $p=1/2$, which leads to a sharper transition than ours (with $p=1$), as shown in Fig.~\ref{fig:w_a}. However, the impact of this parameter on cosmological results is very minor, and we have verified that we obtain the same results as Ref.~\cite{Lin:2020jcb} with $p=1$.

\subsection{Review of the axion-like EDE model}

For comparison, we also consider the axion-like early dark energy (EDE) model~\cite{Karwal:2016vyq,Poulin:2018cxd,Smith:2019ihp}, which corresponds to an extension of the $\Lambda$CDM model, where the existence of an additional subdominant oscillating scalar field $\phi$ is considered. 
The EDE field dynamics is described by the Klein-Gordon equation of motion (at the homogeneous level),
\begin{equation}
    \ddot{\phi} + 3 H \dot{\phi} + V_{n,\phi}(\phi) = 0\,,
\end{equation} 
where $V_n(\phi)$ is a modified axion-like potential defined as
\begin{equation}
    V_n(\phi) = m^2f^2\left[ 1- \cos(\phi/f)   \right]^n.
\end{equation}
$f$ and $m$ correspond to the decay constant and the effective mass of the scalar field, respectively, while the parameter $n$ controls the rate of dilution after the field becomes dynamical.
In the following, we will use the redefined field quantity $\Theta = \phi/f$ for convenience, such that $-\pi \le \Theta \le +\pi$.
At early times, when $H\gg m$, the scalar field $\phi$ is frozen at its initial value since the Hubble friction prevails, which implies that the EDE behaves like a form of dark energy and that its contribution to the total energy density increases relative to the other components. When the Hubble parameter drops below a critical value ($H \sim m$), the field starts evolving toward the minimum of the potential and becomes dynamical. The EDE contribution to the total budget of the Universe is maximum around a critical redshift $z_c$, after which the energy density starts to dilute with an equation-of-state parameter $w_{\rm EDE}(a)$ approximated by~\cite{1983PhRvD..28.1243T,Poulin:2018dzj}:
\begin{equation}
w_{\rm EDE} = \left\{
    \begin{array}{ll}
        -1 \ &\text{if} \ z>z_c, \\
        \frac{n-1}{n+1} \ &\text{if} \ z<z_c.
    \end{array}
\right.
\end{equation}
In the following, we will fix $n=3$ as it was found that the data were relatively insensitive to this parameter provided $2\lesssim n \lesssim 5$ \cite{Smith:2019ihp}, implying that in this specific model $w_{\rm EDE}(z\ll z_c) = 1/2$.
Instead of the theory parameters $f$ and $m$, we make use of $z_c$ and $f_{\rm EDE}(z_c)$, determined through a shooting method \cite{Smith:2019ihp}. We also include the initial field value $\Theta_i$ as a free parameter, whose main role once $f_{\rm EDE}(z_c)$ and $z_c$ are fixed is to set the dynamics of perturbations right around $z_c$, through the EDE sound speed $c_s^2$. Finally, the axion-like EDE model is described by the three following parameters:
\begin{equation}
    \{z_c, \, f_{\rm EDE}(z_c), \, \Theta_i \}\, .
\end{equation}\\

Let us note that the axion-like EDE sound speed $c_s^2(a,k)= \delta P_{\rm EDE}(k,a)/\delta \rho_{\rm EDE}(k,a)$ is scale and time dependent, and is entirely determined by the three EDE parameters specified above.
In the fluid approximation, one can estimate the $a$ and $k$ dependencies of this parameter as~\cite{Poulin:2018dzj,Poulin:2018cxd}
\begin{equation}
	c^2_{s} (a,k)= 
	\begin{cases}
	1 \, , & a \le a_c,  \\
	\dfrac{2a^2(n-1)\varpi^2(a)+k^2}{2a^2(n+1)\varpi^2(a)+k^2}  \,,
	& a > a_c \, ,
	\end{cases}
	\label{eqn:cs2}
\end{equation}
where $\varpi$ corresponds to the angular frequency of the oscillating background field, which has a time dependency fixed by $z_c$, $n$ and $\Theta_i$ (see Ref.~\cite{Poulin:2018dzj}).
Let us note however that the axion-like EDE model we consider in this paper does not rely on this fluid approximation, and instead solves the exact (linearized) Klein-Gordon equation, which is expressed in synchronous gauge as~\cite{Ma:1995ey}:
\begin{equation}
    \delta\phi''_k + 2 H \delta \phi_k' + \left[k^2 + a^2 V_{n,\phi\phi}\right] \delta \phi_k = - h'  \phi'/2 \, ,
\end{equation}
where the prime denotes derivatives with respect to conformal time.

\subsection{Data and analysis methods}
We perform Markov chain Monte Carlo (MCMC) analyses, confronting the ADE model with recent cosmological observations.
To do so, we make use of the Metropolis-Hastings algorithm from the \texttt{MontePython-v3}~\footnote{\url{https://github.com/brinckmann/montepython_public}.} code~\cite{Brinckmann:2018cvx,Audren:2012wb} interfaced with our modified \texttt{CLASS}~\cite{Lesgourgues:2011re,Blas:2011rf} version.~\footnote{\url{https://github.com/PoulinV/AxiCLASS}.}
In this paper, we perform various analyses from a combination of the following datasets:
\begin{itemize}
    \item \textbf{Planck:} The low-$\ell$ CMB temperature and polarization autocorrelations (TT, EE), and the high-$\ell$ TT, TE, EE data~\cite{Planck:2019nip}, as well as the gravitational lensing potential reconstruction from {\it Planck}~2018~\cite{Planck:2018lbu}.
    
    \item \textbf{ext-BAO:} The low-$z$ BAO data gathered from 6dFGS at $z = 0.106$ \cite{Beutler:2011hx}, SDSS DR7 at $z = 0.15$ \cite{Ross:2014qpa}.

    \item \textbf{BOSS BAO/$f\sigma_8$:} BAO measurements, cross-correlated with the redshift space distortion measurements from the CMASS and LOWZ galaxy samples of BOSS DR12 LRG at $z = 0.38$, 0.51, and 0.61 \cite{BOSS:2016wmc}.

    \item \textbf{eBOSS BAO/$f\sigma_8$:} BAO measurements, cross-correlated with the redshift space distortion measurements from the CMASS and LOWZ quasar samples of eBOSS DR16 QSO at $z = 1.48$~\cite{eBOSS:2020yzd}.
    
    \item \textbf{EFTofBOSS:} The EFTofLSS analysis of BOSS DR12 LRG cross-correlated with the reconstructed BAO parameters \cite{Gil-Marin:2015nqa}. The SDSS-III BOSS DR12 galaxy sample data and covariances are described in \cite{BOSS:2016wmc,Kitaura:2015uqa}. The measurements, obtained in \cite{Zhang:2021yna}, are from BOSS catalogs DR12 (v5) combined CMASS-LOWZ~\cite{Reid:2015gra}, and are divided in redshift bins LOWZ, $0.2<z<0.43 \  (z_{\rm eff}=0.32)$, and CMASS, $0.43<z<0.7  \ (z_{\rm eff}=0.57)$, with north and south galactic skies for each, respectively, denoted NGC and SGC. From these data we use the monopole and quadrupole moments of the galaxy power spectrum. The theory prediction and likelihood for the full-modeling information are made available through \texttt{PyBird} \cite{DAmico:2020kxu}, together with the West-coast parametrization~\cite{Nishimichi:2020tvu,Simon:2022lde,Holm:2023laa} as implemented in Ref.~\cite{DAmico:2020kxu}.  In our analyses, we vary 7 EFT parameters per sky cut, namely, 3 bias parameters ($b_1$, $b_3$, and $c_2 \equiv (b_2+b_4)/\sqrt{2}$), 2 counterterm parameters ($c_{\rm ct}$ and $c_{r,1}$), and 2 stochastic parameters ($c_{\epsilon, 0}$ and $c_{\epsilon}^{\rm quad}$). The physical meaning of these parameters and the priors used are described in detail in Ref.~\cite{Holm:2023laa}. Finally, we analyse the BOSS data up to $k_{\rm max}^{\rm CMASS} = 0.23 h \, {\rm Mpc}^{-1}$ for the CMASS sky cut and up to $k_{\rm max}^{\rm LOWZ} = 0.20 h\, {\rm Mpc}^{-1}$ for the LOWZ sky cut (as determined in Ref.~\cite{Colas:2019ret}).
    
    \item \textbf{EFTofeBOSS:} The EFTofLSS analysis \cite{Simon:2022csv} of eBOSS DR16 QSOs \cite{eBOSS:2020yzd}. The QSO catalogs are described in \cite{Ross:2020lqz} and the covariances are built from the EZ-mocks described in \cite{Chuang:2014vfa}. There are about 343 708 quasars selected in the redshift range $0.8<z<2.2$, with $z_{\rm eff}=1.52$, divided into two skies, NGC and SGC~\cite{Beutler:2021eqq,Hou:2020rse}. From these data, we use the monopole and quadrupole moments of the galaxy power spectrum. The theory prediction and likelihood for the full-modeling information are made available through \texttt{PyBird}, together with the West-coast parametrization as implemented in Ref.~\cite{Simon:2022csv}. We use the same EFT parameters as for EFTofBOSS, and we set $k_{\rm max}^{\rm eBOSS} = 0.24 h\,{\rm Mpc}^{-1}$ (as determined in Ref.~\cite{Simon:2022csv}).

    \item \textbf{Pantheon:} The Pantheon catalog of uncalibrated luminosity distance of type Ia supernovae (SNeIa) in the range ${0.01<z<2.3}$~\cite{Scolnic:2017caz}.
    
    \item \textbf{Pantheon+:} The newer Pantheon+ catalog of uncalibrated luminosity distance of SNeIa in the range ${0.001<z<2.26}$~\cite{Brout:2022vxf}. 

    \item \textbf{Pantheon+/S$\boldsymbol{H_0}$ES:} The Pantheon+ catalog cross-correlated with the absolute calibration of the SNeIa from S$H_0$ES~\cite{Riess:2021jrx}.
    
    \item $\boldsymbol{{M_b}}$: Gaussian prior from the most up-to-date late-time measurement of the absolute calibration of the SNeIa from S$H_0$ES, $M_b = -19.253 \pm 0.027$~\cite{Riess:2021jrx}, corresponding to $H_0 = (73.04\pm1.04)$ km/s/Mpc.
\end{itemize}

We choose \textit{Planck} + ext-BAO + BOSS BAO/$f\sigma_8$ + eBOSS BAO/$f\sigma_8$ + Pantheon (optionally with the $M_b$ prior) as our baseline analysis, called, for the sake of simplicity, ``BAO/$f\sigma_8$ + Pan.''
In order to assess the impact of the EFT full-shape analysis of the BOSS and eBOSS data on the ADE resolution of the Hubble tension, we compare the baseline analysis with an equivalent analysis that includes the EFTofBOSS and EFTofeBOSS likelihoods instead of the BOSS and eBOSS BAO/$f\sigma_8$ likelihoods.
This analysis is called ``EFT + Pan.''
Finally, in order to gauge the influence of the new Pantheon data, we replace the Pantheon likelihood with the Pantheon+ likelihood.
This analysis, referred to as ``EFT + PanPlus,'' is compared with the aforementioned EFTofLSS analysis.
In App.~\ref{app:Mb_prior}, we show explicitly that the addition of the $M_b$ prior on top of the Pantheon+ likelihood is equivalent to the use of the full ``Pantheon+/S$H_0$ES'' likelihood as provided in Ref.~\cite{Riess:2021jrx}.\\

For all runs performed, we impose large flat priors on $\{\omega_b,\omega_{\rm cdm},H_0,A_s,n_s,\tau_{\rm reio}\}$, which correspond, respectively, to the dimensionless baryon energy density, the dimensionless cold dark matter energy density, the Hubble parameter today, the variance of curvature perturbations centered on the pivot scale $k_p = 0.05$ Mpc$^{-1}$ (according to the \textit{Planck} convention), the scalar spectral index, and the reionization optical depth.
Regarding the free parameters of the ADE model, we impose logarithmic flat priors on $z_c$, and flat priors on $f_{\rm ADE}(z_c)$ and $w_{\rm ADE}$,
\begin{align*}
    &3 \le \log_{10}(z_c) \le 4.5 \, , \\
    &0\le f_{\rm ADE}(z_c) \le 0.2 \, , \\
    &0 \le w_{f} \le 3.6\, .
\end{align*}
Note that we have verified that a wider prior on $w_{f}$ does not impact our results.
When we compare the ADE model with the axion-like EDE model, we use the following priors for the latter:
\begin{align*}
    &3 \le \log_{10}(z_c) \le 4 \, , \\
    &0\le f_{\rm EDE}(z_c) \le 0.5 \, , \\
    &0 \le \Theta_i \le \pi \,.
\end{align*}
In this paper, we use \textit{Planck} conventions for the treatment of neutrinos; that is, we include two massless and one massive species with $m_{\nu} = 0.06$ eV \cite{Planck:2018vyg}.
In addition, we use \code{Hmcode}~\cite{Mead:2020vgs} to estimate the nonlinear matter clustering solely for the purpose of the CMB lensing.
We define our MCMC chains to be converged when the Gelman-Rubin criterion $R-1 < 0.05$.
Finally, we extract the best-fit parameters from the procedure highlighted in the appendix of Ref.~\cite{Schoneberg:2021qvd}, and we produce our figures thanks to \code{GetDist}~\cite{Lewis:2019xzd}.\\

In this paper, we compare the models with each other using two main metrics.
First, in order to assess the ability of an extended model $\mathcal{M}$ to fit all the cosmological data, we compute the Akaike information criterion (AIC) of this model relative to that of the $\Lambda$CDM. This metric is defined as follows 
\begin{equation}
    \Delta {\rm AIC} = \chi^2_{\rm min, \, \mathcal{M}} - \chi^2_{\rm min, \, \Lambda CDM} + 2\cdot (N_{\mathcal{M}} - N_{\Lambda \rm CDM}) \, ,
\end{equation}
where $\mathcal{M} \in \{{\rm ADE, \, EDE, \,} c_s^2{\rm ADE, \, cADE}\}$, and where $N_{\mathcal{M}}$ stands for the number of free parameters of the model.
This metric enables us to determine whether the fit within a particular model $\mathcal{M}$ significantly improves that of $\Lambda$CDM by penalizing models with a larger number of degrees of freedom.
Second, in order to gauge the ability of the extended model $\mathcal{M}$ to solve the Hubble tension for a given combination of data $\mathcal{D}$ (which does not include the $M_b$ prior), we also compute the residual Hubble tension thanks to the difference of the maximum \textit{a posteriori} (DMAP)~\cite{Raveri:2018wln}, determined by
\begin{equation}
    Q_{\rm DMAP} = \sqrt{\chi^2_{\rm min, \, \mathcal{M}}(\mathcal{D} + M_b) - \chi^2_{\rm min, \, \mathcal{M}}(\mathcal{D})} \, .
\end{equation}
This metric allows us to determine how the addition of the $M_b$ prior to the dataset $\mathcal{D}$ impacts the
fit within a particular model $\mathcal{M}$. Ref.~\cite{Schoneberg:2021qvd} asserts that a model is a good candidate for solving the Hubble tension if it meets these two conditions: $\Delta AIC < -6.91$ and $Q_{\rm DMAP} < 3\sigma$.
Finally, we also consider the Gaussian tension (GT), computed as 
\begin{equation}
    {\rm GT} = \frac{\overline{H_0}(\rm SH_0ES) - \overline{H_0}({\mathcal{D}})}{\sqrt{\sigma_{H_0}^2({\rm SH_0ES}) + \sigma_{H_0}^2({\mathcal{D}})}} \, ,
\end{equation}
where $\overline{H_0}$ and $\sigma_{H_0}$ correspond to the mean and standard deviation of the Hubble parameter today determined from the S$H_0$ES experiment and the dataset $\mathcal{D}$ (within the model $\mathcal{M}$).
The Gaussian tension is certainly the most direct metric for quantifying the Hubble tension, but the main problem with this metric is that it is unable to favor a complex model in which some parameters become irrelevant in the $\Lambda$CDM limit. If a probability density function deviates from Gaussian in a complex model (as is the case for EDE models), only the Gaussian $\Lambda$CDM limit has significant statistical weight~\cite{Schoneberg:2021qvd,Smith:2020rxx}.

\section{Cosmological results}
\label{sec:cosmological_results}

\begin{table*}[]
    \centering
    \scalebox{0.76}{
    \begin{tabular}{|l|c|c|c|c|c|c|}
    
    \hline
      & \multicolumn{2}{|c|}{BAO/$f\sigma_8$ + Pan}  & \multicolumn{2}{|c|}{EFT + Pan}&   \multicolumn{2}{|c|}{EFT + PanPlus}\\
              \hline
              \hline

     $M_b$ prior? & No & Yes & No & Yes & No & Yes\\
        \hline

$f_{\rm ADE}(z_c)$
	 & $< 0.060(0.034)$
	 & $0.081(0.090)\pm 0.018$
	 & $< 0.049(0.010)$
	 & $0.068(0.076)\pm 0.019$
	 & $< 0.036(0.011)$
	 & $0.073(0.082)\pm 0.020$
	 \\
$\log_{10}(z_c)$
	 & $unconst.(3.748)$
	 & $3.655(3.677)\pm 0.093$
	 & $unconst.(3.713)$
	 & $3.676(3.686)^{+0.095}_{-0.120}$
	 & $unconst.(3.957)$
	 & $3.692(3.724)^{+0.098}_{-0.120}$
	 \\
$w_f$
	 & $> 0.49(0.69)$
	 & $0.79(0.76)^{+0.10}_{-0.12}$
	 & $> 0.59(0.70)$
	 & $0.78(0.75)^{+0.11}_{-0.13}$
	 & $> 0.61(0.57)$
	 & $0.76(0.73)\pm 0.13$
	 \\
	 \hline
	 $H_0$
	 & $68.44(69.04)^{+0.47}_{-0.93}$
	 & $71.24(71.49)\pm 0.68$
	 & $68.16(68.26)^{+0.41}_{-0.53}$
	 & $71.01(71.31)\pm 0.73$
	 & $68.03(68.16)^{+0.43}_{-0.53}$
	 & $71.13(71.29)\pm 0.73$
	 \\ 
$\omega_{\rm cdm }$
	 & $0.1212(0.1239)^{+0.0012}_{-0.0030}$
	 & $0.1291(0.1305)\pm 0.0027$
	 & $0.1196(0.1202)^{+0.0009}_{-0.0015}$
	 & $0.1267(0.1280)\pm 0.0027$
	 & $0.1201(0.1211)^{+0.0011}_{-0.0015}$
	 & $0.1278(0.1294)\pm 0.0028$
	 \\
$10^{2}\omega_{b }$
	 & $2.259(2.269)^{+0.016}_{-0.023}$
	 & $2.304(2.309)\pm 0.021$
	 & $2.254(2.252)^{+0.014}_{-0.017}$
	 & $2.303(2.306)\pm 0.022$
	 & $2.250(2.258)\pm 0.018$
	 & $2.306(2.311)\pm 0.022$
	 \\
$10^{9}A_{s }$
	 & $2.123(2.119)\pm 0.030$
	 & $2.159(2.152)\pm 0.031$
	 & $2.111(2.111)\pm 0.030$
	 & $2.148(2.151)^{+0.028}_{-0.032}$
	 & $2.111(2.116)^{+0.028}_{-0.036}$
	 & $2.151(2.150)^{+0.028}_{-0.033}$
	 \\
$n_{s }$
	 & $0.9711(0.9748)^{+0.0041}_{-0.0074}$
	 & $0.9900(0.9925)\pm 0.0061$
	 & $0.9684(0.9686)^{+0.0040}_{-0.0048}$
	 & $0.9878(0.9904)\pm 0.0060$
	 & $0.9679(0.9697)^{+0.0038}_{-0.0047}$
	 & $0.9890(0.9906)\pm 0.0063$
	 \\
$\tau_{\rm reio }$
	 & $0.0583(0.0540)^{+0.0063}_{-0.0075}$
	 & $0.0588(0.0561)^{+0.0066}_{-0.0077}$
	 & $0.0572(0.0573)\pm 0.0070$
	 & $0.0590(0.0584)^{+0.0067}_{-0.0077}$
	 & $0.0570(0.0574)\pm 0.0072$
	 & $0.0585(0.0573)^{+0.0065}_{-0.0077}$
	 \\
	 \hline
	 $S_8$
	 & $0.828(0.834)^{+0.011}_{-0.013}$
	 & $0.843(0.846)\pm 0.013$
	 & $0.820(0.823)\pm 0.010$
	 & $0.832(0.835)\pm 0.012$
	 & $0.825(0.830)^{+0.010}_{-0.011}$
	 & $0.836(0.842)\pm 0.013$
	 \\
$\Omega_{m }$
	 & $0.3083(0.3089)\pm 0.0054$
	 & $0.3011(0.3018)\pm 0.0050$
	 & $0.3074(0.3078)\pm 0.0050$
	 & $0.2983(0.2983)\pm 0.0047$
	 & $0.3096(0.3107)^{+0.0047}_{-0.0053}$
	 & $0.2994(0.3014)\pm 0.0047$
	 \\
  \hline
   \hline
    $\Delta \chi^2_{\rm min}$ & -3.9 & -28.3 & -1.4 & -24.9  & -1.3 & -27.8\\
    $\Delta$AIC & +2.1 & -22.3 & +4.6 & -18.9  & +4.7 & -21.8\\
    \hline
    $Q_{\rm DMAP}$&\multicolumn{2}{|c|}{2.6$\sigma$} &\multicolumn{2}{|c|}{2.9$\sigma$}& \multicolumn{2}{|c|}{3.6$\sigma$}\\
    \hline
    \hline
    $Q_{\rm DMAP}(\rm EDE)$&\multicolumn{2}{|c|}{1.5$\sigma$} &\multicolumn{2}{|c|}{2.4$\sigma$}& \multicolumn{2}{|c|}{2.5$\sigma$}\\
    $Q_{\rm DMAP}(\Lambda \rm CDM)$&\multicolumn{2}{|c|}{5.6$\sigma$} &\multicolumn{2}{|c|}{5.6$\sigma$}& \multicolumn{2}{|c|}{6.3$\sigma$}\\
    \hline

    \end{tabular}}
    \caption{ Mean (best-fit) $\pm 1\sigma$ (or $2\sigma$ for one-sided bounds) of reconstructed parameters in the ADE model confronted to various datasets.
    All datasets include \textit{Planck} + ext-BAO data, while we consider either the BAO/$f\sigma_8$ information or the EFT full-shape analysis for the BOSS and eBOSS data, and we consider either the Pantheon data or the Pantheon+ data (with and without the $M_b$ prior).
    We also display for all datasets the $\Delta\chi^2_{\rm min}$ with respect to $\Lambda$CDM, the associated $\Delta$AIC, as well as the $Q_{\rm DMAP}$. 
    Finally, $Q_{\rm DMAP}(\Lambda \rm CDM)$ and $Q_{\rm DMAP}(\rm EDE)$ corresponds to the $Q_{\rm DMAP}$ of the $\Lambda$CDM and axion-like EDE models for the equivalent datasets.
    }
    \label{tab:cosmoparam}
\end{table*}

In this section, we discuss the cosmological constraints of the ADE and axion-like EDE models and their ability to solve the Hubble tension by successively evaluating the impact of the EFT full-shape analysis of the BOSS and eBOSS data (compared with the standard BAO/$f\sigma_8$ analysis) and the impact of the new Pantheon data (compared with the equivalent older data) on these models.
The cosmological constraints for the ADE model are shown in Tab.~\ref{tab:cosmoparam}, while the $\chi^2_{\rm min}$ values associated with each likelihood are presented in Tab.~\ref{tab:chi2} of App.~\ref{app:chi2}.
In Tab.~\ref{tab:cosmoparam}, we also display the $\Delta\chi^2_{\rm min}$ and the associated $\Delta$AIC with respect to $\Lambda$CDM, as well as the $Q_{\rm DMAP}$ for several combinations of data.\\

Our baseline combination of data, denoted ``BAO/$f\sigma_8$ + Pan,'' refers to \textit{Planck} + ext-BAO + BOSS BAO/$f\sigma_8$ + eBOSS BAO/$f\sigma_8$ + Pantheon, corresponding roughly to that used in Ref.~\cite{Lin:2020jcb}.~\footnote{Note that this analysis used another S$H_0$ES prior, $H_0 = 74.03 \pm 1.42$ km/s/Mpc, from Ref.~\cite{Riess:2019cxk}, and does not take into account the redshift space distortion information (but only the BAO).} For this analysis, combined with the $M_b$ prior, we find $f_{\rm ADE}(z_c) = 0.081\pm 0.018$ and $H_0 = 71.24\pm 0.68$ km/s/Mpc for the ADE model, leading to a residual Hubble tension of $Q_{\rm DMAP} = 2.6 \sigma$ and a preference over $\Lambda$CDM of $\Delta {\rm AIC} = -22.3$ (see Tab.~\ref{tab:cosmoparam}).
Note that this $\chi^2$ improvement is mainly driven by the S$H_0$ES data (as is also the case in the remainder of this paper), implying that this preference over $\Lambda$CDM will no longer be significant if the Hubble tension is due to a systematic error in the data.
Let us underline that with our baseline combination of data, the ADE model satisfies both Ref.~\cite{Schoneberg:2021qvd} conditions.
In addition, we find for the ADE model that ${\rm GT} = 3.7 \sigma$ for our original combination of data.
We are now assessing how the EFTofLSS on the one hand, and the new data from Pantheon+ on the other hand, change these conclusions.

\begin{figure*}
    \centering
    \includegraphics[width=1.3\columnwidth]{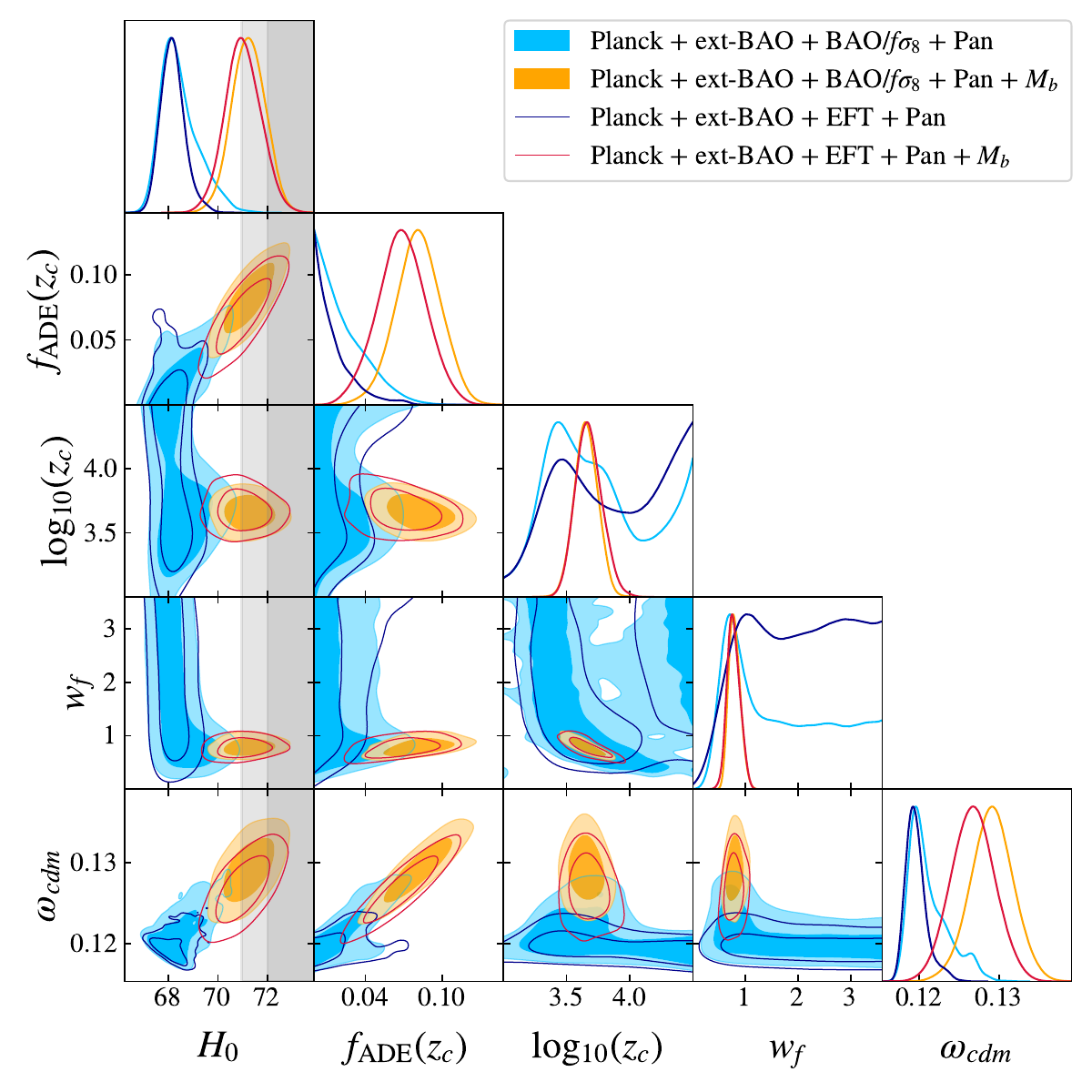}
    \includegraphics[width=1.3\columnwidth]{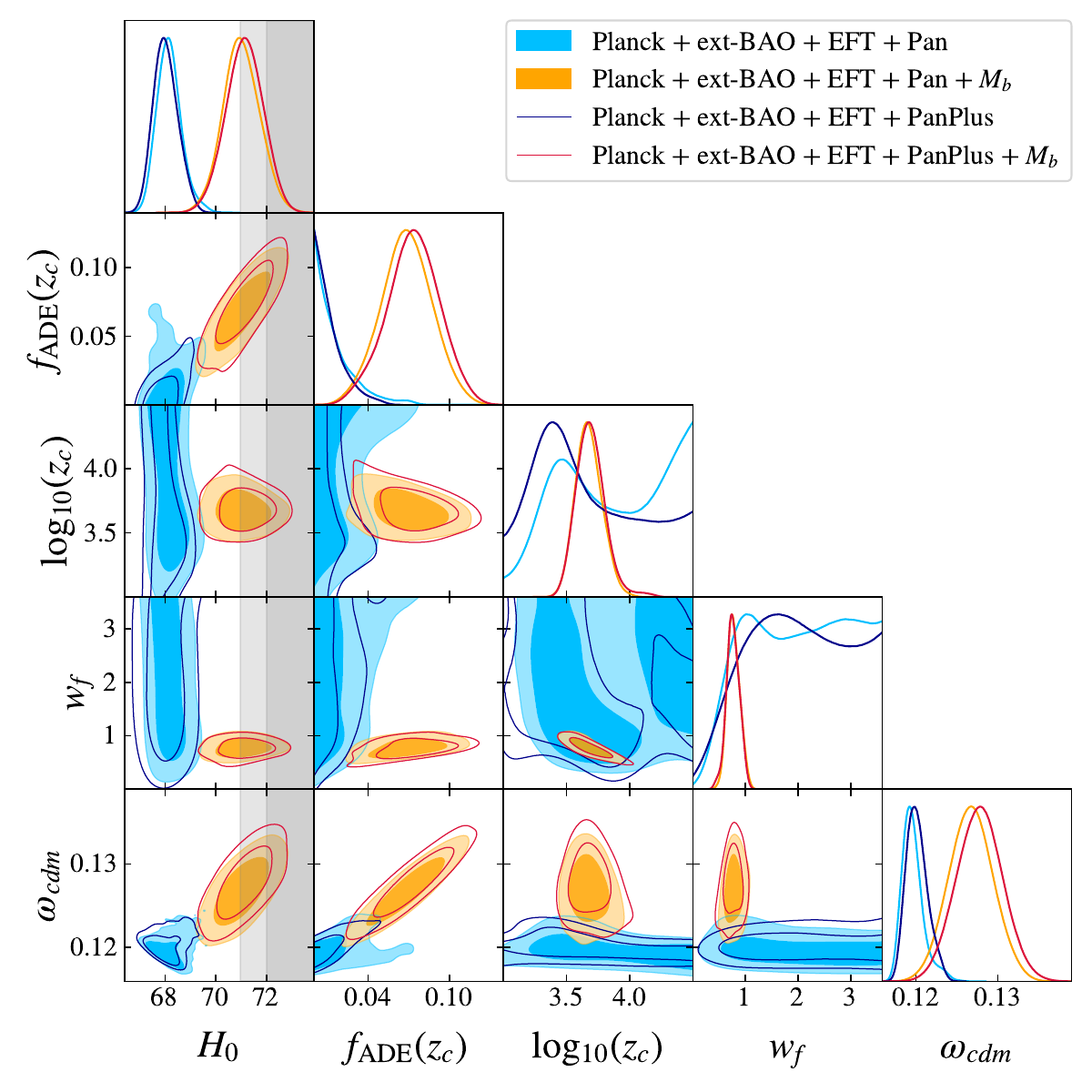}
    \caption{\textit{Top panel}: 2D posterior distributions reconstructed from the BAO/$f\sigma_8$ + Pan dataset compared with the 2D posterior distributions reconstructed from the EFT + Pan dataset, either with or without the $M_b$ prior.
    \textit{Bottom panel}: 2D posterior distributions reconstructed from the EFT + Pan dataset compared with the 2D posterior distributions reconstructed from the EFT + PanPlus dataset, either with or without the $M_b$ prior.
    The gray bands correspond to the $H_0$ constraint associated with the $M_b$ prior, namely, $H_0 = (73.04\pm1.04)$ km/s/Mpc~\cite{Riess:2021jrx}.}
    \label{fig:results}
\end{figure*}

\subsection{Impact of the EFTofLSS analysis}

In the top panel of Fig.~\ref{fig:results}, we show the reconstructed 2D posteriors of the ADE model for the analysis with the BOSS and eBOSS BAO/$f\sigma_8$ likelihoods (namely, the BAO/$f\sigma_8$ + Pan analysis), as well as for the analysis with the EFTofBOSS and EFTofeBOSS likelihoods (namely, the EFT + Pan analysis), either with or without the $M_b$ prior. To isolate the effect of the EFT full-shape analysis, we carry out these analyses using only the older Pantheon data.\\

For the analyses without the $M_b$ prior, the addition of the EFT likelihood has a non-negligible impact on the $f_{\rm ADE}(z_c)$, $w_f$, and $H_0$ constraints. The upper bound of the ADE fractional energy density and the lower bound of $w_f$ are indeed both improved by $\sim 20 \%$, while the standard deviation of $H_0$ is reduced by $\sim 35 \%$. \\

When we consider the $M_b$ prior, EFTofBOSS and EFTofeBOSS do not improve the parameter constraints of this model over the BAO/$f\sigma_8$ information.
However, these likelihoods shift $f_{\rm ADE}(z_c)$ and $H_0$ toward smaller values of $0.7\sigma$ and $0.3\sigma$,~\footnote{Since we are considering here the same experiments (with different methods for extracting cosmological constraints), we use the following metric: $2\cdot(\theta_i - \theta_j)/(\sigma_{\theta, i} + \sigma_{\theta, j})$, where $\theta_i$ and $\sigma_{\theta, i}$ are, respectively, the mean value and the standard deviation of the parameter $\theta$ for the dataset $i$.} respectively.
The EFT full-shape analysis of the BOSS and eBOSS data therefore slightly reduces the ability of this model to resolve the Hubble tension, and the $Q_{\rm DMAP}$ changes from $2.6\sigma$ to $2.9\sigma$ when EFT likelihoods are considered (see Tab.~\ref{tab:cosmoparam}). In particular, the $\chi^2_{\rm min}$ associated with the $M_b$ prior is degraded by $1.0$ compared to the BAO/$f\sigma_8$ analysis.
In addition, the preference for this model over the $\Lambda$CDM model is slightly reduced, given that the $\Delta {\rm AIC}$ changes from $-22.3$ to $-18.9$ when the EFT likelihood is added (see Tab.~\ref{tab:cosmoparam}).
Note that at this point, the ADE model still satisfies both conditions of Ref.~\cite{Schoneberg:2021qvd}, even though $Q_{\rm DMAP} \sim 3\sigma$.
However, the Gaussian tension changes from $3.7\sigma$ to $4.3\sigma$ when EFT likelihoods are considered, which can be explained by the fact that the $f_{\rm ADE}(z_c)$ parameter is better constrained by the EFT + Pan dataset.\\

For the axion-like EDE case, we find for the equivalent analyses (see Tab.~\ref{tab:chi2} of App.~\ref{app:chi2}) that the $Q_{\rm DMAP}$ changes from $1.5\sigma$ to $2.4\sigma$, and that the $\Delta {\rm AIC}$ changes from $-29.1$ to $-22.9$, when EFT likelihoods are added.~\footnote{The similar analysis in Ref.~\cite{Simon:2022adh}, which does not include the eBOSS data, determined that $Q_{\rm DMAP} = 2.0 \sigma$ for the BAO/$f\sigma_8$ + Pan analysis and that $Q_{\rm DMAP} = 2.1 \sigma$ for the EFT + Pan analysis (see Tab.~8 in Ref.~\cite{Simon:2022adh}).
This difference is due solely to the eBOSS data: the $\chi^2$ of the eBOSS BAO/$f\sigma_8$ likelihood is improved when the $M_b$ prior is added (which decreases the $Q_{\rm DMAP}$ of the BAO/$f\sigma_8$ + Pan analysis), while the $\chi^2$ is degraded for the EFTofeBOSS likelihood when the $M_b$ prior is added (which increases the $Q_{\rm DMAP}$ of the EFT + Pan analysis).}
The ADE model slightly better supports the addition of the EFT likelihood compared to the EDE model, insofar as the $Q_{\rm DMAP}$ and $\Delta {\rm AIC}$ are more stable (see Tab.~\ref{tab:cosmoparam}).
However, the EDE model remains a better model to solve the Hubble tension, with $Q_{\rm DMAP}=2.4\sigma$ for the EFT + Pan analysis, compared to $Q_{\rm DMAP}=2.9\sigma$ for the ADE model, and has a better fit to the data when the $M_b$ prior is added, with $\Delta {\rm AIC}= -22.9$, compared to $\Delta {\rm AIC}= -18.9$ for the ADE model.
For a detailed discussion of the EFTofLSS impact on the EDE model in the framework of the BOSS data, please refer to Ref.~\cite{Simon:2022adh}.~\footnote{Note that Ref~\cite{Simon:2022adh} used an $H_0$ prior equivalent to the $M_b$ prior, and did not consider the EFTofeBOSS likelihood (as well as the eBOSS BAO/$f\sigma_8$ likelihood). We leave a detailed evaluation of the impact of eBOSS data on the EDE model for future work.}

\subsection{Impact of the Pantheon+ data}

Let us now turn to the impact of the latest Pantheon data, namely, the Pantheon+ data, on the ability of these models to resolve the Hubble tension. In the bottom panel of Fig.~\ref{fig:results}, we show the reconstructed 2D posteriors of the ADE model for the analyses with the old Pantheon data (\textit{i.e.}, the EFT + Pan analysis), as well as for the analyses with the updated data (\textit{i.e.}, the EFT + PanPlus analysis). To isolate the effect of the Pantheon+ data, we carry out these analyses using only the EFT full-shape analysis of the BOSS and eBOSS data.\\

The analysis with the Pantheon+ data, but without any S$H_0$ES prior, improves significantly the $95 \%$ C.L. constraints on $f_{\rm ADE}(z_c)$ by $\sim 30\%$.
This implies that $H_0$ is shifted down by $0.2\sigma$~\footnote{Since we are considering here different experiments, we use the following metric: $(\theta_i - \theta_j)/\sqrt{\sigma_{\theta, i}^2 + \sigma_{\theta, j}^2}$, where $\theta_i$ and $\sigma_{\theta, i}$ are, respectively, the mean value and the standard deviation of the parameter $\theta$ for the dataset $i$.} compared to the analysis with the old Pantheon data.
Although the ADE model prefers a higher value of  $\omega_{\rm cdm}$  than $\Lambda$CDM (because ADE slows down the evolution
of the growing modes), the larger $\Omega_m$ favored by the Pantheon+ data ($\Omega_m = 0.334 \pm 0.018$~\cite{Brout:2022vxf}) leads to a large $\omega_{\rm cdm} = \Omega_{\rm cdm} \cdot h^2$ which is not sufficiently compensated for by ADE.
Then, to offset the high value of $\Omega_m$, the current Hubble parameter decreases slightly, as well as $f_{\rm ADE}(z_c)$, since the latter is positively correlated with $H_0$.\\

When the $M_b$ prior is included, nonzero contributions of ADE are favored. One may have expected that the tighter constraints from Pantheon+ may reduce the contribution of $f_{\rm ADE}$ and the value of $H_0$. These are in fact stable when compared to analyses with the older Pantheon data, with similar error bars between the EFT + Pan + $M_b$ and EFT + PanPlus + $M_b$ analyses.
Thus, if we rely solely on the posterior distributions, we could argue that the Pantheon+ data do not change the conclusion about the ADE resolution of the Hubble tension.
However, it turns out that the ADE model is not able to accommodate at the same time the large values of $H_0$ and $\Omega_m$ that are favored by the Pantheon+ data once they are calibrated with $M_b$.
Indeed, the best-fit value $H_0 = 71.29$ km/s/Mpc is $1.7\sigma$ lower than the S$H_0$ES constraint [$H_0 = (73.04\pm1.04)$ km/s/Mpc], while the best-fit value $\Omega_m = 0.3014$ is $1.8\sigma$ lower than the Pantheon+ constraint ($\Omega_m = 0.334 \pm 0.018$).
Therefore, the ADE model does not provide a good fit to the $M_b$ prior ($\chi^2_{M_b}= 6.42$ as shown in Tab.~\ref{tab:chi2} of App.~\ref{app:chi2}), while the fit to the Pantheon+ data is worse (by +1.6) with the inclusion of the $M_b$ prior.
These degradations of $\chi^2_{\rm min}$~\footnote{Let us note that the $\chi^2_{\rm min}$ of the other likelihoods are stable between the Pantheon and Pantheon+ analyses, and therefore play no role in the change in $Q_{\rm DMAP}$ between these two analyses.} imply that the $Q_{\rm DMAP}$ changes from $2.9\sigma$ ($5.6\sigma$ for $\Lambda$CDM) to $3.6\sigma$ ($6.3\sigma$ for $\Lambda$CDM) when we consider the Pantheon+ data (see Tab.~\ref{tab:cosmoparam}), which severely limits the ability of this model to resolve the $H_0$ tension.
One of the two criteria of Ref.~\cite{Schoneberg:2021qvd}, namely, $Q_{\rm DMAP} < 3\sigma$, is indeed no longer fulfilled.
However, while the Pantheon+ data and the $M_b$ prior from Ref.~\cite{Riess:2021jrx} seriously restrict the ability of the ADE model to resolve the Hubble tension, these data improve the preference for this model over $\Lambda$CDM, since the $\Delta {\rm AIC}$ changes from $-18.9$ to $-21.8$. We nevertheless caution overinterpreting this preference, given that the $Q_{\rm DMAP}$ indicates that  combining these datasets is not statistically consistent.
In addition, the Gaussian tension ${\rm GT = 4.4\sigma}$ is stable with respect to the EFT + Pan dataset.~\footnote{Note that for the same dataset, we obtain ${\rm GT = 3.8\sigma}$ for the axion-like EDE model and ${\rm GT = 4.8\sigma}$ for the $\Lambda$CDM model.}\\

\begin{figure*}
    \centering
    \includegraphics[width=2.25\columnwidth]{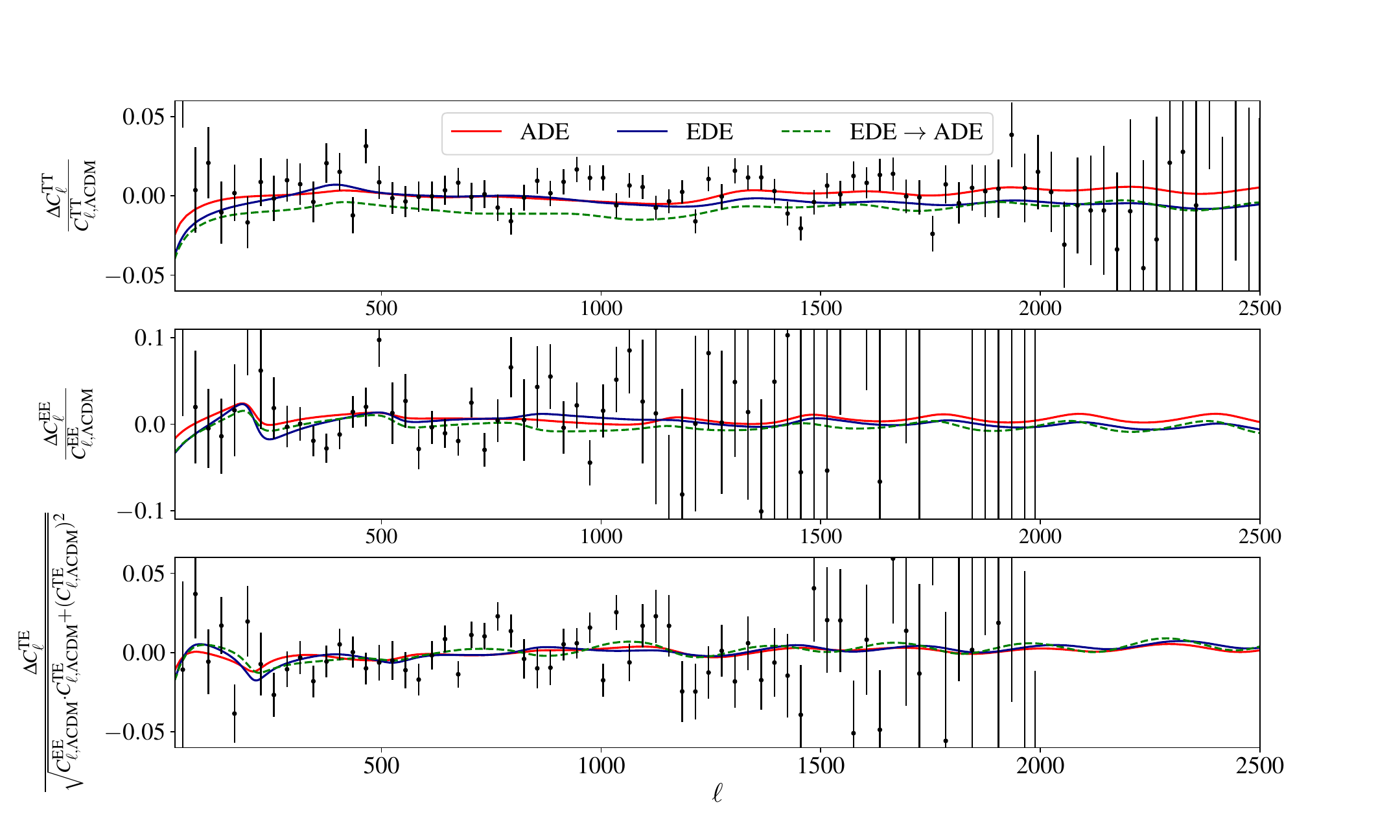}
    \caption{CMB power spectra residuals with respect to $\Lambda$CDM for the ADE (red) and axion-like EDE (black) models.
    All cosmological parameters of the $\Lambda$CDM, ADE, and axion-like EDE models have been set to their EFT + PanPlus  best-fits, while the displayed data (normalized to the $\Lambda$CDM best-fit) correspond to the \textit{Planck} 2018 data~\cite{Planck:2019nip}.
    Finally, for the plot titled ``EDE $\to$ ADE,'' we set the $\Lambda$CDM parameters to the axion-like EDE best-fit, while the $z_c$ and $w_f$ parameters are set to the ADE best-fit.
    The last ADE parameter, namely, $f_{\rm ADE}(z_c) = 0.095$, is determined such that the values of $100\theta_s = 1.042$ and $r_s = 140.53$ Mpc are the same as for the EDE best-fit.}
    \label{fig:CMB_power_spectra}
\end{figure*}

\begin{figure*}
    \centering
    \includegraphics[width=1\columnwidth]{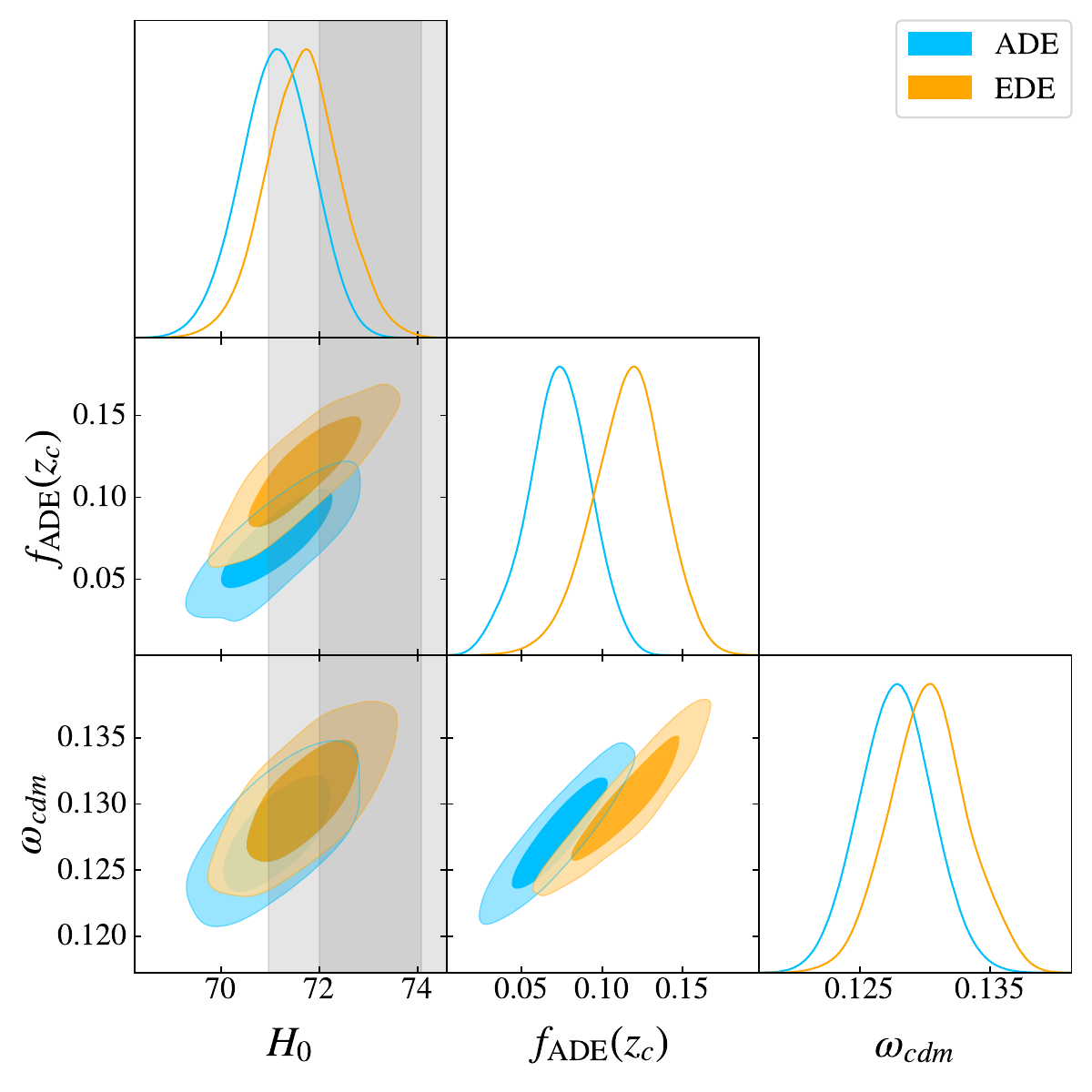}
    \includegraphics[width=1\columnwidth]{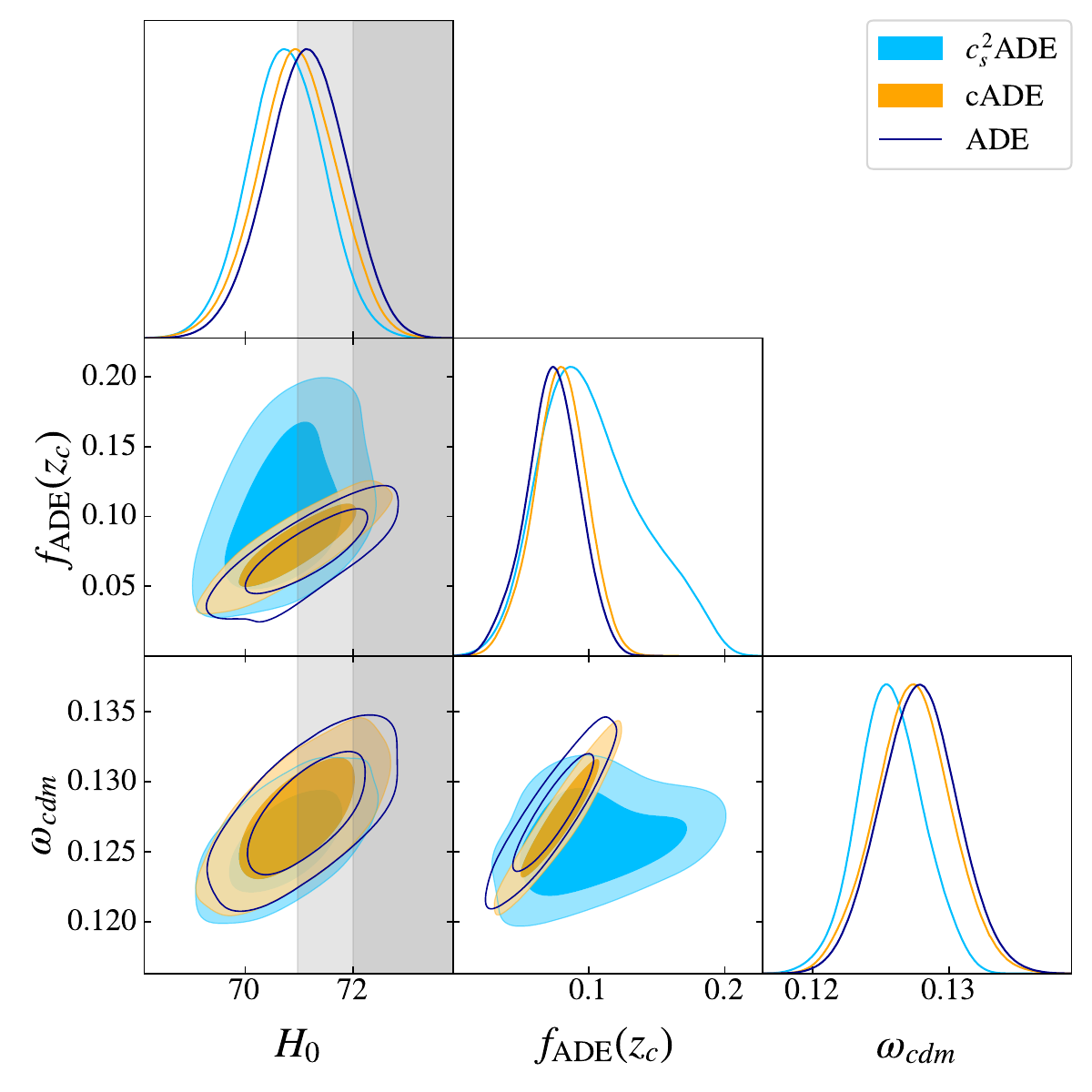}
    \caption{\textit{Left panel}: 2D posterior distributions reconstructed from the \textit{Planck} + ext-BAO + EFT + PanPlus + $M_b$ dataset, \textit{i.e.}, the most up-to-date dataset, for our baseline ADE model and the standard axion-like EDE model.
    \textit{Right panel}: 2D posterior distributions reconstructed from the \textit{Planck} + ext-BAO + EFT + PanPlus + $M_b$ dataset for the $c_s^2$ADE model (namely, our baseline ADE model with the variation of $c_s^2$), the cADE model (namely, our baseline ADE model with $c_s^2 = w_f = 1$), and our baseline ADE model.
    The gray bands correspond to the $H_0$ constraint associated with the $M_b$ prior, namely, $H_0 = (73.04\pm1.04)$ km/s/Mpc~\cite{Riess:2021jrx}.}
    \label{fig:results_extensions}
\end{figure*}

In the left panel of Fig.~\ref{fig:results_extensions}, we show the 2D posterior distributions of the axion-like EDE model reconstructed from the EFT + PanPlus + $M_b$ dataset, while the associated cosmological constraints are displayed in Tab.~\ref{tab:cosmoparam_extensions}.
For the axion-like EDE case, we find that the $Q_{\rm DMAP}$ changes from $2.4\sigma$ to $2.5\sigma$, and that the $\Delta {\rm AIC}$ changes from $-22.9$ to $-29.1$, between the old and the new Pantheon data analysis (see Tab.~\ref{tab:chi2} of App.~\ref{app:chi2} for the individual $\chi^2_{\rm min}$).
This model better supports these new data, since the $Q_{\rm DMAP}$ is stable (and especially the $\chi^2_{\rm min}$ of the S$H_0$ES prior), while the $\Delta {\rm AIC}$, as in the case of the ADE model, decreases significantly.
Whereas with the addition of the EFT data we had a slight preference for EDE over ADE, with the Pantheon+ data the preference for this model becomes clearly apparent: in the axion-like EDE model, $H_0 = 71.67 \pm 0.77$ km/s/Mpc with $Q_{\rm DMAP}=2.5\sigma$, while in the ADE model, $H_0 = 71.13 \pm 0.73$ km/s/Mpc with $Q_{\rm DMAP}=3.6\sigma$. In addition, the axion-like EDE model provides a better overall fit than the ADE model, with $\Delta AIC ({\rm EDE - ADE}) = + 7.3$.
The two main contributions to this difference come from the \textit{Planck} data (and in particular the high-$\ell$ TTTEEE likelihood), where $\Delta \chi^2 ({\rm EDE - ADE}) = + 3.7$, and from the S$H_0$ES prior, where $\Delta \chi^2 ({\rm EDE - ADE}) = +2.7$.
The axion-like EDE model is capable of better compensating the effect of large values of $H_0$ and $\Omega_{\rm m}$ (and therefore $\omega_{\rm cdm}$) on the CMB compared to the ADE model.\\

In order to understand why the axion-like EDE model performs better than ADE, we plot in Fig.~\ref{fig:CMB_power_spectra} the CMB power spectra residuals with respect to the $\Lambda$CDM best-fit for these two models.
In this figure, we also plot (in green dashed) the CMB power spectra residuals of the ADE model, where we set the $\Lambda$CDM parameters to the axion-like EDE best-fit, and the $z_c$ and $w_f$ parameters to the ADE best-fit.
The last ADE parameter, namely, $f_{\rm ADE}(z_c)$, is determined such that the values of the angular acoustic scale at recombination $\theta_*$ and the comoving sound horizon at recombination $r_*$ are the same as for the EDE best-fit.
In other words, this plot would represent the best-fit of the ADE model if the latter could reduce the Hubble tension to the same level as the axion-like EDE model.
In this figure, the main difference between the ADE and ADE $\to$ EDE plots stems from the suppression (particularly at low $\ell$) of the $C_{\ell}^{\rm TT}$ power spectrum for the EDE $\to$ ADE analysis.
This suppression typically corresponds to the effect of a large value of $\omega_{\rm cdm}$ (and also $n_s$), showing that the ADE model is not able to compensate for a high value of $\Omega_{\rm cdm}h^2$ in the same way as the axion-like EDE model.
This is explained by the fact that the EDE model allows the sound speed to decrease in the $k$ range associated with $\ell< 500$, making it easier to compensate for the effect of increasing $\Omega_{\rm cdm}h^2$ in the low-$\ell$ TT power spectrum.
Let us note that the effect of the increase in $\Omega_{\rm cdm}h^2$ is more significant for the modes that have reentered the horizon at the time when $f_{\rm ADE}$ is decreasing, and therefore no longer significantly suppresses the evolution of the growing modes.
In order to compensate for this effect, it is therefore helpful to decrease $c_s^2$ for $l<500$, insofar as a reduction in this parameter leads to an enhancement in the Weyl potential (see Ref.~\cite{Lin:2019qug}).
Note that these results are compatible with Ref.~\cite{Lin:2019qug}, but interestingly the limitation in the value of $\Omega_{\rm cdm}h^2$ does not arise from the CMB polarization as in that reference (which considered \textit{Planck} 2015 data), but from the CMB temperature.

\section{Model variations}
\label{sec:discussion}

\begin{table*}[]
    \centering
    \scalebox{0.8}{
    \begin{tabular}{|l|c|c|c|}
    
    \hline
      & EDE  & $c_s^2$ADE & cADE \\
              \hline
              \hline

$f_{\rm ADE/EDE}(z_c)$
	 & $0.116(0.128)^{+0.023}_{-0.021}$
	 & $0.103(0.080)^{+0.028}_{-0.046}$
  & $0.079(0.087)\pm 0.019$

	 \\
$\log_{10}(z_c)$
	 & $3.69(3.84)^{+0.20}_{-0.16}$
	 & $3.61(3.73)^{+0.12}_{-0.10}$
  & $3.540(3.532)\pm 0.058$

	 \\
$\Theta_i$
	 & $2.77(2.88)^{+0.15}_{-0.072}$
	 & --
  & --

	 \\
$w_f$
	 & --
	 & $unconst.(0.71)$
  & --

	 \\
$c_s^2$
	 & --
	 & $> 0.701(0.72)$
  & --

	 \\
	 \hline
	 $H_0$
	 & $71.67(71.84)\pm 0.77$
	 & $70.76(71.23)\pm 0.70$
  & $70.95(71.23)\pm 0.73$

	 \\ 
$\omega_{\rm cdm }$
	 & $0.1303(0.1309)\pm 0.0030$
	 & $0.1257(0.1294)\pm 0.0024$
  & $0.1273(0.1286)\pm 0.0028$

	 \\
$10^{2}\omega_{b }$
	 & $2.294(2.312)\pm 0.024$
	 & $2.304(2.310)\pm 0.020$
  & $2.305(2.308)\pm 0.021$

	 \\
$10^{9}A_{s }$
	 & $2.149(2.143)^{+0.027}_{-0.034}$
	 & $2.152(2.150)^{+0.030}_{-0.036}$
  & $2.149(2.158)\pm 0.031$

	 \\
$n_{s }$
	 & $0.9898(0.9951)\pm 0.0061$
	 & $0.9882(0.9902)\pm 0.0065$
  & $0.9849(0.9874)\pm 0.0057$

	 \\
$\tau_{\rm reio }$
	 & $0.0590(0.0590)^{+0.0063}_{-0.0079}$
	 & $0.0592(0.0573)^{+0.0069}_{-0.0080}$
  & $0.0573(0.0583)^{+0.0066}_{-0.0078}$

	 \\
	 \hline
	 $S_8$
	 & $0.836(0.836)\pm 0.011$
	 & $0.831(0.842)\pm 0.012$
  & $0.836(0.841)\pm 0.012$

	 \\
$\Omega_{m }$
	 & $0.2995(0.2997)\pm 0.0047$
	 & $0.2985(0.3018)\pm 0.0049$
  & $0.3000(0.3001)\pm 0.0047$

	 \\
  \hline
   \hline
    $\Delta \chi^2_{\rm min}$ & $-35.1$ & $-27.9$ & $-24.1$\\
    $\Delta$AIC & $-29.1$ & $-19.9$ & $-20.1$ \\
    \hline
    $Q_{\rm DMAP}$ & $2.5\sigma$ & $3.6\sigma$ &  $3.9\sigma$ \\
    \hline

    \end{tabular}}
    \caption{ Mean (best-fit) $\pm 1\sigma$ (or $2\sigma$ for one-sided bounds) of reconstructed parameters in the EDE, $c_s^2$ADE, and cADE models confronted to the \textit{Planck} + ext-BAO + EFT + PanPlus + $M_b$ dataset, \textit{i.e.}, the most up-to-date dataset.
    We also display for each model the $\Delta\chi^2_{\rm min}$ with respect to $\Lambda$CDM, the associated $\Delta$AIC, as well as the $Q_{\rm DMAP}$.
    }
    \label{tab:cosmoparam_extensions}
\end{table*}

\subsection{Variation of $c_s^2$}

\begin{figure}
    \centering
    \includegraphics[width=1\columnwidth]{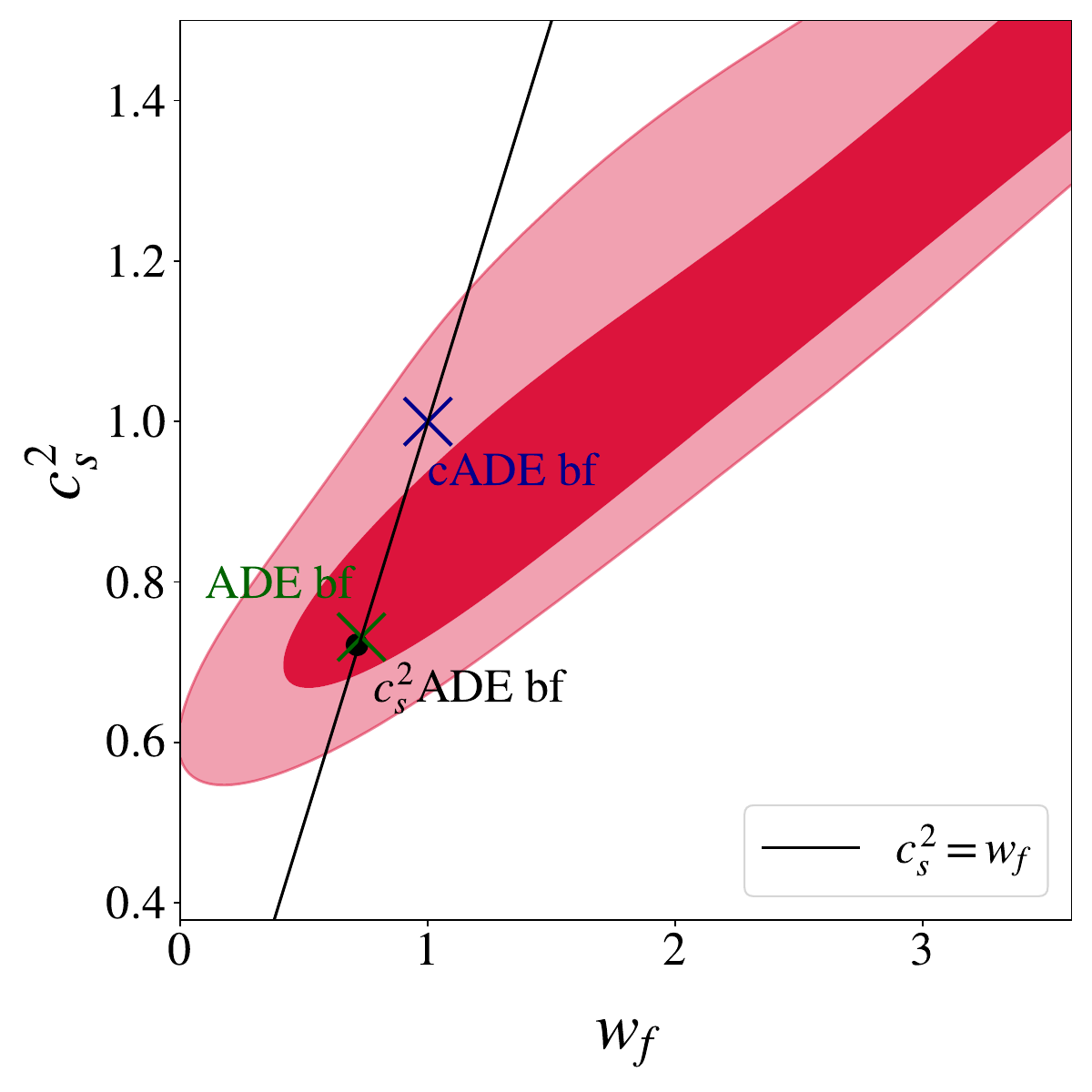}
    \caption{2D posterior distribution of the $c_s^2 - w_f$ plane reconstructed from the \textit{Planck} + ext-BAO + EFT + PanPlus + $M_b$ dataset for the $c_s^2$ADE model.
    The solid line corresponds to $c_s^2 = w_f$, while the blue and green crosses correspond, respectively, to the cADE model and the best-fit of our baseline ADE model.
    The black circle represents the best-fit of the $c_s^2$ADE model.}
    \label{fig:cs2}
\end{figure}

In the previous sections, we fixed $c_s^2(a_c) = w_f$ instead of varying these two parameters independently.
In the right panel of Fig.~\ref{fig:results_extensions}, we show the 2D posterior distributions reconstructed from the EFT + PanPlus + $M_b$ dataset for our baseline ADE model by relaxing this assumption, while in Tab.~\ref{tab:cosmoparam_extensions} we display the associated cosmological constraints.
To do so, we have applied the prior of Refs.~\cite{Lin:2019qug,Lin:2020jcb} to $c_s^2$, namely,
\begin{align*}
    &0 \le c_s^2 \le 1.5 \, .
\end{align*}
In the following, we simply call this extended model ``$c_s^2$ADE,'' for which we still consider that $p=1$.
Interestingly, and in line with Ref.~\cite{Lin:2019qug}, the assumption $c_s^2 = w_f$ does not change our conclusions, especially regarding the Hubble tension: we obtain $Q_{\rm DMAP} = 3.6 \sigma$, which is similar to that of our baseline ADE model (see Tab.~\ref{tab:chi2} of App.~\ref{app:chi2} for the $\chi^2$ values). 
In this specific case, we obtain $H_0 = 70.76\pm 0.70$ km/s/Mpc, which is $0.5 \sigma$ lower than the $H_0$ value from our baseline ADE model. This is due to projection effects caused by the non-Gaussian posteriors of $c_s^2$ and $w_f$, and we notice that the best-fit value ($H_0 = 71.23$ km/s/Mpc) is very close to that of the ADE model.
Thus, the relaxation of this hypothesis does not resolve the Hubble tension, while the $\Delta AIC$ worsens somewhat in this model because of the additional parameter ($\Delta AIC = -20.1$ instead of $-21.8$ for our baseline ADE model).
In addition, as shown in Fig.~\ref{fig:cs2}, the best-fit point of the ADE model in the $c_s^2 - w_f$ plane lies in the $68\%$ C.L. reconstructed from the $c_s^2$ADE model, and is very close to the best-fit point of this model. This implies that setting $c_s^2 = w_f$ is a good approximation around the best-fit of the $c_s^2$ADE model.

\subsection{The cADE model}

Refs.~\cite{Lin:2019qug} and~\cite{Lin:2020jcb} showed that the special case where $c_s^2=w_f=1$ made it possible to solve the Hubble tension. In this particular model, called ``cADE,'' the ADE component is a canonical scalar which goes from a frozen phase ($w=-1$) to a kinetion phase ($w=1$) around matter-radiation equality.
This model is particularly interesting because it allows the Hubble tension to be resolved with only two more parameters than the $\Lambda$CDM model [namely, $f_{\rm ADE} (z_c)$ and $\log_{10}(z_c)$].
However, while in Ref.~\cite{Lin:2019qug} the case $c_s^2=w_f=1$ is within the $68\%$ C.L. of the $c_s^2$ and $w_f$ parameters (see Fig.~1 of this reference), one can see in Fig.~\ref{fig:cs2} that this particular case is no longer located in the $1\sigma$ region.~\footnote{Let us note that Refs.~\cite{Lin:2019qug,Lin:2020jcb} set $p=1/2$, while we set $p=1$, but this difference does not change the results.}
In the right panel of Fig.~\ref{fig:results_extensions}, we display the 2D posterior distributions of the cADE model reconstructed from the EFT + PanPlus + $M_b$ dataset,  while in Tab.~\ref{tab:cosmoparam_extensions} we display the associated cosmological constraints.
We can clearly see that this particular model is unable to resolve the Hubble tension with current data, since we obtain $H_0 = 70.95\pm 0.73$ km/s/Mpc and $f_{\rm ADE}(z_c) = 0.079\pm 0.019$, with a $Q_{\rm DMAP} = 3.9\sigma$ (compared to $Q_{\rm DMAP} = 3.6\sigma$ for our baseline ADE model).

\section{Conclusion}

In this paper, we have updated the constraints on the acoustic dark energy and axion-like early dark energy models by first assessing the impact of the EFT full-shape analysis applied to the BOSS LRG and eBOSS QSO data, and second the impact of the latest Pantheon+ data.
\begin{itemize}
    \item When we consider the full-shape analysis of the BOSS and eBOSS data, combined with \textit{Planck}, ext-BAO measurements, Pantheon data from~\cite{Scolnic:2017caz}, and S$H_0$ES data from~\cite{Riess:2021jrx}, we obtain $H_0 = 71.01 \pm 0.73$ km/s/Mpc with a residual Hubble tension of $2.9\sigma$ (compared to $2.4\sigma$ for the axion-like EDE model and $5.6\sigma$ for the $\Lambda$CDM model).
    \item We have demonstrated that the EFTofLSS analysis slightly reduces the ability of this model to resolve the Hubble tension compared to the BAO/$f\sigma_8$ analysis, which has a residual tension of $2.6\sigma$ (with $H_0 = 71.24 \pm 0.68$ km/s/Mpc). 
    \item  Although the axion-like EDE model remains a better solution to the Hubble tension after using the EFTofBOSS and EFTofeBOSS likelihoods, we have shown that the EFTofLSS analysis has a stronger impact on this model.
    \item Importantly, when we replace the Pantheon data with the Pantheons+ data from~\cite{Brout:2022vxf}, the ADE model no longer resolves the Hubble tension at a suitable level, leading to a $3.6\sigma$ residual tension (compared to $2.5\sigma$ for the EDE model and $6.3\sigma$ for the $\Lambda$CDM model).
\item 
Whereas with the EFTofLSS analysis we had only a slight preference for EDE over ADE, with the new data from Pantheon+ and S$H_0$ES, the preference for this model became clearly apparent, due to the fact that axion-like EDE manages to compensate a higher $\Omega_{\rm cdm} h^2$ in \textit{Planck} data thanks to the scale dependence of the sound speed.\\
\item 
Finally, we have verified that relaxing the assumption $c_s^2 = w_f$ does not alter our conclusions, justifying this choice. 
In addition, for the cADE model (where $c_s^2 = w_f = 1$), we have obtained $H_0 = 70.95 \pm 0.73$ km/s/Mpc with a $Q_{\rm DMAP} = 3.9\sigma$, implying that one can no longer solve the Hubble tension with this constrained ADE model, contrary to previous results~\cite{Lin:2019qug, Lin:2020jcb}.\\

\end{itemize}

Let us add a few words about the $S_8 \equiv \sigma_8 \cdot \sqrt{\Omega_m/0.3}$ tension (see \textit{e.g.}, Ref.~\cite{Abdalla:2022yfr} for a review). EDE-like models are known to slightly increase the amplitude of fluctuations $\sigma_8$ with respect to $\Lambda$CDM~\cite{Poulin:2018cxd,Hill:2020osr,Vagnozzi:2021gjh} due to an increase in $\omega_{\rm cdm}$ and $n_s$.
In particular, increasing $\omega_{\rm cdm}$ brings forward matter-radiation equality $a_{\rm eq}$, leaving more time for growing modes (that are subhorizon at $a_{\rm eq}$) to evolve in the matter era.
Considering our most up-to-date dataset (\textit{i.e.}, EFT+PanPlus+$M_b$), we have obtained a Gaussian tension~\footnote{ We use here the Gaussian metric $(\theta_i - \theta_j)/\sqrt{\sigma_{\theta, i}^2 + \sigma_{\theta, j}^2}$, where $\theta_i$ and $\sigma_{\theta, i}$ are, respectively, the mean value and the standard deviation of the parameter $\theta$ for the dataset $i$. For the week lensing determination of the $S_8$ parameter, we use the simple weighted mean and uncertainty of $S_8^{\rm GT} = 0.766^{+0.020}_{-0.014}$ from the combination of KiDS-1000$\times$dFLensS$+$BOSS, $S_8 = 0.769^{+0.016}_{-0.012}$~\cite{Heymans:2020gsg}, and DES-Y3, $S_8 = 0.775^{+0.026}_{-0.024}$~\cite{DES:2021wwk}.} on $S_8$ of $3.2\sigma$, $3.5\sigma$ and $3.8\sigma$ for the $\Lambda$CDM, ADE and axion-like EDE models, respectively. It is interesting to note that the better the model is able to resolve the Hubble tension, the higher the $S_8$ tension. In order to resolve these two tensions simultaneously in the context of EDE cosmologies, it is therefore necessary to find a mechanism that reduces the growth of small-scale modes, as could be achieved by an interaction between EDE and DM~\cite{Liu:2023haw}.\\

In this paper, we have shown that the new data from Pantheon and S$H_0$ES, and to a lesser extent the EFTofLSS applied to the BOSS and eBOSS data, can have a decisive impact on models which aim to resolve the Hubble tension.
We leave for future work the study of the impact on the Hubble tension of such an analysis applied to other early dark energy models, such as new early dark energy~\cite{Niedermann:2019olb,Niedermann:2020dwg}, rock ``n'' roll dark energy ~\cite{Agrawal:2019lmo}, or early modified gravity~\cite{Braglia:2020auw,FrancoAbellan:2023gec}.

\begin{acknowledgements}

The author would like to warmly thank Vivian Poulin and Tristan L. Smith for their comments and insights throughout the project.
Much of this work was carried out during a four-week visit to the \textit{Center for Theoretical physics} (CTP) at the \textit{Massachusetts Institute of Technology} (MIT).
The author would therefore like to express his gratitude to the members of the CTP, and in particular to Tracy Slatyer, for their hospitality and kindness.
These results have been made possible thanks to LUPM's cloud computing infrastructure founded by Ocevu labex, and France-Grilles.
This project has received support from the European Union’s Horizon 2020 research and innovation program under the Marie Skodowska-Curie grant agreement No 860881-HIDDeN. This project has also received funding from the European Research Council (ERC) under the
European Union’s HORIZON-ERC-2022 (Grant agreement No. 101076865).
\end{acknowledgements}

\appendix

\section{$M_b$ prior}\label{app:Mb_prior}

\begin{figure*}
    \centering
    \includegraphics[width=1.3\columnwidth]{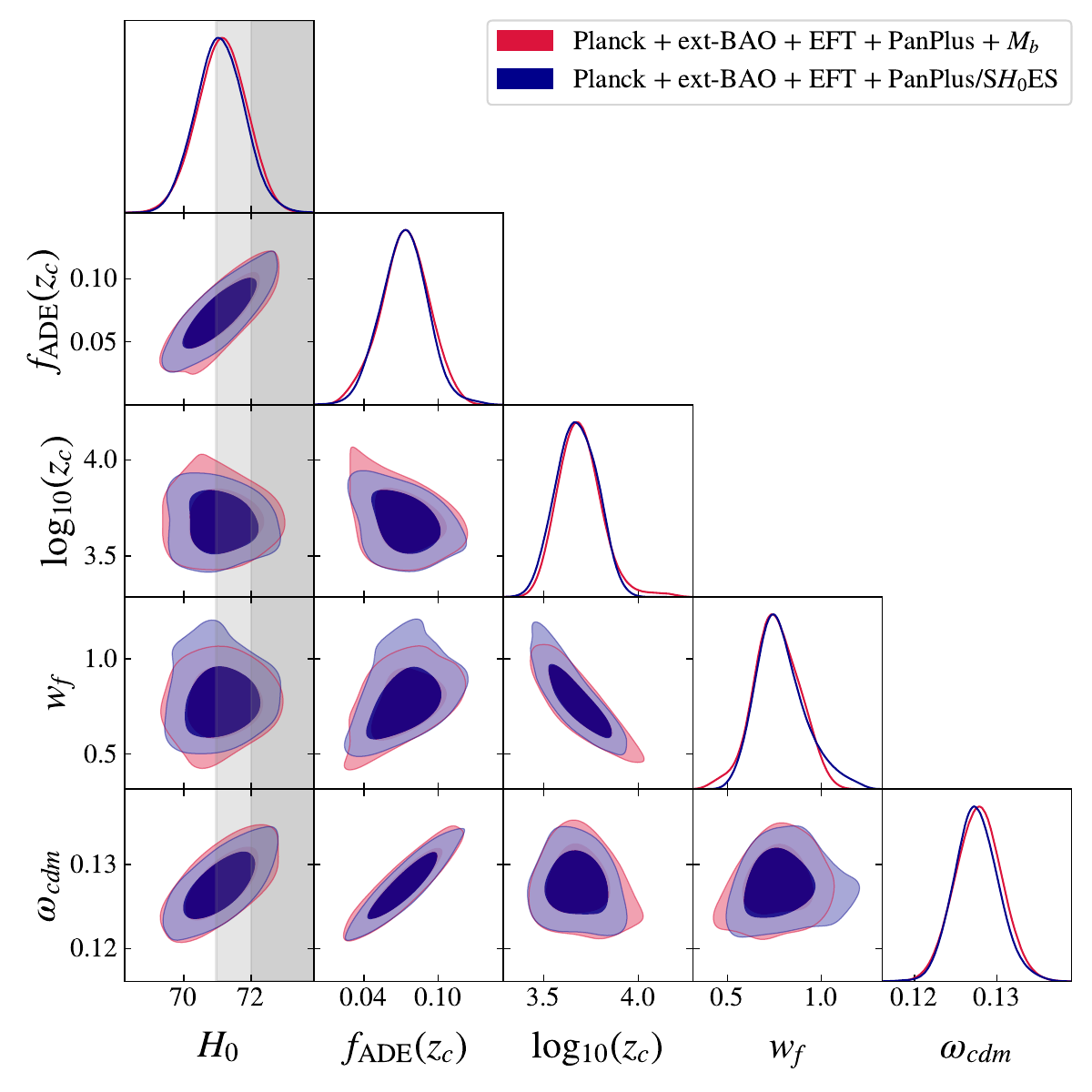}
    \caption{2D posterior distributions reconstructed from \textit{Planck} + ext-BAO + EFT, either with the $M_b$ prior on top of the Pantheon+ likelihood, or with the cross-correlation between the Pantheon+ data and the S$H_0$ES data (namely, the Pantheon+/S$H_0$ES likelihood) as provided in Ref.~\cite{Riess:2021jrx}. The gray band corresponds to the $H_0$ constraint associated with the $M_b$ prior, namely, $H_0 = (73.04\pm1.04)$ km/s/Mpc~\cite{Riess:2021jrx}.}
    \label{fig:Mb_prior}
\end{figure*}

In this appendix, we show explicitly, thanks to Fig~\ref{fig:Mb_prior}, that the addition of the $M_b$ prior~\cite{Riess:2021jrx} on top of the Pantheon+ likelihood is equivalent to the use of the full Pantheon+/S$H_0$ES likelihood as provided in Ref.~\cite{Riess:2021jrx}.
Since the constraints are similar, we have chosen to show in this paper the results with the $M_b$ prior, for the sake of convenience, in order to determine the $Q_{\rm DMAP}$ values easily.

\section{$\chi^2$ table}\label{app:chi2}

In this appendix, we report the best-fit $\chi^2$ per experiment for the $\Lambda$CDM model, the ADE model, as well as the axion-like EDE model for several combinations of data. 

\begin{table*}[]
    \centering
    \begin{tabular}{|c|c|c |c c c c c c c c |}
        \hline
        Data & Model & $\chi^2$ tot & P18TTTEE & P18lens & ext-BAO & BOSS & eBOSS & Pan & $M_b$ & PanPlus/S$H_0$ES \\
        \hline \rule{0pt}{3ex}
        \multirow{3}{*}{BAO/$f\sigma_8$+Pan} & $\Lambda$CDM & 3816.39 & 2763.03 & 8.87 & 1.38 & 6.15 & 9.88 & 1027.07 & -- & --\\
        & ADE & 3812.50 & 2759.48 & 9.14 &  1.30 & 6.55 & 8.92 & 1027.10 & -- & --\\
        & EDE & 3809.87 & 2757.41 & 9.70 &  1.64 & 6.30 & 7.89 & 1026.93 & -- & --\\
        \hline \rule{0pt}{3ex}
        \multirow{3}{*}{BAO/$f\sigma_8$+Pan+$M_b$} & $\Lambda$CDM & 3847.33 & 2765.48 & 9.12 & 1.84 & 5.91 & 9.14 & 1026.89 & 28.94 & --\\
        & ADE & 3819.06 & 2763.73 & 10.09 &  1.77 & 6.95 & 6.32 & 1026.89 & 3.32 & --\\
        & EDE & 3812.26 & 2759.10 & 9.89 &  1.91 & 6.97 & 6.21 & 1026.87 & 1.31 & --\\
        \hline \rule{0pt}{3ex}
        \multirow{3}{*}{EFT+Pan} & $\Lambda$CDM & 4020.07 & 2762.14 & 8.87 &  1.25 &  160.20 & 60.44 & 1027.17 & -- & --\\
        & ADE & 4018.67 & 2761.21 & 8.99 &  1.40 &   159.63 & 60.39 & 1027.06 & -- & --\\
        & EDE & 4017.09 &  2759.20 & 9.29 & 1.60  &  159.54 & 60.51 & 1026.95 & -- & --\\
        \hline \rule{0pt}{3ex}
        \multirow{3}{*}{EFT+Pan+$M_b$}  & $\Lambda$CDM & 4051.76 & 2764.81 & 9.13 & 1.78 &  158.30 & 61.16 & 1026.90 & 29.61 & --\\
        & ADE & 4026.87 &   2763.76 & 9.62 &  2.11 &  159.94 & 60.26 & 1026.86 & 4.33 & --\\
        & EDE & 4022.83 &  2758.51 & 9.60 &  1.99  & 160.36  & 61.64 & 1026.87 & 3.86 & --\\
        \hline \rule{0pt}{3ex}
        \multirow{5}{*}{EFT+PanPlus} & $\Lambda$CDM & 4404.28 & 2762.12 & 8.78 & 1.20 &  160.44 & 60.43 & 1411.31 & -- & --\\
        & cADE & 4404.07 &  2761.57 & 8.86 & 1.20 &   160.68 & 60.46 & 1411.31 & -- & --\\
        & ADE & 4402.96 & 2760.49 & 8.89 & 1.21 & 160.78 & 60.23 & 1411.36 & -- & --\\
        & $c_s^2$ADE & 4402.93 & 2760.47 & 8.90 & 1.21 &  160.76 & 60.23 & 1411.37 & -- & --\\
        & EDE & 4402.54 &  2758.74 & 9.02 & 1.38 &  160.38 & 61.21 & 1411.82 & -- & --\\ 
        \hline \rule{0pt}{3ex}
        \multirow{5}{*}{EFT+PanPlus+$M_b$} & $\Lambda$CDM & 4443.78 &   2766.95  & 9.69 & 1.95 & 158.36 & 60.63 & 1413.17 & 33.02 & --  \\
        & cADE & 4419.67 & 2766.20 & 9.67 & 1.92 &  160.68  & 61.17 & 1413.27 & 6.76 & --  \\
        & ADE & 4415.94 & 2763.42 & 9.82 & 1.81 &  160.52  & 60.95 & 1412.99 & 6.42 & --  \\
        & $c_s^2$ADE & 4415.89 & 2763.16 & 9.82 & 1.77 & 160.58  & 60.96 & 1412.91 & 6.70 & --  \\
        & EDE & 4408.67 &  2759.71 & 10.05 & 1.96 & 159.64 & 60.24 & 1413.35 & 3.72 & --  \\
        \hline \rule{0pt}{3ex}
        \multirow{2}{*}{EFT+PanPlus/S$H_0$ES} & $\Lambda$CDM & 4318.12 &  2767.42 & 9.24 & 2.20 & 158.01 & 60.39 & -- & -- & 1320.85  \\
        & ADE & 4292.12 &  2763.74 & 9.77 & 1.94 &  160.31 & 60.20 & -- & -- & 1296.17 \\
        \hline
    \end{tabular}
    \caption{Table of best-fit $\chi^2$ of the different models considered in this work for various combinations of likelihood.
    All datasets include \textit{Planck} + ext-BAO data.
    Note that the columns ``BOSS'' and ``eBOSS'' refer either to the BAO/$f\sigma_8$ analysis or to the EFT full-shape analysis.
    Similarly, the column ``Pan'' refers to either Pantheon data or Pantheon+ data.
    Finally, ``PanPlus/S$H_0$ES'' corresponds to the full Pantheon+/S$H_0$ES likelihood as provided in Ref.~\cite{Riess:2021jrx}.}
    \label{tab:chi2}
\end{table*}

\bibliography{biblio}

\begin{thebibliography}{121}%
\makeatletter
\providecommand \@ifxundefined [1]{%
 \@ifx{#1\undefined}
}%
\providecommand \@ifnum [1]{%
 \ifnum #1\expandafter \@firstoftwo
 \else \expandafter \@secondoftwo
 \fi
}%
\providecommand \@ifx [1]{%
 \ifx #1\expandafter \@firstoftwo
 \else \expandafter \@secondoftwo
 \fi
}%
\providecommand \natexlab [1]{#1}%
\providecommand \enquote  [1]{``#1''}%
\providecommand \bibnamefont  [1]{#1}%
\providecommand \bibfnamefont [1]{#1}%
\providecommand \citenamefont [1]{#1}%
\providecommand \href@noop [0]{\@secondoftwo}%
\providecommand \href [0]{\begingroup \@sanitize@url \@href}%
\providecommand \@href[1]{\@@startlink{#1}\@@href}%
\providecommand \@@href[1]{\endgroup#1\@@endlink}%
\providecommand \@sanitize@url [0]{\catcode `\\12\catcode `\$12\catcode
  `\&12\catcode `\#12\catcode `\^12\catcode `\_12\catcode `\%12\relax}%
\providecommand \@@startlink[1]{}%
\providecommand \@@endlink[0]{}%
\providecommand \url  [0]{\begingroup\@sanitize@url \@url }%
\providecommand \@url [1]{\endgroup\@href {#1}{\urlprefix }}%
\providecommand \urlprefix  [0]{URL }%
\providecommand \Eprint [0]{\href }%
\providecommand \doibase [0]{http://dx.doi.org/}%
\providecommand \selectlanguage [0]{\@gobble}%
\providecommand \bibinfo  [0]{\@secondoftwo}%
\providecommand \bibfield  [0]{\@secondoftwo}%
\providecommand \translation [1]{[#1]}%
\providecommand \BibitemOpen [0]{}%
\providecommand \bibitemStop [0]{}%
\providecommand \bibitemNoStop [0]{.\EOS\space}%
\providecommand \EOS [0]{\spacefactor3000\relax}%
\providecommand \BibitemShut  [1]{\csname bibitem#1\endcsname}%
\let\auto@bib@innerbib\@empty
\bibitem [{\citenamefont {Aghanim}\ \emph
  {et~al.}(2020{\natexlab{a}})\citenamefont {Aghanim} \emph
  {et~al.}}]{Planck:2018vyg}%
  \BibitemOpen
  \bibfield  {author} {\bibinfo {author} {\bibfnamefont {N.}~\bibnamefont
  {Aghanim}} \emph {et~al.} (\bibinfo {collaboration} {Planck}),\ }\bibfield
  {title} {\enquote {\bibinfo {title} {{Planck 2018 results. VI. Cosmological
  parameters}},}\ }\href {\doibase 10.1051/0004-6361/201833910} {\bibfield
  {journal} {\bibinfo  {journal} {Astron. Astrophys.}\ }\textbf {\bibinfo
  {volume} {641}},\ \bibinfo {pages} {A6} (\bibinfo {year}
  {2020}{\natexlab{a}})},\ \bibinfo {note} {[Erratum: Astron.Astrophys. 652, C4
  (2021)]},\ \Eprint {http://arxiv.org/abs/1807.06209} {arXiv:1807.06209
  [astro-ph.CO]} \BibitemShut {NoStop}%
\bibitem [{\citenamefont {Riess}\ \emph {et~al.}(2021)\citenamefont {Riess}
  \emph {et~al.}}]{Riess:2021jrx}%
  \BibitemOpen
  \bibfield  {author} {\bibinfo {author} {\bibfnamefont {Adam~G.}\ \bibnamefont
  {Riess}} \emph {et~al.},\ }\bibfield  {title} {\enquote {\bibinfo {title} {{A
  Comprehensive Measurement of the Local Value of the Hubble Constant with 1
  km/s/Mpc Uncertainty from the Hubble Space Telescope and the SH0ES Team}},}\
  }\href@noop {} {\  (\bibinfo {year} {2021})},\ \Eprint
  {http://arxiv.org/abs/2112.04510} {arXiv:2112.04510 [astro-ph.CO]}
  \BibitemShut {NoStop}%
\bibitem [{\citenamefont {Riess}\ \emph {et~al.}(2022)\citenamefont {Riess},
  \citenamefont {Breuval}, \citenamefont {Yuan}, \citenamefont {Casertano},
  \citenamefont {Macri}, \citenamefont {Bowers}, \citenamefont {Scolnic},
  \citenamefont {Cantat-Gaudin}, \citenamefont {Anderson},\ and\ \citenamefont
  {Reyes}}]{Riess:2022mme}%
  \BibitemOpen
  \bibfield  {author} {\bibinfo {author} {\bibfnamefont {Adam~G.}\ \bibnamefont
  {Riess}}, \bibinfo {author} {\bibfnamefont {Louise}\ \bibnamefont {Breuval}},
  \bibinfo {author} {\bibfnamefont {Wenlong}\ \bibnamefont {Yuan}}, \bibinfo
  {author} {\bibfnamefont {Stefano}\ \bibnamefont {Casertano}}, \bibinfo
  {author} {\bibfnamefont {Lucas~M.}\ \bibnamefont {Macri}}, \bibinfo {author}
  {\bibfnamefont {J.~Bradley}\ \bibnamefont {Bowers}}, \bibinfo {author}
  {\bibfnamefont {Dan}\ \bibnamefont {Scolnic}}, \bibinfo {author}
  {\bibfnamefont {Tristan}\ \bibnamefont {Cantat-Gaudin}}, \bibinfo {author}
  {\bibfnamefont {Richard~I.}\ \bibnamefont {Anderson}}, \ and\ \bibinfo
  {author} {\bibfnamefont {Mauricio~Cruz}\ \bibnamefont {Reyes}},\ }\bibfield
  {title} {\enquote {\bibinfo {title} {{Cluster Cepheids with High Precision
  Gaia Parallaxes, Low Zero-point Uncertainties, and Hubble Space Telescope
  Photometry}},}\ }\href {\doibase 10.3847/1538-4357/ac8f24} {\bibfield
  {journal} {\bibinfo  {journal} {Astrophys. J.}\ }\textbf {\bibinfo {volume}
  {938}},\ \bibinfo {pages} {36} (\bibinfo {year} {2022})},\ \Eprint
  {http://arxiv.org/abs/2208.01045} {arXiv:2208.01045 [astro-ph.CO]}
  \BibitemShut {NoStop}%
\bibitem [{\citenamefont {Dainotti}\ \emph {et~al.}(2021)\citenamefont
  {Dainotti}, \citenamefont {De~Simone}, \citenamefont {Schiavone},
  \citenamefont {Montani}, \citenamefont {Rinaldi},\ and\ \citenamefont
  {Lambiase}}]{Dainotti:2021pqg}%
  \BibitemOpen
  \bibfield  {author} {\bibinfo {author} {\bibfnamefont {Maria~Giovanna}\
  \bibnamefont {Dainotti}}, \bibinfo {author} {\bibfnamefont {Biagio}\
  \bibnamefont {De~Simone}}, \bibinfo {author} {\bibfnamefont {Tiziano}\
  \bibnamefont {Schiavone}}, \bibinfo {author} {\bibfnamefont {Giovanni}\
  \bibnamefont {Montani}}, \bibinfo {author} {\bibfnamefont {Enrico}\
  \bibnamefont {Rinaldi}}, \ and\ \bibinfo {author} {\bibfnamefont {Gaetano}\
  \bibnamefont {Lambiase}},\ }\bibfield  {title} {\enquote {\bibinfo {title}
  {{On the Hubble constant tension in the SNe Ia Pantheon sample}},}\ }\href
  {\doibase 10.3847/1538-4357/abeb73} {\bibfield  {journal} {\bibinfo
  {journal} {Astrophys. J.}\ }\textbf {\bibinfo {volume} {912}},\ \bibinfo
  {pages} {150} (\bibinfo {year} {2021})},\ \Eprint
  {http://arxiv.org/abs/2103.02117} {arXiv:2103.02117 [astro-ph.CO]}
  \BibitemShut {NoStop}%
\bibitem [{\citenamefont {Dainotti}\ \emph {et~al.}(2022)\citenamefont
  {Dainotti}, \citenamefont {De~Simone}, \citenamefont {Schiavone},
  \citenamefont {Montani}, \citenamefont {Rinaldi}, \citenamefont {Lambiase},
  \citenamefont {Bogdan},\ and\ \citenamefont {Ugale}}]{Dainotti:2022bzg}%
  \BibitemOpen
  \bibfield  {author} {\bibinfo {author} {\bibfnamefont {Maria~Giovanna}\
  \bibnamefont {Dainotti}}, \bibinfo {author} {\bibfnamefont {Biagio}\
  \bibnamefont {De~Simone}}, \bibinfo {author} {\bibfnamefont {Tiziano}\
  \bibnamefont {Schiavone}}, \bibinfo {author} {\bibfnamefont {Giovanni}\
  \bibnamefont {Montani}}, \bibinfo {author} {\bibfnamefont {Enrico}\
  \bibnamefont {Rinaldi}}, \bibinfo {author} {\bibfnamefont {Gaetano}\
  \bibnamefont {Lambiase}}, \bibinfo {author} {\bibfnamefont {Malgorzata}\
  \bibnamefont {Bogdan}}, \ and\ \bibinfo {author} {\bibfnamefont {Sahil}\
  \bibnamefont {Ugale}},\ }\bibfield  {title} {\enquote {\bibinfo {title} {{On
  the Evolution of the Hubble Constant with the SNe Ia Pantheon Sample and
  Baryon Acoustic Oscillations: A Feasibility Study for GRB-Cosmology in
  2030}},}\ }\href {\doibase 10.3390/galaxies10010024} {\bibfield  {journal}
  {\bibinfo  {journal} {Galaxies}\ }\textbf {\bibinfo {volume} {10}},\ \bibinfo
  {pages} {24} (\bibinfo {year} {2022})},\ \Eprint
  {http://arxiv.org/abs/2201.09848} {arXiv:2201.09848 [astro-ph.CO]}
  \BibitemShut {NoStop}%
\bibitem [{\citenamefont {Mortsell}\ \emph
  {et~al.}(2022{\natexlab{a}})\citenamefont {Mortsell}, \citenamefont {Goobar},
  \citenamefont {Johansson},\ and\ \citenamefont {Dhawan}}]{Mortsell:2021nzg}%
  \BibitemOpen
  \bibfield  {author} {\bibinfo {author} {\bibfnamefont {Edvard}\ \bibnamefont
  {Mortsell}}, \bibinfo {author} {\bibfnamefont {Ariel}\ \bibnamefont
  {Goobar}}, \bibinfo {author} {\bibfnamefont {Joel}\ \bibnamefont
  {Johansson}}, \ and\ \bibinfo {author} {\bibfnamefont {Suhail}\ \bibnamefont
  {Dhawan}},\ }\bibfield  {title} {\enquote {\bibinfo {title} {{Sensitivity of
  the Hubble Constant Determination to Cepheid Calibration}},}\ }\href
  {\doibase 10.3847/1538-4357/ac756e} {\bibfield  {journal} {\bibinfo
  {journal} {Astrophys. J.}\ }\textbf {\bibinfo {volume} {933}},\ \bibinfo
  {pages} {212} (\bibinfo {year} {2022}{\natexlab{a}})},\ \Eprint
  {http://arxiv.org/abs/2105.11461} {arXiv:2105.11461 [astro-ph.CO]}
  \BibitemShut {NoStop}%
\bibitem [{\citenamefont {Mortsell}\ \emph
  {et~al.}(2022{\natexlab{b}})\citenamefont {Mortsell}, \citenamefont {Goobar},
  \citenamefont {Johansson},\ and\ \citenamefont {Dhawan}}]{Mortsell:2021tcx}%
  \BibitemOpen
  \bibfield  {author} {\bibinfo {author} {\bibfnamefont {Edvard}\ \bibnamefont
  {Mortsell}}, \bibinfo {author} {\bibfnamefont {Ariel}\ \bibnamefont
  {Goobar}}, \bibinfo {author} {\bibfnamefont {Joel}\ \bibnamefont
  {Johansson}}, \ and\ \bibinfo {author} {\bibfnamefont {Suhail}\ \bibnamefont
  {Dhawan}},\ }\bibfield  {title} {\enquote {\bibinfo {title} {{The Hubble
  Tension Revisited: Additional Local Distance Ladder Uncertainties}},}\ }\href
  {\doibase 10.3847/1538-4357/ac7c19} {\bibfield  {journal} {\bibinfo
  {journal} {Astrophys. J.}\ }\textbf {\bibinfo {volume} {935}},\ \bibinfo
  {pages} {58} (\bibinfo {year} {2022}{\natexlab{b}})},\ \Eprint
  {http://arxiv.org/abs/2106.09400} {arXiv:2106.09400 [astro-ph.CO]}
  \BibitemShut {NoStop}%
\bibitem [{\citenamefont {Follin}\ and\ \citenamefont
  {Knox}(2018)}]{Follin:2017ljs}%
  \BibitemOpen
  \bibfield  {author} {\bibinfo {author} {\bibfnamefont {Brent}\ \bibnamefont
  {Follin}}\ and\ \bibinfo {author} {\bibfnamefont {Lloyd}\ \bibnamefont
  {Knox}},\ }\bibfield  {title} {\enquote {\bibinfo {title} {{Insensitivity of
  the distance ladder Hubble constant determination to Cepheid calibration
  modelling choices}},}\ }\href {\doibase 10.1093/mnras/sty720} {\bibfield
  {journal} {\bibinfo  {journal} {Mon. Not. Roy. Astron. Soc.}\ }\textbf
  {\bibinfo {volume} {477}},\ \bibinfo {pages} {4534--4542} (\bibinfo {year}
  {2018})},\ \Eprint {http://arxiv.org/abs/1707.01175} {arXiv:1707.01175
  [astro-ph.CO]} \BibitemShut {NoStop}%
\bibitem [{\citenamefont {Brout}\ and\ \citenamefont
  {Scolnic}(2021)}]{Brout:2020msh}%
  \BibitemOpen
  \bibfield  {author} {\bibinfo {author} {\bibfnamefont {Dillon}\ \bibnamefont
  {Brout}}\ and\ \bibinfo {author} {\bibfnamefont {Daniel}\ \bibnamefont
  {Scolnic}},\ }\bibfield  {title} {\enquote {\bibinfo {title}
  {{It\textquoteright{}s Dust: Solving the Mysteries of the Intrinsic Scatter
  and Host-galaxy Dependence of Standardized Type Ia Supernova
  Brightnesses}},}\ }\href {\doibase 10.3847/1538-4357/abd69b} {\bibfield
  {journal} {\bibinfo  {journal} {Astrophys. J.}\ }\textbf {\bibinfo {volume}
  {909}},\ \bibinfo {pages} {26} (\bibinfo {year} {2021})},\ \Eprint
  {http://arxiv.org/abs/2004.10206} {arXiv:2004.10206 [astro-ph.CO]}
  \BibitemShut {NoStop}%
\bibitem [{\citenamefont {Bernal}\ \emph {et~al.}(2016)\citenamefont {Bernal},
  \citenamefont {Verde},\ and\ \citenamefont {Riess}}]{Bernal:2016gxb}%
  \BibitemOpen
  \bibfield  {author} {\bibinfo {author} {\bibfnamefont {Jose~Luis}\
  \bibnamefont {Bernal}}, \bibinfo {author} {\bibfnamefont {Licia}\
  \bibnamefont {Verde}}, \ and\ \bibinfo {author} {\bibfnamefont {Adam~G.}\
  \bibnamefont {Riess}},\ }\bibfield  {title} {\enquote {\bibinfo {title} {{The
  trouble with $H_0$}},}\ }\href {\doibase 10.1088/1475-7516/2016/10/019}
  {\bibfield  {journal} {\bibinfo  {journal} {JCAP}\ }\textbf {\bibinfo
  {volume} {1610}},\ \bibinfo {pages} {019} (\bibinfo {year} {2016})},\ \Eprint
  {http://arxiv.org/abs/1607.05617} {arXiv:1607.05617 [astro-ph.CO]}
  \BibitemShut {NoStop}%
\bibitem [{\citenamefont {Aylor}\ \emph {et~al.}(2019)\citenamefont {Aylor},
  \citenamefont {Joy}, \citenamefont {Knox}, \citenamefont {Millea},
  \citenamefont {Raghunathan},\ and\ \citenamefont {Wu}}]{Aylor:2018drw}%
  \BibitemOpen
  \bibfield  {author} {\bibinfo {author} {\bibfnamefont {Kevin}\ \bibnamefont
  {Aylor}}, \bibinfo {author} {\bibfnamefont {MacKenzie}\ \bibnamefont {Joy}},
  \bibinfo {author} {\bibfnamefont {Lloyd}\ \bibnamefont {Knox}}, \bibinfo
  {author} {\bibfnamefont {Marius}\ \bibnamefont {Millea}}, \bibinfo {author}
  {\bibfnamefont {Srinivasan}\ \bibnamefont {Raghunathan}}, \ and\ \bibinfo
  {author} {\bibfnamefont {W.~L.~Kimmy}\ \bibnamefont {Wu}},\ }\bibfield
  {title} {\enquote {\bibinfo {title} {{Sounds Discordant: Classical Distance
  Ladder \& $\Lambda$CDM -based Determinations of the Cosmological Sound
  Horizon}},}\ }\href {\doibase 10.3847/1538-4357/ab0898} {\bibfield  {journal}
  {\bibinfo  {journal} {Astrophys. J.}\ }\textbf {\bibinfo {volume} {874}},\
  \bibinfo {pages} {4} (\bibinfo {year} {2019})},\ \Eprint
  {http://arxiv.org/abs/1811.00537} {arXiv:1811.00537 [astro-ph.CO]}
  \BibitemShut {NoStop}%
\bibitem [{\citenamefont {Knox}\ and\ \citenamefont
  {Millea}(2020)}]{Knox:2019rjx}%
  \BibitemOpen
  \bibfield  {author} {\bibinfo {author} {\bibfnamefont {Lloyd}\ \bibnamefont
  {Knox}}\ and\ \bibinfo {author} {\bibfnamefont {Marius}\ \bibnamefont
  {Millea}},\ }\bibfield  {title} {\enquote {\bibinfo {title} {{Hubble constant
  hunter\textquoteright{}s guide}},}\ }\href {\doibase
  10.1103/PhysRevD.101.043533} {\bibfield  {journal} {\bibinfo  {journal}
  {Phys. Rev. D}\ }\textbf {\bibinfo {volume} {101}},\ \bibinfo {pages}
  {043533} (\bibinfo {year} {2020})},\ \Eprint
  {http://arxiv.org/abs/1908.03663} {arXiv:1908.03663 [astro-ph.CO]}
  \BibitemShut {NoStop}%
\bibitem [{\citenamefont {Camarena}\ and\ \citenamefont
  {Marra}(2021)}]{Camarena:2021jlr}%
  \BibitemOpen
  \bibfield  {author} {\bibinfo {author} {\bibfnamefont {David}\ \bibnamefont
  {Camarena}}\ and\ \bibinfo {author} {\bibfnamefont {Valerio}\ \bibnamefont
  {Marra}},\ }\bibfield  {title} {\enquote {\bibinfo {title} {{On the use of
  the local prior on the absolute magnitude of Type Ia supernovae in
  cosmological inference}},}\ }\href {\doibase 10.1093/mnras/stab1200}
  {\bibfield  {journal} {\bibinfo  {journal} {Mon. Not. Roy. Astron. Soc.}\
  }\textbf {\bibinfo {volume} {504}},\ \bibinfo {pages} {5164--5171} (\bibinfo
  {year} {2021})},\ \Eprint {http://arxiv.org/abs/2101.08641} {arXiv:2101.08641
  [astro-ph.CO]} \BibitemShut {NoStop}%
\bibitem [{\citenamefont {{Efstathiou}}(2021)}]{Efstathiou:2021ocp}%
  \BibitemOpen
  \bibfield  {author} {\bibinfo {author} {\bibfnamefont {George}\ \bibnamefont
  {{Efstathiou}}},\ }\bibfield  {title} {\enquote {\bibinfo {title} {{To H0 or
  not to H0?}}}\ }\href@noop {} {\bibfield  {journal} {\bibinfo  {journal}
  {arXiv e-prints}\ } (\bibinfo {year} {2021})},\ \Eprint
  {http://arxiv.org/abs/2103.08723} {arXiv:2103.08723 [astro-ph.CO]}
  \BibitemShut {NoStop}%
\bibitem [{\citenamefont {Sch\"oneberg}\ \emph {et~al.}(2021)\citenamefont
  {Sch\"oneberg}, \citenamefont {Franco~Abell\'an}, \citenamefont
  {P\'erez~S\'anchez}, \citenamefont {Witte}, \citenamefont {Poulin},\ and\
  \citenamefont {Lesgourgues}}]{Schoneberg:2021qvd}%
  \BibitemOpen
  \bibfield  {author} {\bibinfo {author} {\bibfnamefont {Nils}\ \bibnamefont
  {Sch\"oneberg}}, \bibinfo {author} {\bibfnamefont {Guillermo}\ \bibnamefont
  {Franco~Abell\'an}}, \bibinfo {author} {\bibfnamefont {Andrea}\ \bibnamefont
  {P\'erez~S\'anchez}}, \bibinfo {author} {\bibfnamefont {Samuel~J.}\
  \bibnamefont {Witte}}, \bibinfo {author} {\bibfnamefont {Vivian}\
  \bibnamefont {Poulin}}, \ and\ \bibinfo {author} {\bibfnamefont {Julien}\
  \bibnamefont {Lesgourgues}},\ }\bibfield  {title} {\enquote {\bibinfo {title}
  {{The $H_0$ Olympics: A fair ranking of proposed models}},}\ }\href@noop {}
  {\  (\bibinfo {year} {2021})},\ \Eprint {http://arxiv.org/abs/2107.10291}
  {arXiv:2107.10291 [astro-ph.CO]} \BibitemShut {NoStop}%
\bibitem [{\citenamefont {Karwal}\ and\ \citenamefont
  {Kamionkowski}(2016)}]{Karwal:2016vyq}%
  \BibitemOpen
  \bibfield  {author} {\bibinfo {author} {\bibfnamefont {Tanvi}\ \bibnamefont
  {Karwal}}\ and\ \bibinfo {author} {\bibfnamefont {Marc}\ \bibnamefont
  {Kamionkowski}},\ }\bibfield  {title} {\enquote {\bibinfo {title} {{Dark
  energy at early times, the Hubble parameter, and the string axiverse}},}\
  }\href {\doibase 10.1103/PhysRevD.94.103523} {\bibfield  {journal} {\bibinfo
  {journal} {Phys. Rev.}\ }\textbf {\bibinfo {volume} {D94}},\ \bibinfo {pages}
  {103523} (\bibinfo {year} {2016})},\ \Eprint
  {http://arxiv.org/abs/1608.01309} {arXiv:1608.01309 [astro-ph.CO]}
  \BibitemShut {NoStop}%
\bibitem [{\citenamefont {Poulin}\ \emph {et~al.}(2019)\citenamefont {Poulin},
  \citenamefont {Smith}, \citenamefont {Karwal},\ and\ \citenamefont
  {Kamionkowski}}]{Poulin:2018cxd}%
  \BibitemOpen
  \bibfield  {author} {\bibinfo {author} {\bibfnamefont {Vivian}\ \bibnamefont
  {Poulin}}, \bibinfo {author} {\bibfnamefont {Tristan~L.}\ \bibnamefont
  {Smith}}, \bibinfo {author} {\bibfnamefont {Tanvi}\ \bibnamefont {Karwal}}, \
  and\ \bibinfo {author} {\bibfnamefont {Marc}\ \bibnamefont {Kamionkowski}},\
  }\bibfield  {title} {\enquote {\bibinfo {title} {{Early Dark Energy Can
  Resolve The Hubble Tension}},}\ }\href {\doibase
  10.1103/PhysRevLett.122.221301} {\bibfield  {journal} {\bibinfo  {journal}
  {Phys. Rev. Lett.}\ }\textbf {\bibinfo {volume} {122}},\ \bibinfo {pages}
  {221301} (\bibinfo {year} {2019})},\ \Eprint
  {http://arxiv.org/abs/1811.04083} {arXiv:1811.04083 [astro-ph.CO]}
  \BibitemShut {NoStop}%
\bibitem [{\citenamefont {Smith}\ \emph {et~al.}(2020)\citenamefont {Smith},
  \citenamefont {Poulin},\ and\ \citenamefont {Amin}}]{Smith:2019ihp}%
  \BibitemOpen
  \bibfield  {author} {\bibinfo {author} {\bibfnamefont {Tristan~L.}\
  \bibnamefont {Smith}}, \bibinfo {author} {\bibfnamefont {Vivian}\
  \bibnamefont {Poulin}}, \ and\ \bibinfo {author} {\bibfnamefont {Mustafa~A.}\
  \bibnamefont {Amin}},\ }\bibfield  {title} {\enquote {\bibinfo {title}
  {{Oscillating scalar fields and the Hubble tension: a resolution with novel
  signatures}},}\ }\href {\doibase 10.1103/PhysRevD.101.063523} {\bibfield
  {journal} {\bibinfo  {journal} {Phys. Rev. D}\ }\textbf {\bibinfo {volume}
  {101}},\ \bibinfo {pages} {063523} (\bibinfo {year} {2020})},\ \Eprint
  {http://arxiv.org/abs/1908.06995} {arXiv:1908.06995 [astro-ph.CO]}
  \BibitemShut {NoStop}%
\bibitem [{\citenamefont {Niedermann}\ and\ \citenamefont
  {Sloth}(2019)}]{Niedermann:2019olb}%
  \BibitemOpen
  \bibfield  {author} {\bibinfo {author} {\bibfnamefont {Florian}\ \bibnamefont
  {Niedermann}}\ and\ \bibinfo {author} {\bibfnamefont {Martin~S.}\
  \bibnamefont {Sloth}},\ }\bibfield  {title} {\enquote {\bibinfo {title} {{New
  Early Dark Energy}},}\ }\href@noop {} {\  (\bibinfo {year} {2019})},\ \Eprint
  {http://arxiv.org/abs/1910.10739} {arXiv:1910.10739 [astro-ph.CO]}
  \BibitemShut {NoStop}%
\bibitem [{\citenamefont {Niedermann}\ and\ \citenamefont
  {Sloth}(2020)}]{Niedermann:2020dwg}%
  \BibitemOpen
  \bibfield  {author} {\bibinfo {author} {\bibfnamefont {Florian}\ \bibnamefont
  {Niedermann}}\ and\ \bibinfo {author} {\bibfnamefont {Martin~S.}\
  \bibnamefont {Sloth}},\ }\bibfield  {title} {\enquote {\bibinfo {title}
  {{Resolving the Hubble Tension with New Early Dark Energy}},}\ }\href@noop {}
  {\  (\bibinfo {year} {2020})},\ \Eprint {http://arxiv.org/abs/2006.06686}
  {arXiv:2006.06686 [astro-ph.CO]} \BibitemShut {NoStop}%
\bibitem [{\citenamefont {Ye}\ and\ \citenamefont {Piao}(2020)}]{Ye:2020btb}%
  \BibitemOpen
  \bibfield  {author} {\bibinfo {author} {\bibfnamefont {Gen}\ \bibnamefont
  {Ye}}\ and\ \bibinfo {author} {\bibfnamefont {Yun-Song}\ \bibnamefont
  {Piao}},\ }\bibfield  {title} {\enquote {\bibinfo {title} {{Is the Hubble
  tension a hint of AdS phase around recombination?}}}\ }\href {\doibase
  10.1103/PhysRevD.101.083507} {\bibfield  {journal} {\bibinfo  {journal}
  {Phys. Rev. D}\ }\textbf {\bibinfo {volume} {101}},\ \bibinfo {pages}
  {083507} (\bibinfo {year} {2020})},\ \Eprint
  {http://arxiv.org/abs/2001.02451} {arXiv:2001.02451 [astro-ph.CO]}
  \BibitemShut {NoStop}%
\bibitem [{\citenamefont {Agrawal}\ \emph {et~al.}(2019)\citenamefont
  {Agrawal}, \citenamefont {Cyr-Racine}, \citenamefont {Pinner},\ and\
  \citenamefont {Randall}}]{Agrawal:2019lmo}%
  \BibitemOpen
  \bibfield  {author} {\bibinfo {author} {\bibfnamefont {Prateek}\ \bibnamefont
  {Agrawal}}, \bibinfo {author} {\bibfnamefont {Francis-Yan}\ \bibnamefont
  {Cyr-Racine}}, \bibinfo {author} {\bibfnamefont {David}\ \bibnamefont
  {Pinner}}, \ and\ \bibinfo {author} {\bibfnamefont {Lisa}\ \bibnamefont
  {Randall}},\ }\bibfield  {title} {\enquote {\bibinfo {title} {{Rock 'n' Roll
  Solutions to the Hubble Tension}},}\ }\href@noop {} {\  (\bibinfo {year}
  {2019})},\ \Eprint {http://arxiv.org/abs/1904.01016} {arXiv:1904.01016
  [astro-ph.CO]} \BibitemShut {NoStop}%
\bibitem [{\citenamefont {Berghaus}\ and\ \citenamefont
  {Karwal}(2020)}]{Berghaus:2019cls}%
  \BibitemOpen
  \bibfield  {author} {\bibinfo {author} {\bibfnamefont {Kim~V.}\ \bibnamefont
  {Berghaus}}\ and\ \bibinfo {author} {\bibfnamefont {Tanvi}\ \bibnamefont
  {Karwal}},\ }\bibfield  {title} {\enquote {\bibinfo {title} {{Thermal
  Friction as a Solution to the Hubble Tension}},}\ }\href {\doibase
  10.1103/PhysRevD.101.083537} {\bibfield  {journal} {\bibinfo  {journal}
  {Phys. Rev. D}\ }\textbf {\bibinfo {volume} {101}},\ \bibinfo {pages}
  {083537} (\bibinfo {year} {2020})},\ \Eprint
  {http://arxiv.org/abs/1911.06281} {arXiv:1911.06281 [astro-ph.CO]}
  \BibitemShut {NoStop}%
\bibitem [{\citenamefont {Braglia}\ \emph {et~al.}(2020)\citenamefont
  {Braglia}, \citenamefont {Emond}, \citenamefont {Finelli}, \citenamefont
  {Gumrukcuoglu},\ and\ \citenamefont {Koyama}}]{Braglia:2020bym}%
  \BibitemOpen
  \bibfield  {author} {\bibinfo {author} {\bibfnamefont {Matteo}\ \bibnamefont
  {Braglia}}, \bibinfo {author} {\bibfnamefont {William~T.}\ \bibnamefont
  {Emond}}, \bibinfo {author} {\bibfnamefont {Fabio}\ \bibnamefont {Finelli}},
  \bibinfo {author} {\bibfnamefont {A.~Emir}\ \bibnamefont {Gumrukcuoglu}}, \
  and\ \bibinfo {author} {\bibfnamefont {Kazuya}\ \bibnamefont {Koyama}},\
  }\bibfield  {title} {\enquote {\bibinfo {title} {{Unified framework for early
  dark energy from $\alpha$-attractors}},}\ }\href {\doibase
  10.1103/PhysRevD.102.083513} {\bibfield  {journal} {\bibinfo  {journal}
  {Phys. Rev. D}\ }\textbf {\bibinfo {volume} {102}},\ \bibinfo {pages}
  {083513} (\bibinfo {year} {2020})},\ \Eprint
  {http://arxiv.org/abs/2005.14053} {arXiv:2005.14053 [astro-ph.CO]}
  \BibitemShut {NoStop}%
\bibitem [{\citenamefont {Braglia}\ \emph {et~al.}(2021)\citenamefont
  {Braglia}, \citenamefont {Ballardini}, \citenamefont {Finelli},\ and\
  \citenamefont {Koyama}}]{Braglia:2020auw}%
  \BibitemOpen
  \bibfield  {author} {\bibinfo {author} {\bibfnamefont {Matteo}\ \bibnamefont
  {Braglia}}, \bibinfo {author} {\bibfnamefont {Mario}\ \bibnamefont
  {Ballardini}}, \bibinfo {author} {\bibfnamefont {Fabio}\ \bibnamefont
  {Finelli}}, \ and\ \bibinfo {author} {\bibfnamefont {Kazuya}\ \bibnamefont
  {Koyama}},\ }\bibfield  {title} {\enquote {\bibinfo {title} {{Early modified
  gravity in light of the $H_0$ tension and LSS data}},}\ }\href {\doibase
  10.1103/PhysRevD.103.043528} {\bibfield  {journal} {\bibinfo  {journal}
  {Phys. Rev. D}\ }\textbf {\bibinfo {volume} {103}},\ \bibinfo {pages}
  {043528} (\bibinfo {year} {2021})},\ \Eprint
  {http://arxiv.org/abs/2011.12934} {arXiv:2011.12934 [astro-ph.CO]}
  \BibitemShut {NoStop}%
\bibitem [{\citenamefont {Gonzalez}\ \emph {et~al.}(2020)\citenamefont
  {Gonzalez}, \citenamefont {Hertzberg},\ and\ \citenamefont
  {Rompineve}}]{Gonzalez:2020fdy}%
  \BibitemOpen
  \bibfield  {author} {\bibinfo {author} {\bibfnamefont {Mark}\ \bibnamefont
  {Gonzalez}}, \bibinfo {author} {\bibfnamefont {Mark~P.}\ \bibnamefont
  {Hertzberg}}, \ and\ \bibinfo {author} {\bibfnamefont {Fabrizio}\
  \bibnamefont {Rompineve}},\ }\bibfield  {title} {\enquote {\bibinfo {title}
  {{Ultralight Scalar Decay and the Hubble Tension}},}\ }\href@noop {} {\
  (\bibinfo {year} {2020})},\ \Eprint {http://arxiv.org/abs/2006.13959}
  {arXiv:2006.13959 [astro-ph.CO]} \BibitemShut {NoStop}%
\bibitem [{\citenamefont {Rezazadeh}\ \emph {et~al.}(2022)\citenamefont
  {Rezazadeh}, \citenamefont {Ashoorioon},\ and\ \citenamefont
  {Grin}}]{Rezazadeh:2022lsf}%
  \BibitemOpen
  \bibfield  {author} {\bibinfo {author} {\bibfnamefont {K.}~\bibnamefont
  {Rezazadeh}}, \bibinfo {author} {\bibfnamefont {A.}~\bibnamefont
  {Ashoorioon}}, \ and\ \bibinfo {author} {\bibfnamefont {D.}~\bibnamefont
  {Grin}},\ }\bibfield  {title} {\enquote {\bibinfo {title} {{Cascading Dark
  Energy}},}\ }\href@noop {} {\  (\bibinfo {year} {2022})},\ \Eprint
  {http://arxiv.org/abs/2208.07631} {arXiv:2208.07631 [astro-ph.CO]}
  \BibitemShut {NoStop}%
\bibitem [{\citenamefont {Herold}\ \emph {et~al.}(2021)\citenamefont {Herold},
  \citenamefont {Ferreira},\ and\ \citenamefont {Komatsu}}]{Herold:2021ksg}%
  \BibitemOpen
  \bibfield  {author} {\bibinfo {author} {\bibfnamefont {Laura}\ \bibnamefont
  {Herold}}, \bibinfo {author} {\bibfnamefont {Elisa G.~M.}\ \bibnamefont
  {Ferreira}}, \ and\ \bibinfo {author} {\bibfnamefont {Eiichiro}\ \bibnamefont
  {Komatsu}},\ }\bibfield  {title} {\enquote {\bibinfo {title} {{New constraint
  on Early Dark Energy from Planck and BOSS data using the profile
  likelihood}},}\ }\href@noop {} {\  (\bibinfo {year} {2021})},\ \Eprint
  {http://arxiv.org/abs/2112.12140} {arXiv:2112.12140 [astro-ph.CO]}
  \BibitemShut {NoStop}%
\bibitem [{\citenamefont {G\'omez-Valent}(2022)}]{Gomez-Valent:2022hkb}%
  \BibitemOpen
  \bibfield  {author} {\bibinfo {author} {\bibfnamefont {Adri\`a}\ \bibnamefont
  {G\'omez-Valent}},\ }\bibfield  {title} {\enquote {\bibinfo {title} {{Fast
  test to assess the impact of marginalization in Monte~Carlo analyses and its
  application to cosmology}},}\ }\href {\doibase 10.1103/PhysRevD.106.063506}
  {\bibfield  {journal} {\bibinfo  {journal} {Phys. Rev. D}\ }\textbf {\bibinfo
  {volume} {106}},\ \bibinfo {pages} {063506} (\bibinfo {year} {2022})},\
  \Eprint {http://arxiv.org/abs/2203.16285} {arXiv:2203.16285 [astro-ph.CO]}
  \BibitemShut {NoStop}%
\bibitem [{\citenamefont {Poulin}\ \emph {et~al.}(2021)\citenamefont {Poulin},
  \citenamefont {Smith},\ and\ \citenamefont {Bartlett}}]{Poulin:2021bjr}%
  \BibitemOpen
  \bibfield  {author} {\bibinfo {author} {\bibfnamefont {Vivian}\ \bibnamefont
  {Poulin}}, \bibinfo {author} {\bibfnamefont {Tristan~L.}\ \bibnamefont
  {Smith}}, \ and\ \bibinfo {author} {\bibfnamefont {Alexa}\ \bibnamefont
  {Bartlett}},\ }\bibfield  {title} {\enquote {\bibinfo {title} {{Dark Energy
  at early times and ACT: a larger Hubble constant without late-time
  priors}},}\ }\href@noop {} {\  (\bibinfo {year} {2021})},\ \Eprint
  {http://arxiv.org/abs/2109.06229} {arXiv:2109.06229 [astro-ph.CO]}
  \BibitemShut {NoStop}%
\bibitem [{\citenamefont {Hill}\ \emph {et~al.}(2020)\citenamefont {Hill},
  \citenamefont {McDonough}, \citenamefont {Toomey},\ and\ \citenamefont
  {Alexander}}]{Hill:2020osr}%
  \BibitemOpen
  \bibfield  {author} {\bibinfo {author} {\bibfnamefont {J.~Colin}\
  \bibnamefont {Hill}}, \bibinfo {author} {\bibfnamefont {Evan}\ \bibnamefont
  {McDonough}}, \bibinfo {author} {\bibfnamefont {Michael~W.}\ \bibnamefont
  {Toomey}}, \ and\ \bibinfo {author} {\bibfnamefont {Stephon}\ \bibnamefont
  {Alexander}},\ }\bibfield  {title} {\enquote {\bibinfo {title} {{Early dark
  energy does not restore cosmological concordance}},}\ }\href {\doibase
  10.1103/PhysRevD.102.043507} {\bibfield  {journal} {\bibinfo  {journal}
  {Phys. Rev. D}\ }\textbf {\bibinfo {volume} {102}},\ \bibinfo {pages}
  {043507} (\bibinfo {year} {2020})},\ \Eprint
  {http://arxiv.org/abs/2003.07355} {arXiv:2003.07355 [astro-ph.CO]}
  \BibitemShut {NoStop}%
\bibitem [{\citenamefont {La~Posta}\ \emph {et~al.}(2021)\citenamefont
  {La~Posta}, \citenamefont {Louis}, \citenamefont {Garrido},\ and\
  \citenamefont {Hill}}]{LaPosta:2021pgm}%
  \BibitemOpen
  \bibfield  {author} {\bibinfo {author} {\bibfnamefont {Adrien}\ \bibnamefont
  {La~Posta}}, \bibinfo {author} {\bibfnamefont {Thibaut}\ \bibnamefont
  {Louis}}, \bibinfo {author} {\bibfnamefont {Xavier}\ \bibnamefont {Garrido}},
  \ and\ \bibinfo {author} {\bibfnamefont {J.~Colin}\ \bibnamefont {Hill}},\
  }\bibfield  {title} {\enquote {\bibinfo {title} {{Constraints on
  Pre-Recombination Early Dark Energy from SPT-3G Public Data}},}\ }\href@noop
  {} {\  (\bibinfo {year} {2021})},\ \Eprint {http://arxiv.org/abs/2112.10754}
  {arXiv:2112.10754 [astro-ph.CO]} \BibitemShut {NoStop}%
\bibitem [{\citenamefont {Reeves}\ \emph {et~al.}(2023)\citenamefont {Reeves},
  \citenamefont {Herold}, \citenamefont {Vagnozzi}, \citenamefont {Sherwin},\
  and\ \citenamefont {Ferreira}}]{Reeves:2022aoi}%
  \BibitemOpen
  \bibfield  {author} {\bibinfo {author} {\bibfnamefont {Alexander}\
  \bibnamefont {Reeves}}, \bibinfo {author} {\bibfnamefont {Laura}\
  \bibnamefont {Herold}}, \bibinfo {author} {\bibfnamefont {Sunny}\
  \bibnamefont {Vagnozzi}}, \bibinfo {author} {\bibfnamefont {Blake~D.}\
  \bibnamefont {Sherwin}}, \ and\ \bibinfo {author} {\bibfnamefont {Elisa
  G.~M.}\ \bibnamefont {Ferreira}},\ }\bibfield  {title} {\enquote {\bibinfo
  {title} {{Restoring cosmological concordance with early dark energy and
  massive neutrinos?}}}\ }\href {\doibase 10.1093/mnras/stad317} {\bibfield
  {journal} {\bibinfo  {journal} {Mon. Not. Roy. Astron. Soc.}\ }\textbf
  {\bibinfo {volume} {520}},\ \bibinfo {pages} {3688--3695} (\bibinfo {year}
  {2023})},\ \Eprint {http://arxiv.org/abs/2207.01501} {arXiv:2207.01501
  [astro-ph.CO]} \BibitemShut {NoStop}%
\bibitem [{\citenamefont {Herold}\ and\ \citenamefont
  {Ferreira}(2023)}]{Herold:2022iib}%
  \BibitemOpen
  \bibfield  {author} {\bibinfo {author} {\bibfnamefont {Laura}\ \bibnamefont
  {Herold}}\ and\ \bibinfo {author} {\bibfnamefont {Elisa G.~M.}\ \bibnamefont
  {Ferreira}},\ }\bibfield  {title} {\enquote {\bibinfo {title} {{Resolving the
  Hubble tension with early dark energy}},}\ }\href {\doibase
  10.1103/PhysRevD.108.043513} {\bibfield  {journal} {\bibinfo  {journal}
  {Phys. Rev. D}\ }\textbf {\bibinfo {volume} {108}},\ \bibinfo {pages}
  {043513} (\bibinfo {year} {2023})},\ \Eprint
  {http://arxiv.org/abs/2210.16296} {arXiv:2210.16296 [astro-ph.CO]}
  \BibitemShut {NoStop}%
\bibitem [{\citenamefont {Eskilt}\ \emph {et~al.}(2023)\citenamefont {Eskilt},
  \citenamefont {Herold}, \citenamefont {Komatsu}, \citenamefont {Murai},
  \citenamefont {Namikawa},\ and\ \citenamefont {Naokawa}}]{Eskilt:2023nxm}%
  \BibitemOpen
  \bibfield  {author} {\bibinfo {author} {\bibfnamefont {Johannes~R.}\
  \bibnamefont {Eskilt}}, \bibinfo {author} {\bibfnamefont {Laura}\
  \bibnamefont {Herold}}, \bibinfo {author} {\bibfnamefont {Eiichiro}\
  \bibnamefont {Komatsu}}, \bibinfo {author} {\bibfnamefont {Kai}\ \bibnamefont
  {Murai}}, \bibinfo {author} {\bibfnamefont {Toshiya}\ \bibnamefont
  {Namikawa}}, \ and\ \bibinfo {author} {\bibfnamefont {Fumihiro}\ \bibnamefont
  {Naokawa}},\ }\bibfield  {title} {\enquote {\bibinfo {title} {{Constraints on
  Early Dark Energy from Isotropic Cosmic Birefringence}},}\ }\href {\doibase
  10.1103/PhysRevLett.131.121001} {\bibfield  {journal} {\bibinfo  {journal}
  {Phys. Rev. Lett.}\ }\textbf {\bibinfo {volume} {131}},\ \bibinfo {pages}
  {121001} (\bibinfo {year} {2023})},\ \Eprint
  {http://arxiv.org/abs/2303.15369} {arXiv:2303.15369 [astro-ph.CO]}
  \BibitemShut {NoStop}%
\bibitem [{\citenamefont {Murgia}\ \emph {et~al.}(2021)\citenamefont {Murgia},
  \citenamefont {Abell\'an},\ and\ \citenamefont {Poulin}}]{Murgia:2020ryi}%
  \BibitemOpen
  \bibfield  {author} {\bibinfo {author} {\bibfnamefont {Riccardo}\
  \bibnamefont {Murgia}}, \bibinfo {author} {\bibfnamefont {Guillermo~F.}\
  \bibnamefont {Abell\'an}}, \ and\ \bibinfo {author} {\bibfnamefont {Vivian}\
  \bibnamefont {Poulin}},\ }\bibfield  {title} {\enquote {\bibinfo {title}
  {{Early dark energy resolution to the Hubble tension in light of weak lensing
  surveys and lensing anomalies}},}\ }\href {\doibase
  10.1103/PhysRevD.103.063502} {\bibfield  {journal} {\bibinfo  {journal}
  {Phys. Rev. D}\ }\textbf {\bibinfo {volume} {103}},\ \bibinfo {pages}
  {063502} (\bibinfo {year} {2021})},\ \Eprint
  {http://arxiv.org/abs/2009.10733} {arXiv:2009.10733 [astro-ph.CO]}
  \BibitemShut {NoStop}%
\bibitem [{\citenamefont {Goldstein}\ \emph {et~al.}(2023)\citenamefont
  {Goldstein}, \citenamefont {Hill}, \citenamefont {Ir\v{s}i\v{c}},\ and\
  \citenamefont {Sherwin}}]{Goldstein:2023gnw}%
  \BibitemOpen
  \bibfield  {author} {\bibinfo {author} {\bibfnamefont {Samuel}\ \bibnamefont
  {Goldstein}}, \bibinfo {author} {\bibfnamefont {J.~Colin}\ \bibnamefont
  {Hill}}, \bibinfo {author} {\bibfnamefont {Vid}\ \bibnamefont
  {Ir\v{s}i\v{c}}}, \ and\ \bibinfo {author} {\bibfnamefont {Blake~D.}\
  \bibnamefont {Sherwin}},\ }\bibfield  {title} {\enquote {\bibinfo {title}
  {{Canonical Hubble-Tension-Resolving Early Dark Energy Cosmologies Are
  Inconsistent with the Lyman-\ensuremath{\alpha} Forest}},}\ }\href {\doibase
  10.1103/PhysRevLett.131.201001} {\bibfield  {journal} {\bibinfo  {journal}
  {Phys. Rev. Lett.}\ }\textbf {\bibinfo {volume} {131}},\ \bibinfo {pages}
  {201001} (\bibinfo {year} {2023})},\ \Eprint
  {http://arxiv.org/abs/2303.00746} {arXiv:2303.00746 [astro-ph.CO]}
  \BibitemShut {NoStop}%
\bibitem [{\citenamefont {Poulin}\ \emph {et~al.}(2023)\citenamefont {Poulin},
  \citenamefont {Smith},\ and\ \citenamefont {Karwal}}]{Poulin:2023lkg}%
  \BibitemOpen
  \bibfield  {author} {\bibinfo {author} {\bibfnamefont {Vivian}\ \bibnamefont
  {Poulin}}, \bibinfo {author} {\bibfnamefont {Tristan~L.}\ \bibnamefont
  {Smith}}, \ and\ \bibinfo {author} {\bibfnamefont {Tanvi}\ \bibnamefont
  {Karwal}},\ }\bibfield  {title} {\enquote {\bibinfo {title} {{The Ups and
  Downs of Early Dark Energy solutions to the Hubble tension: a review of
  models, hints and constraints circa 2023}},}\ }\href@noop {} {\  (\bibinfo
  {year} {2023})},\ \Eprint {http://arxiv.org/abs/2302.09032} {arXiv:2302.09032
  [astro-ph.CO]} \BibitemShut {NoStop}%
\bibitem [{\citenamefont {Di~Valentino}\ \emph {et~al.}(2021)\citenamefont
  {Di~Valentino}, \citenamefont {Mena}, \citenamefont {Pan}, \citenamefont
  {Visinelli}, \citenamefont {Yang}, \citenamefont {Melchiorri}, \citenamefont
  {Mota}, \citenamefont {Riess},\ and\ \citenamefont
  {Silk}}]{DiValentino:2021izs}%
  \BibitemOpen
  \bibfield  {author} {\bibinfo {author} {\bibfnamefont {Eleonora}\
  \bibnamefont {Di~Valentino}}, \bibinfo {author} {\bibfnamefont {Olga}\
  \bibnamefont {Mena}}, \bibinfo {author} {\bibfnamefont {Supriya}\
  \bibnamefont {Pan}}, \bibinfo {author} {\bibfnamefont {Luca}\ \bibnamefont
  {Visinelli}}, \bibinfo {author} {\bibfnamefont {Weiqiang}\ \bibnamefont
  {Yang}}, \bibinfo {author} {\bibfnamefont {Alessandro}\ \bibnamefont
  {Melchiorri}}, \bibinfo {author} {\bibfnamefont {David~F.}\ \bibnamefont
  {Mota}}, \bibinfo {author} {\bibfnamefont {Adam~G.}\ \bibnamefont {Riess}}, \
  and\ \bibinfo {author} {\bibfnamefont {Joseph}\ \bibnamefont {Silk}},\
  }\bibfield  {title} {\enquote {\bibinfo {title} {{In the realm of the Hubble
  tension\textemdash{}a review of solutions}},}\ }\href {\doibase
  10.1088/1361-6382/ac086d} {\bibfield  {journal} {\bibinfo  {journal} {Class.
  Quant. Grav.}\ }\textbf {\bibinfo {volume} {38}},\ \bibinfo {pages} {153001}
  (\bibinfo {year} {2021})},\ \Eprint {http://arxiv.org/abs/2103.01183}
  {arXiv:2103.01183 [astro-ph.CO]} \BibitemShut {NoStop}%
\bibitem [{\citenamefont {Lin}\ \emph {et~al.}(2019)\citenamefont {Lin},
  \citenamefont {Benevento}, \citenamefont {Hu},\ and\ \citenamefont
  {Raveri}}]{Lin:2019qug}%
  \BibitemOpen
  \bibfield  {author} {\bibinfo {author} {\bibfnamefont {Meng-Xiang}\
  \bibnamefont {Lin}}, \bibinfo {author} {\bibfnamefont {Giampaolo}\
  \bibnamefont {Benevento}}, \bibinfo {author} {\bibfnamefont {Wayne}\
  \bibnamefont {Hu}}, \ and\ \bibinfo {author} {\bibfnamefont {Marco}\
  \bibnamefont {Raveri}},\ }\bibfield  {title} {\enquote {\bibinfo {title}
  {{Acoustic Dark Energy: Potential Conversion of the Hubble Tension}},}\
  }\href {\doibase 10.1103/PhysRevD.100.063542} {\bibfield  {journal} {\bibinfo
   {journal} {Phys. Rev.}\ }\textbf {\bibinfo {volume} {D100}},\ \bibinfo
  {pages} {063542} (\bibinfo {year} {2019})},\ \Eprint
  {http://arxiv.org/abs/1905.12618} {arXiv:1905.12618 [astro-ph.CO]}
  \BibitemShut {NoStop}%
\bibitem [{\citenamefont {Lin}\ \emph {et~al.}(2020)\citenamefont {Lin},
  \citenamefont {Hu},\ and\ \citenamefont {Raveri}}]{Lin:2020jcb}%
  \BibitemOpen
  \bibfield  {author} {\bibinfo {author} {\bibfnamefont {Meng-Xiang}\
  \bibnamefont {Lin}}, \bibinfo {author} {\bibfnamefont {Wayne}\ \bibnamefont
  {Hu}}, \ and\ \bibinfo {author} {\bibfnamefont {Marco}\ \bibnamefont
  {Raveri}},\ }\bibfield  {title} {\enquote {\bibinfo {title} {{Testing $H_0$
  in Acoustic Dark Energy with Planck and ACT Polarization}},}\ }\href
  {\doibase 10.1103/PhysRevD.102.123523} {\bibfield  {journal} {\bibinfo
  {journal} {Phys. Rev. D}\ }\textbf {\bibinfo {volume} {102}},\ \bibinfo
  {pages} {123523} (\bibinfo {year} {2020})},\ \Eprint
  {http://arxiv.org/abs/2009.08974} {arXiv:2009.08974 [astro-ph.CO]}
  \BibitemShut {NoStop}%
\bibitem [{\citenamefont {Alam}\ \emph {et~al.}(2017)\citenamefont {Alam} \emph
  {et~al.}}]{BOSS:2016wmc}%
  \BibitemOpen
  \bibfield  {author} {\bibinfo {author} {\bibfnamefont {Shadab}\ \bibnamefont
  {Alam}} \emph {et~al.} (\bibinfo {collaboration} {BOSS}),\ }\bibfield
  {title} {\enquote {\bibinfo {title} {{The clustering of galaxies in the
  completed SDSS-III Baryon Oscillation Spectroscopic Survey: cosmological
  analysis of the DR12 galaxy sample}},}\ }\href {\doibase
  10.1093/mnras/stx721} {\bibfield  {journal} {\bibinfo  {journal} {Mon. Not.
  Roy. Astron. Soc.}\ }\textbf {\bibinfo {volume} {470}},\ \bibinfo {pages}
  {2617--2652} (\bibinfo {year} {2017})},\ \Eprint
  {http://arxiv.org/abs/1607.03155} {arXiv:1607.03155 [astro-ph.CO]}
  \BibitemShut {NoStop}%
\bibitem [{\citenamefont {Alam}\ \emph {et~al.}(2021)\citenamefont {Alam} \emph
  {et~al.}}]{eBOSS:2020yzd}%
  \BibitemOpen
  \bibfield  {author} {\bibinfo {author} {\bibfnamefont {Shadab}\ \bibnamefont
  {Alam}} \emph {et~al.} (\bibinfo {collaboration} {eBOSS}),\ }\bibfield
  {title} {\enquote {\bibinfo {title} {{Completed SDSS-IV extended Baryon
  Oscillation Spectroscopic Survey: Cosmological implications from two decades
  of spectroscopic surveys at the Apache Point Observatory}},}\ }\href
  {\doibase 10.1103/PhysRevD.103.083533} {\bibfield  {journal} {\bibinfo
  {journal} {Phys. Rev. D}\ }\textbf {\bibinfo {volume} {103}},\ \bibinfo
  {pages} {083533} (\bibinfo {year} {2021})},\ \Eprint
  {http://arxiv.org/abs/2007.08991} {arXiv:2007.08991 [astro-ph.CO]}
  \BibitemShut {NoStop}%
\bibitem [{\citenamefont {Brout}\ \emph {et~al.}(2022)\citenamefont {Brout}
  \emph {et~al.}}]{Brout:2022vxf}%
  \BibitemOpen
  \bibfield  {author} {\bibinfo {author} {\bibfnamefont {Dillon}\ \bibnamefont
  {Brout}} \emph {et~al.},\ }\bibfield  {title} {\enquote {\bibinfo {title}
  {{The Pantheon+ Analysis: Cosmological Constraints}},}\ }\href@noop {} {\
  (\bibinfo {year} {2022})},\ \Eprint {http://arxiv.org/abs/2202.04077}
  {arXiv:2202.04077 [astro-ph.CO]} \BibitemShut {NoStop}%
\bibitem [{\citenamefont {Carrasco}\ \emph {et~al.}(2012)\citenamefont
  {Carrasco}, \citenamefont {Hertzberg},\ and\ \citenamefont
  {Senatore}}]{Carrasco:2012cv}%
  \BibitemOpen
  \bibfield  {author} {\bibinfo {author} {\bibfnamefont {John Joseph~M.}\
  \bibnamefont {Carrasco}}, \bibinfo {author} {\bibfnamefont {Mark~P.}\
  \bibnamefont {Hertzberg}}, \ and\ \bibinfo {author} {\bibfnamefont
  {Leonardo}\ \bibnamefont {Senatore}},\ }\bibfield  {title} {\enquote
  {\bibinfo {title} {{The Effective Field Theory of Cosmological Large Scale
  Structures}},}\ }\href {\doibase 10.1007/JHEP09(2012)082} {\bibfield
  {journal} {\bibinfo  {journal} {JHEP}\ }\textbf {\bibinfo {volume} {09}},\
  \bibinfo {pages} {082} (\bibinfo {year} {2012})},\ \Eprint
  {http://arxiv.org/abs/1206.2926} {arXiv:1206.2926 [astro-ph.CO]} \BibitemShut
  {NoStop}%
\bibitem [{\citenamefont {Baumann}\ \emph {et~al.}(2012)\citenamefont
  {Baumann}, \citenamefont {Nicolis}, \citenamefont {Senatore},\ and\
  \citenamefont {Zaldarriaga}}]{Baumann:2010tm}%
  \BibitemOpen
  \bibfield  {author} {\bibinfo {author} {\bibfnamefont {Daniel}\ \bibnamefont
  {Baumann}}, \bibinfo {author} {\bibfnamefont {Alberto}\ \bibnamefont
  {Nicolis}}, \bibinfo {author} {\bibfnamefont {Leonardo}\ \bibnamefont
  {Senatore}}, \ and\ \bibinfo {author} {\bibfnamefont {Matias}\ \bibnamefont
  {Zaldarriaga}},\ }\bibfield  {title} {\enquote {\bibinfo {title}
  {{Cosmological Non-Linearities as an Effective Fluid}},}\ }\href {\doibase
  10.1088/1475-7516/2012/07/051} {\bibfield  {journal} {\bibinfo  {journal}
  {JCAP}\ }\textbf {\bibinfo {volume} {07}},\ \bibinfo {pages} {051} (\bibinfo
  {year} {2012})},\ \Eprint {http://arxiv.org/abs/1004.2488} {arXiv:1004.2488
  [astro-ph.CO]} \BibitemShut {NoStop}%
\bibitem [{\citenamefont {Porto}\ \emph {et~al.}(2014)\citenamefont {Porto},
  \citenamefont {Senatore},\ and\ \citenamefont {Zaldarriaga}}]{Porto:2013qua}%
  \BibitemOpen
  \bibfield  {author} {\bibinfo {author} {\bibfnamefont {Rafael~A.}\
  \bibnamefont {Porto}}, \bibinfo {author} {\bibfnamefont {Leonardo}\
  \bibnamefont {Senatore}}, \ and\ \bibinfo {author} {\bibfnamefont {Matias}\
  \bibnamefont {Zaldarriaga}},\ }\bibfield  {title} {\enquote {\bibinfo {title}
  {{The Lagrangian-space Effective Field Theory of Large Scale Structures}},}\
  }\href {\doibase 10.1088/1475-7516/2014/05/022} {\bibfield  {journal}
  {\bibinfo  {journal} {JCAP}\ }\textbf {\bibinfo {volume} {05}},\ \bibinfo
  {pages} {022} (\bibinfo {year} {2014})},\ \Eprint
  {http://arxiv.org/abs/1311.2168} {arXiv:1311.2168 [astro-ph.CO]} \BibitemShut
  {NoStop}%
\bibitem [{\citenamefont {Pajer}\ and\ \citenamefont
  {Zaldarriaga}(2013)}]{Pajer:2013jj}%
  \BibitemOpen
  \bibfield  {author} {\bibinfo {author} {\bibfnamefont {Enrico}\ \bibnamefont
  {Pajer}}\ and\ \bibinfo {author} {\bibfnamefont {Matias}\ \bibnamefont
  {Zaldarriaga}},\ }\bibfield  {title} {\enquote {\bibinfo {title} {{On the
  Renormalization of the Effective Field Theory of Large Scale Structures}},}\
  }\href {\doibase 10.1088/1475-7516/2013/08/037} {\bibfield  {journal}
  {\bibinfo  {journal} {JCAP}\ }\textbf {\bibinfo {volume} {08}},\ \bibinfo
  {pages} {037} (\bibinfo {year} {2013})},\ \Eprint
  {http://arxiv.org/abs/1301.7182} {arXiv:1301.7182 [astro-ph.CO]} \BibitemShut
  {NoStop}%
\bibitem [{\citenamefont {Abolhasani}\ \emph {et~al.}(2016)\citenamefont
  {Abolhasani}, \citenamefont {Mirbabayi},\ and\ \citenamefont
  {Pajer}}]{Abolhasani:2015mra}%
  \BibitemOpen
  \bibfield  {author} {\bibinfo {author} {\bibfnamefont {Ali~Akbar}\
  \bibnamefont {Abolhasani}}, \bibinfo {author} {\bibfnamefont {Mehrdad}\
  \bibnamefont {Mirbabayi}}, \ and\ \bibinfo {author} {\bibfnamefont {Enrico}\
  \bibnamefont {Pajer}},\ }\bibfield  {title} {\enquote {\bibinfo {title}
  {{Systematic Renormalization of the Effective Theory of Large Scale
  Structure}},}\ }\href {\doibase 10.1088/1475-7516/2016/05/063} {\bibfield
  {journal} {\bibinfo  {journal} {JCAP}\ }\textbf {\bibinfo {volume} {05}},\
  \bibinfo {pages} {063} (\bibinfo {year} {2016})},\ \Eprint
  {http://arxiv.org/abs/1509.07886} {arXiv:1509.07886 [hep-th]} \BibitemShut
  {NoStop}%
\bibitem [{\citenamefont {Senatore}\ and\ \citenamefont
  {Zaldarriaga}(2014)}]{Senatore:2014vja}%
  \BibitemOpen
  \bibfield  {author} {\bibinfo {author} {\bibfnamefont {Leonardo}\
  \bibnamefont {Senatore}}\ and\ \bibinfo {author} {\bibfnamefont {Matias}\
  \bibnamefont {Zaldarriaga}},\ }\bibfield  {title} {\enquote {\bibinfo {title}
  {{Redshift Space Distortions in the Effective Field Theory of Large Scale
  Structures}},}\ }\href@noop {} {\  (\bibinfo {year} {2014})},\ \Eprint
  {http://arxiv.org/abs/1409.1225} {arXiv:1409.1225 [astro-ph.CO]} \BibitemShut
  {NoStop}%
\bibitem [{\citenamefont {Baldauf}\ \emph {et~al.}(2015)\citenamefont
  {Baldauf}, \citenamefont {Mirbabayi}, \citenamefont {Simonovi\'c},\ and\
  \citenamefont {Zaldarriaga}}]{Baldauf:2015xfa}%
  \BibitemOpen
  \bibfield  {author} {\bibinfo {author} {\bibfnamefont {Tobias}\ \bibnamefont
  {Baldauf}}, \bibinfo {author} {\bibfnamefont {Mehrdad}\ \bibnamefont
  {Mirbabayi}}, \bibinfo {author} {\bibfnamefont {Marko}\ \bibnamefont
  {Simonovi\'c}}, \ and\ \bibinfo {author} {\bibfnamefont {Matias}\
  \bibnamefont {Zaldarriaga}},\ }\bibfield  {title} {\enquote {\bibinfo {title}
  {{Equivalence Principle and the Baryon Acoustic Peak}},}\ }\href {\doibase
  10.1103/PhysRevD.92.043514} {\bibfield  {journal} {\bibinfo  {journal} {Phys.
  Rev. D}\ }\textbf {\bibinfo {volume} {92}},\ \bibinfo {pages} {043514}
  (\bibinfo {year} {2015})},\ \Eprint {http://arxiv.org/abs/1504.04366}
  {arXiv:1504.04366 [astro-ph.CO]} \BibitemShut {NoStop}%
\bibitem [{\citenamefont {Senatore}\ and\ \citenamefont
  {Zaldarriaga}(2015)}]{Senatore:2014via}%
  \BibitemOpen
  \bibfield  {author} {\bibinfo {author} {\bibfnamefont {Leonardo}\
  \bibnamefont {Senatore}}\ and\ \bibinfo {author} {\bibfnamefont {Matias}\
  \bibnamefont {Zaldarriaga}},\ }\bibfield  {title} {\enquote {\bibinfo {title}
  {{The IR-resummed Effective Field Theory of Large Scale Structures}},}\
  }\href {\doibase 10.1088/1475-7516/2015/02/013} {\bibfield  {journal}
  {\bibinfo  {journal} {JCAP}\ }\textbf {\bibinfo {volume} {02}},\ \bibinfo
  {pages} {013} (\bibinfo {year} {2015})},\ \Eprint
  {http://arxiv.org/abs/1404.5954} {arXiv:1404.5954 [astro-ph.CO]} \BibitemShut
  {NoStop}%
\bibitem [{\citenamefont {Senatore}\ and\ \citenamefont
  {Trevisan}(2018)}]{Senatore:2017pbn}%
  \BibitemOpen
  \bibfield  {author} {\bibinfo {author} {\bibfnamefont {Leonardo}\
  \bibnamefont {Senatore}}\ and\ \bibinfo {author} {\bibfnamefont {Gabriele}\
  \bibnamefont {Trevisan}},\ }\bibfield  {title} {\enquote {\bibinfo {title}
  {{On the IR-Resummation in the EFTofLSS}},}\ }\href {\doibase
  10.1088/1475-7516/2018/05/019} {\bibfield  {journal} {\bibinfo  {journal}
  {JCAP}\ }\textbf {\bibinfo {volume} {05}},\ \bibinfo {pages} {019} (\bibinfo
  {year} {2018})},\ \Eprint {http://arxiv.org/abs/1710.02178} {arXiv:1710.02178
  [astro-ph.CO]} \BibitemShut {NoStop}%
\bibitem [{\citenamefont {Lewandowski}\ and\ \citenamefont
  {Senatore}(2020)}]{Lewandowski:2018ywf}%
  \BibitemOpen
  \bibfield  {author} {\bibinfo {author} {\bibfnamefont {Matthew}\ \bibnamefont
  {Lewandowski}}\ and\ \bibinfo {author} {\bibfnamefont {Leonardo}\
  \bibnamefont {Senatore}},\ }\bibfield  {title} {\enquote {\bibinfo {title}
  {{An analytic implementation of the IR-resummation for the BAO peak}},}\
  }\href {\doibase 10.1088/1475-7516/2020/03/018} {\bibfield  {journal}
  {\bibinfo  {journal} {JCAP}\ }\textbf {\bibinfo {volume} {03}},\ \bibinfo
  {pages} {018} (\bibinfo {year} {2020})},\ \Eprint
  {http://arxiv.org/abs/1810.11855} {arXiv:1810.11855 [astro-ph.CO]}
  \BibitemShut {NoStop}%
\bibitem [{\citenamefont {Blas}\ \emph {et~al.}(2016)\citenamefont {Blas},
  \citenamefont {Garny}, \citenamefont {Ivanov},\ and\ \citenamefont
  {Sibiryakov}}]{Blas:2016sfa}%
  \BibitemOpen
  \bibfield  {author} {\bibinfo {author} {\bibfnamefont {Diego}\ \bibnamefont
  {Blas}}, \bibinfo {author} {\bibfnamefont {Mathias}\ \bibnamefont {Garny}},
  \bibinfo {author} {\bibfnamefont {Mikhail~M.}\ \bibnamefont {Ivanov}}, \ and\
  \bibinfo {author} {\bibfnamefont {Sergey}\ \bibnamefont {Sibiryakov}},\
  }\bibfield  {title} {\enquote {\bibinfo {title} {{Time-Sliced Perturbation
  Theory II: Baryon Acoustic Oscillations and Infrared Resummation}},}\ }\href
  {\doibase 10.1088/1475-7516/2016/07/028} {\bibfield  {journal} {\bibinfo
  {journal} {JCAP}\ }\textbf {\bibinfo {volume} {07}},\ \bibinfo {pages} {028}
  (\bibinfo {year} {2016})},\ \Eprint {http://arxiv.org/abs/1605.02149}
  {arXiv:1605.02149 [astro-ph.CO]} \BibitemShut {NoStop}%
\bibitem [{\citenamefont {Carrasco}\ \emph
  {et~al.}(2014{\natexlab{a}})\citenamefont {Carrasco}, \citenamefont
  {Foreman}, \citenamefont {Green},\ and\ \citenamefont
  {Senatore}}]{Carrasco:2013sva}%
  \BibitemOpen
  \bibfield  {author} {\bibinfo {author} {\bibfnamefont {John Joseph~M.}\
  \bibnamefont {Carrasco}}, \bibinfo {author} {\bibfnamefont {Simon}\
  \bibnamefont {Foreman}}, \bibinfo {author} {\bibfnamefont {Daniel}\
  \bibnamefont {Green}}, \ and\ \bibinfo {author} {\bibfnamefont {Leonardo}\
  \bibnamefont {Senatore}},\ }\bibfield  {title} {\enquote {\bibinfo {title}
  {{The 2-loop matter power spectrum and the IR-safe integrand}},}\ }\href
  {\doibase 10.1088/1475-7516/2014/07/056} {\bibfield  {journal} {\bibinfo
  {journal} {JCAP}\ }\textbf {\bibinfo {volume} {07}},\ \bibinfo {pages} {056}
  (\bibinfo {year} {2014}{\natexlab{a}})},\ \Eprint
  {http://arxiv.org/abs/1304.4946} {arXiv:1304.4946 [astro-ph.CO]} \BibitemShut
  {NoStop}%
\bibitem [{\citenamefont {Carrasco}\ \emph
  {et~al.}(2014{\natexlab{b}})\citenamefont {Carrasco}, \citenamefont
  {Foreman}, \citenamefont {Green},\ and\ \citenamefont
  {Senatore}}]{Carrasco:2013mua}%
  \BibitemOpen
  \bibfield  {author} {\bibinfo {author} {\bibfnamefont {John Joseph~M.}\
  \bibnamefont {Carrasco}}, \bibinfo {author} {\bibfnamefont {Simon}\
  \bibnamefont {Foreman}}, \bibinfo {author} {\bibfnamefont {Daniel}\
  \bibnamefont {Green}}, \ and\ \bibinfo {author} {\bibfnamefont {Leonardo}\
  \bibnamefont {Senatore}},\ }\bibfield  {title} {\enquote {\bibinfo {title}
  {{The Effective Field Theory of Large Scale Structures at Two Loops}},}\
  }\href {\doibase 10.1088/1475-7516/2014/07/057} {\bibfield  {journal}
  {\bibinfo  {journal} {JCAP}\ }\textbf {\bibinfo {volume} {07}},\ \bibinfo
  {pages} {057} (\bibinfo {year} {2014}{\natexlab{b}})},\ \Eprint
  {http://arxiv.org/abs/1310.0464} {arXiv:1310.0464 [astro-ph.CO]} \BibitemShut
  {NoStop}%
\bibitem [{\citenamefont {Senatore}(2015)}]{Senatore:2014eva}%
  \BibitemOpen
  \bibfield  {author} {\bibinfo {author} {\bibfnamefont {Leonardo}\
  \bibnamefont {Senatore}},\ }\bibfield  {title} {\enquote {\bibinfo {title}
  {{Bias in the Effective Field Theory of Large Scale Structures}},}\ }\href
  {\doibase 10.1088/1475-7516/2015/11/007} {\bibfield  {journal} {\bibinfo
  {journal} {JCAP}\ }\textbf {\bibinfo {volume} {11}},\ \bibinfo {pages} {007}
  (\bibinfo {year} {2015})},\ \Eprint {http://arxiv.org/abs/1406.7843}
  {arXiv:1406.7843 [astro-ph.CO]} \BibitemShut {NoStop}%
\bibitem [{\citenamefont {Mirbabayi}\ \emph {et~al.}(2015)\citenamefont
  {Mirbabayi}, \citenamefont {Schmidt},\ and\ \citenamefont
  {Zaldarriaga}}]{Mirbabayi:2014zca}%
  \BibitemOpen
  \bibfield  {author} {\bibinfo {author} {\bibfnamefont {Mehrdad}\ \bibnamefont
  {Mirbabayi}}, \bibinfo {author} {\bibfnamefont {Fabian}\ \bibnamefont
  {Schmidt}}, \ and\ \bibinfo {author} {\bibfnamefont {Matias}\ \bibnamefont
  {Zaldarriaga}},\ }\bibfield  {title} {\enquote {\bibinfo {title} {{Biased
  Tracers and Time Evolution}},}\ }\href {\doibase
  10.1088/1475-7516/2015/07/030} {\bibfield  {journal} {\bibinfo  {journal}
  {JCAP}\ }\textbf {\bibinfo {volume} {07}},\ \bibinfo {pages} {030} (\bibinfo
  {year} {2015})},\ \Eprint {http://arxiv.org/abs/1412.5169} {arXiv:1412.5169
  [astro-ph.CO]} \BibitemShut {NoStop}%
\bibitem [{\citenamefont {Angulo}\ \emph {et~al.}(2015)\citenamefont {Angulo},
  \citenamefont {Fasiello}, \citenamefont {Senatore},\ and\ \citenamefont
  {Vlah}}]{Angulo:2015eqa}%
  \BibitemOpen
  \bibfield  {author} {\bibinfo {author} {\bibfnamefont {Raul}\ \bibnamefont
  {Angulo}}, \bibinfo {author} {\bibfnamefont {Matteo}\ \bibnamefont
  {Fasiello}}, \bibinfo {author} {\bibfnamefont {Leonardo}\ \bibnamefont
  {Senatore}}, \ and\ \bibinfo {author} {\bibfnamefont {Zvonimir}\ \bibnamefont
  {Vlah}},\ }\bibfield  {title} {\enquote {\bibinfo {title} {{On the Statistics
  of Biased Tracers in the Effective Field Theory of Large Scale
  Structures}},}\ }\href {\doibase 10.1088/1475-7516/2015/09/029,
  10.1088/1475-7516/2015/9/029} {\bibfield  {journal} {\bibinfo  {journal}
  {JCAP}\ }\textbf {\bibinfo {volume} {1509}},\ \bibinfo {pages} {029}
  (\bibinfo {year} {2015})},\ \Eprint {http://arxiv.org/abs/1503.08826}
  {arXiv:1503.08826 [astro-ph.CO]} \BibitemShut {NoStop}%
\bibitem [{\citenamefont {Fujita}\ \emph {et~al.}(2020)\citenamefont {Fujita},
  \citenamefont {Mauerhofer}, \citenamefont {Senatore}, \citenamefont {Vlah},\
  and\ \citenamefont {Angulo}}]{Fujita:2016dne}%
  \BibitemOpen
  \bibfield  {author} {\bibinfo {author} {\bibfnamefont {Tomohiro}\
  \bibnamefont {Fujita}}, \bibinfo {author} {\bibfnamefont {Valentin}\
  \bibnamefont {Mauerhofer}}, \bibinfo {author} {\bibfnamefont {Leonardo}\
  \bibnamefont {Senatore}}, \bibinfo {author} {\bibfnamefont {Zvonimir}\
  \bibnamefont {Vlah}}, \ and\ \bibinfo {author} {\bibfnamefont {Raul}\
  \bibnamefont {Angulo}},\ }\bibfield  {title} {\enquote {\bibinfo {title}
  {{Very Massive Tracers and Higher Derivative Biases}},}\ }\href {\doibase
  10.1088/1475-7516/2020/01/009} {\bibfield  {journal} {\bibinfo  {journal}
  {JCAP}\ }\textbf {\bibinfo {volume} {01}},\ \bibinfo {pages} {009} (\bibinfo
  {year} {2020})},\ \Eprint {http://arxiv.org/abs/1609.00717} {arXiv:1609.00717
  [astro-ph.CO]} \BibitemShut {NoStop}%
\bibitem [{\citenamefont {Perko}\ \emph {et~al.}(2016)\citenamefont {Perko},
  \citenamefont {Senatore}, \citenamefont {Jennings},\ and\ \citenamefont
  {Wechsler}}]{Perko:2016puo}%
  \BibitemOpen
  \bibfield  {author} {\bibinfo {author} {\bibfnamefont {Ashley}\ \bibnamefont
  {Perko}}, \bibinfo {author} {\bibfnamefont {Leonardo}\ \bibnamefont
  {Senatore}}, \bibinfo {author} {\bibfnamefont {Elise}\ \bibnamefont
  {Jennings}}, \ and\ \bibinfo {author} {\bibfnamefont {Risa~H.}\ \bibnamefont
  {Wechsler}},\ }\bibfield  {title} {\enquote {\bibinfo {title} {{Biased
  Tracers in Redshift Space in the EFT of Large-Scale Structure}},}\
  }\href@noop {} {\  (\bibinfo {year} {2016})},\ \Eprint
  {http://arxiv.org/abs/1610.09321} {arXiv:1610.09321 [astro-ph.CO]}
  \BibitemShut {NoStop}%
\bibitem [{\citenamefont {Nadler}\ \emph {et~al.}(2018)\citenamefont {Nadler},
  \citenamefont {Perko},\ and\ \citenamefont {Senatore}}]{Nadler:2017qto}%
  \BibitemOpen
  \bibfield  {author} {\bibinfo {author} {\bibfnamefont {Ethan~O.}\
  \bibnamefont {Nadler}}, \bibinfo {author} {\bibfnamefont {Ashley}\
  \bibnamefont {Perko}}, \ and\ \bibinfo {author} {\bibfnamefont {Leonardo}\
  \bibnamefont {Senatore}},\ }\bibfield  {title} {\enquote {\bibinfo {title}
  {{On the Bispectra of Very Massive Tracers in the Effective Field Theory of
  Large-Scale Structure}},}\ }\href {\doibase 10.1088/1475-7516/2018/02/058}
  {\bibfield  {journal} {\bibinfo  {journal} {JCAP}\ }\textbf {\bibinfo
  {volume} {02}},\ \bibinfo {pages} {058} (\bibinfo {year} {2018})},\ \Eprint
  {http://arxiv.org/abs/1710.10308} {arXiv:1710.10308 [astro-ph.CO]}
  \BibitemShut {NoStop}%
\bibitem [{\citenamefont {D'Amico}\ \emph
  {et~al.}(2020{\natexlab{a}})\citenamefont {D'Amico}, \citenamefont {Gleyzes},
  \citenamefont {Kokron}, \citenamefont {Markovic}, \citenamefont {Senatore},
  \citenamefont {Zhang}, \citenamefont {Beutler},\ and\ \citenamefont
  {Gil-Marín}}]{DAmico:2019fhj}%
  \BibitemOpen
  \bibfield  {author} {\bibinfo {author} {\bibfnamefont {Guido}\ \bibnamefont
  {D'Amico}}, \bibinfo {author} {\bibfnamefont {Jérôme}\ \bibnamefont
  {Gleyzes}}, \bibinfo {author} {\bibfnamefont {Nickolas}\ \bibnamefont
  {Kokron}}, \bibinfo {author} {\bibfnamefont {Katarina}\ \bibnamefont
  {Markovic}}, \bibinfo {author} {\bibfnamefont {Leonardo}\ \bibnamefont
  {Senatore}}, \bibinfo {author} {\bibfnamefont {Pierre}\ \bibnamefont
  {Zhang}}, \bibinfo {author} {\bibfnamefont {Florian}\ \bibnamefont
  {Beutler}}, \ and\ \bibinfo {author} {\bibfnamefont {Héctor}\ \bibnamefont
  {Gil-Marín}},\ }\bibfield  {title} {\enquote {\bibinfo {title} {{The
  Cosmological Analysis of the SDSS/BOSS data from the Effective Field Theory
  of Large-Scale Structure}},}\ }\href {\doibase 10.1088/1475-7516/2020/05/005}
  {\bibfield  {journal} {\bibinfo  {journal} {JCAP}\ }\textbf {\bibinfo
  {volume} {05}},\ \bibinfo {pages} {005} (\bibinfo {year}
  {2020}{\natexlab{a}})},\ \Eprint {http://arxiv.org/abs/1909.05271}
  {arXiv:1909.05271 [astro-ph.CO]} \BibitemShut {NoStop}%
\bibitem [{\citenamefont {Simon}\ \emph
  {et~al.}(2023{\natexlab{a}})\citenamefont {Simon}, \citenamefont {Zhang},\
  and\ \citenamefont {Poulin}}]{Simon:2022csv}%
  \BibitemOpen
  \bibfield  {author} {\bibinfo {author} {\bibfnamefont {Th\'eo}\ \bibnamefont
  {Simon}}, \bibinfo {author} {\bibfnamefont {Pierre}\ \bibnamefont {Zhang}}, \
  and\ \bibinfo {author} {\bibfnamefont {Vivian}\ \bibnamefont {Poulin}},\
  }\bibfield  {title} {\enquote {\bibinfo {title} {{Cosmological inference from
  the EFTofLSS: the eBOSS QSO full-shape analysis}},}\ }\href {\doibase
  10.1088/1475-7516/2023/07/041} {\bibfield  {journal} {\bibinfo  {journal}
  {JCAP}\ }\textbf {\bibinfo {volume} {07}},\ \bibinfo {pages} {041} (\bibinfo
  {year} {2023}{\natexlab{a}})},\ \Eprint {http://arxiv.org/abs/2210.14931}
  {arXiv:2210.14931 [astro-ph.CO]} \BibitemShut {NoStop}%
\bibitem [{\citenamefont {Ivanov}\ \emph
  {et~al.}(2020{\natexlab{a}})\citenamefont {Ivanov}, \citenamefont
  {Simonovi\'c},\ and\ \citenamefont {Zaldarriaga}}]{Ivanov:2019pdj}%
  \BibitemOpen
  \bibfield  {author} {\bibinfo {author} {\bibfnamefont {Mikhail~M.}\
  \bibnamefont {Ivanov}}, \bibinfo {author} {\bibfnamefont {Marko}\
  \bibnamefont {Simonovi\'c}}, \ and\ \bibinfo {author} {\bibfnamefont
  {Matias}\ \bibnamefont {Zaldarriaga}},\ }\bibfield  {title} {\enquote
  {\bibinfo {title} {{Cosmological Parameters from the BOSS Galaxy Power
  Spectrum}},}\ }\href {\doibase 10.1088/1475-7516/2020/05/042} {\bibfield
  {journal} {\bibinfo  {journal} {JCAP}\ }\textbf {\bibinfo {volume} {05}},\
  \bibinfo {pages} {042} (\bibinfo {year} {2020}{\natexlab{a}})},\ \Eprint
  {http://arxiv.org/abs/1909.05277} {arXiv:1909.05277 [astro-ph.CO]}
  \BibitemShut {NoStop}%
\bibitem [{\citenamefont {Colas}\ \emph {et~al.}(2020)\citenamefont {Colas},
  \citenamefont {D'amico}, \citenamefont {Senatore}, \citenamefont {Zhang},\
  and\ \citenamefont {Beutler}}]{Colas:2019ret}%
  \BibitemOpen
  \bibfield  {author} {\bibinfo {author} {\bibfnamefont {Thomas}\ \bibnamefont
  {Colas}}, \bibinfo {author} {\bibfnamefont {Guido}\ \bibnamefont {D'amico}},
  \bibinfo {author} {\bibfnamefont {Leonardo}\ \bibnamefont {Senatore}},
  \bibinfo {author} {\bibfnamefont {Pierre}\ \bibnamefont {Zhang}}, \ and\
  \bibinfo {author} {\bibfnamefont {Florian}\ \bibnamefont {Beutler}},\
  }\bibfield  {title} {\enquote {\bibinfo {title} {{Efficient Cosmological
  Analysis of the SDSS/BOSS data from the Effective Field Theory of Large-Scale
  Structure}},}\ }\href {\doibase 10.1088/1475-7516/2020/06/001} {\bibfield
  {journal} {\bibinfo  {journal} {JCAP}\ }\textbf {\bibinfo {volume} {06}},\
  \bibinfo {pages} {001} (\bibinfo {year} {2020})},\ \Eprint
  {http://arxiv.org/abs/1909.07951} {arXiv:1909.07951 [astro-ph.CO]}
  \BibitemShut {NoStop}%
\bibitem [{\citenamefont {D'Amico}\ \emph
  {et~al.}(2020{\natexlab{b}})\citenamefont {D'Amico}, \citenamefont
  {Senatore},\ and\ \citenamefont {Zhang}}]{DAmico:2020kxu}%
  \BibitemOpen
  \bibfield  {author} {\bibinfo {author} {\bibfnamefont {Guido}\ \bibnamefont
  {D'Amico}}, \bibinfo {author} {\bibfnamefont {Leonardo}\ \bibnamefont
  {Senatore}}, \ and\ \bibinfo {author} {\bibfnamefont {Pierre}\ \bibnamefont
  {Zhang}},\ }\bibfield  {title} {\enquote {\bibinfo {title} {{Limits on $w$CDM
  from the EFTofLSS with the PyBird code}},}\ }\href@noop {} {\  (\bibinfo
  {year} {2020}{\natexlab{b}})},\ \Eprint {http://arxiv.org/abs/2003.07956}
  {arXiv:2003.07956 [astro-ph.CO]} \BibitemShut {NoStop}%
\bibitem [{\citenamefont {D'Amico}\ \emph
  {et~al.}(2020{\natexlab{c}})\citenamefont {D'Amico}, \citenamefont {Donath},
  \citenamefont {Senatore},\ and\ \citenamefont {Zhang}}]{DAmico:2020tty}%
  \BibitemOpen
  \bibfield  {author} {\bibinfo {author} {\bibfnamefont {Guido}\ \bibnamefont
  {D'Amico}}, \bibinfo {author} {\bibfnamefont {Yaniv}\ \bibnamefont {Donath}},
  \bibinfo {author} {\bibfnamefont {Leonardo}\ \bibnamefont {Senatore}}, \ and\
  \bibinfo {author} {\bibfnamefont {Pierre}\ \bibnamefont {Zhang}},\ }\bibfield
   {title} {\enquote {\bibinfo {title} {{Limits on Clustering and Smooth
  Quintessence from the EFTofLSS}},}\ }\href@noop {} {\  (\bibinfo {year}
  {2020}{\natexlab{c}})},\ \Eprint {http://arxiv.org/abs/2012.07554}
  {arXiv:2012.07554 [astro-ph.CO]} \BibitemShut {NoStop}%
\bibitem [{\citenamefont {Simon}\ \emph {et~al.}(2022)\citenamefont {Simon},
  \citenamefont {Franco~Abell\'an}, \citenamefont {Du}, \citenamefont
  {Poulin},\ and\ \citenamefont {Tsai}}]{Simon:2022ftd}%
  \BibitemOpen
  \bibfield  {author} {\bibinfo {author} {\bibfnamefont {Th\'eo}\ \bibnamefont
  {Simon}}, \bibinfo {author} {\bibfnamefont {Guillermo}\ \bibnamefont
  {Franco~Abell\'an}}, \bibinfo {author} {\bibfnamefont {Peizhi}\ \bibnamefont
  {Du}}, \bibinfo {author} {\bibfnamefont {Vivian}\ \bibnamefont {Poulin}}, \
  and\ \bibinfo {author} {\bibfnamefont {Yuhsin}\ \bibnamefont {Tsai}},\
  }\bibfield  {title} {\enquote {\bibinfo {title} {{Constraining decaying dark
  matter with BOSS data and the effective field theory of large-scale
  structures}},}\ }\href {\doibase 10.1103/PhysRevD.106.023516} {\bibfield
  {journal} {\bibinfo  {journal} {Phys. Rev. D}\ }\textbf {\bibinfo {volume}
  {106}},\ \bibinfo {pages} {023516} (\bibinfo {year} {2022})},\ \Eprint
  {http://arxiv.org/abs/2203.07440} {arXiv:2203.07440 [astro-ph.CO]}
  \BibitemShut {NoStop}%
\bibitem [{\citenamefont {Simon}\ \emph
  {et~al.}(2023{\natexlab{b}})\citenamefont {Simon}, \citenamefont {Zhang},
  \citenamefont {Poulin},\ and\ \citenamefont {Smith}}]{Simon:2022lde}%
  \BibitemOpen
  \bibfield  {author} {\bibinfo {author} {\bibfnamefont {Th\'eo}\ \bibnamefont
  {Simon}}, \bibinfo {author} {\bibfnamefont {Pierre}\ \bibnamefont {Zhang}},
  \bibinfo {author} {\bibfnamefont {Vivian}\ \bibnamefont {Poulin}}, \ and\
  \bibinfo {author} {\bibfnamefont {Tristan~L.}\ \bibnamefont {Smith}},\
  }\bibfield  {title} {\enquote {\bibinfo {title} {{Consistency of effective
  field theory analyses of the BOSS power spectrum}},}\ }\href {\doibase
  10.1103/PhysRevD.107.123530} {\bibfield  {journal} {\bibinfo  {journal}
  {Phys. Rev. D}\ }\textbf {\bibinfo {volume} {107}},\ \bibinfo {pages}
  {123530} (\bibinfo {year} {2023}{\natexlab{b}})},\ \Eprint
  {http://arxiv.org/abs/2208.05929} {arXiv:2208.05929 [astro-ph.CO]}
  \BibitemShut {NoStop}%
\bibitem [{\citenamefont {Chen}\ \emph {et~al.}(2022)\citenamefont {Chen},
  \citenamefont {Vlah},\ and\ \citenamefont {White}}]{Chen:2021wdi}%
  \BibitemOpen
  \bibfield  {author} {\bibinfo {author} {\bibfnamefont {Shi-Fan}\ \bibnamefont
  {Chen}}, \bibinfo {author} {\bibfnamefont {Zvonimir}\ \bibnamefont {Vlah}}, \
  and\ \bibinfo {author} {\bibfnamefont {Martin}\ \bibnamefont {White}},\
  }\bibfield  {title} {\enquote {\bibinfo {title} {{A new analysis of galaxy
  2-point functions in the BOSS survey, including full-shape information and
  post-reconstruction BAO}},}\ }\href {\doibase 10.1088/1475-7516/2022/02/008}
  {\bibfield  {journal} {\bibinfo  {journal} {JCAP}\ }\textbf {\bibinfo
  {volume} {02}},\ \bibinfo {pages} {008} (\bibinfo {year} {2022})},\ \Eprint
  {http://arxiv.org/abs/2110.05530} {arXiv:2110.05530 [astro-ph.CO]}
  \BibitemShut {NoStop}%
\bibitem [{\citenamefont {Zhang}\ \emph {et~al.}(2022)\citenamefont {Zhang},
  \citenamefont {D'Amico}, \citenamefont {Senatore}, \citenamefont {Zhao},\
  and\ \citenamefont {Cai}}]{Zhang:2021yna}%
  \BibitemOpen
  \bibfield  {author} {\bibinfo {author} {\bibfnamefont {Pierre}\ \bibnamefont
  {Zhang}}, \bibinfo {author} {\bibfnamefont {Guido}\ \bibnamefont {D'Amico}},
  \bibinfo {author} {\bibfnamefont {Leonardo}\ \bibnamefont {Senatore}},
  \bibinfo {author} {\bibfnamefont {Cheng}\ \bibnamefont {Zhao}}, \ and\
  \bibinfo {author} {\bibfnamefont {Yifu}\ \bibnamefont {Cai}},\ }\bibfield
  {title} {\enquote {\bibinfo {title} {{BOSS Correlation Function analysis from
  the Effective Field Theory of Large-Scale Structure}},}\ }\href {\doibase
  10.1088/1475-7516/2022/02/036} {\bibfield  {journal} {\bibinfo  {journal}
  {JCAP}\ }\textbf {\bibinfo {volume} {02}},\ \bibinfo {pages} {036} (\bibinfo
  {year} {2022})},\ \Eprint {http://arxiv.org/abs/2110.07539} {arXiv:2110.07539
  [astro-ph.CO]} \BibitemShut {NoStop}%
\bibitem [{\citenamefont {Philcox}\ and\ \citenamefont
  {Ivanov}(2022)}]{Philcox:2021kcw}%
  \BibitemOpen
  \bibfield  {author} {\bibinfo {author} {\bibfnamefont {Oliver H.~E.}\
  \bibnamefont {Philcox}}\ and\ \bibinfo {author} {\bibfnamefont {Mikhail~M.}\
  \bibnamefont {Ivanov}},\ }\bibfield  {title} {\enquote {\bibinfo {title}
  {{BOSS DR12 full-shape cosmology: \ensuremath{\Lambda}CDM constraints from
  the large-scale galaxy power spectrum and bispectrum monopole}},}\ }\href
  {\doibase 10.1103/PhysRevD.105.043517} {\bibfield  {journal} {\bibinfo
  {journal} {Phys. Rev. D}\ }\textbf {\bibinfo {volume} {105}},\ \bibinfo
  {pages} {043517} (\bibinfo {year} {2022})},\ \Eprint
  {http://arxiv.org/abs/2112.04515} {arXiv:2112.04515 [astro-ph.CO]}
  \BibitemShut {NoStop}%
\bibitem [{\citenamefont {Kumar}\ \emph {et~al.}(2022)\citenamefont {Kumar},
  \citenamefont {Nunes},\ and\ \citenamefont {Yadav}}]{Kumar:2022vee}%
  \BibitemOpen
  \bibfield  {author} {\bibinfo {author} {\bibfnamefont {Suresh}\ \bibnamefont
  {Kumar}}, \bibinfo {author} {\bibfnamefont {Rafael~C.}\ \bibnamefont
  {Nunes}}, \ and\ \bibinfo {author} {\bibfnamefont {Priya}\ \bibnamefont
  {Yadav}},\ }\bibfield  {title} {\enquote {\bibinfo {title} {{Updating
  non-standard neutrinos properties with Planck-CMB data and full-shape
  analysis of BOSS and eBOSS galaxies}},}\ }\href {\doibase
  10.1088/1475-7516/2022/09/060} {\bibfield  {journal} {\bibinfo  {journal}
  {JCAP}\ }\textbf {\bibinfo {volume} {09}},\ \bibinfo {pages} {060} (\bibinfo
  {year} {2022})},\ \Eprint {http://arxiv.org/abs/2205.04292} {arXiv:2205.04292
  [astro-ph.CO]} \BibitemShut {NoStop}%
\bibitem [{\citenamefont {Nunes}\ \emph {et~al.}(2022)\citenamefont {Nunes},
  \citenamefont {Vagnozzi}, \citenamefont {Kumar}, \citenamefont
  {Di~Valentino},\ and\ \citenamefont {Mena}}]{Nunes:2022bhn}%
  \BibitemOpen
  \bibfield  {author} {\bibinfo {author} {\bibfnamefont {Rafael~C.}\
  \bibnamefont {Nunes}}, \bibinfo {author} {\bibfnamefont {Sunny}\ \bibnamefont
  {Vagnozzi}}, \bibinfo {author} {\bibfnamefont {Suresh}\ \bibnamefont
  {Kumar}}, \bibinfo {author} {\bibfnamefont {Eleonora}\ \bibnamefont
  {Di~Valentino}}, \ and\ \bibinfo {author} {\bibfnamefont {Olga}\ \bibnamefont
  {Mena}},\ }\bibfield  {title} {\enquote {\bibinfo {title} {{New tests of dark
  sector interactions from the full-shape galaxy power spectrum}},}\ }\href
  {\doibase 10.1103/PhysRevD.105.123506} {\bibfield  {journal} {\bibinfo
  {journal} {Phys. Rev. D}\ }\textbf {\bibinfo {volume} {105}},\ \bibinfo
  {pages} {123506} (\bibinfo {year} {2022})},\ \Eprint
  {http://arxiv.org/abs/2203.08093} {arXiv:2203.08093 [astro-ph.CO]}
  \BibitemShut {NoStop}%
\bibitem [{\citenamefont {Lagu\"e}\ \emph {et~al.}(2021)\citenamefont
  {Lagu\"e}, \citenamefont {Bond}, \citenamefont {Hlo\v{z}ek}, \citenamefont
  {Rogers}, \citenamefont {Marsh},\ and\ \citenamefont {Grin}}]{Lague:2021frh}%
  \BibitemOpen
  \bibfield  {author} {\bibinfo {author} {\bibfnamefont {Alex}\ \bibnamefont
  {Lagu\"e}}, \bibinfo {author} {\bibfnamefont {J.~Richard}\ \bibnamefont
  {Bond}}, \bibinfo {author} {\bibfnamefont {Ren\'ee}\ \bibnamefont
  {Hlo\v{z}ek}}, \bibinfo {author} {\bibfnamefont {Keir~K.}\ \bibnamefont
  {Rogers}}, \bibinfo {author} {\bibfnamefont {David J.~E.}\ \bibnamefont
  {Marsh}}, \ and\ \bibinfo {author} {\bibfnamefont {Daniel}\ \bibnamefont
  {Grin}},\ }\bibfield  {title} {\enquote {\bibinfo {title} {{Constraining
  Ultralight Axions with Galaxy Surveys}},}\ }\href@noop {} {\  (\bibinfo
  {year} {2021})},\ \Eprint {http://arxiv.org/abs/2104.07802} {arXiv:2104.07802
  [astro-ph.CO]} \BibitemShut {NoStop}%
\bibitem [{\citenamefont {Philcox}\ \emph {et~al.}(2020)\citenamefont
  {Philcox}, \citenamefont {Sherwin}, \citenamefont {Farren},\ and\
  \citenamefont {Baxter}}]{Philcox:2020xbv}%
  \BibitemOpen
  \bibfield  {author} {\bibinfo {author} {\bibfnamefont {Oliver~H.E.}\
  \bibnamefont {Philcox}}, \bibinfo {author} {\bibfnamefont {Blake~D.}\
  \bibnamefont {Sherwin}}, \bibinfo {author} {\bibfnamefont {Gerrit~S.}\
  \bibnamefont {Farren}}, \ and\ \bibinfo {author} {\bibfnamefont {Eric~J.}\
  \bibnamefont {Baxter}},\ }\bibfield  {title} {\enquote {\bibinfo {title}
  {{Determining the Hubble Constant without the Sound Horizon: Measurements
  from Galaxy Surveys}},}\ }\href@noop {} {\  (\bibinfo {year} {2020})},\
  \Eprint {http://arxiv.org/abs/2008.08084} {arXiv:2008.08084 [astro-ph.CO]}
  \BibitemShut {NoStop}%
\bibitem [{\citenamefont {Smith}\ \emph {et~al.}(2022)\citenamefont {Smith},
  \citenamefont {Poulin},\ and\ \citenamefont {Simon}}]{Smith:2022iax}%
  \BibitemOpen
  \bibfield  {author} {\bibinfo {author} {\bibfnamefont {Tristan~L.}\
  \bibnamefont {Smith}}, \bibinfo {author} {\bibfnamefont {Vivian}\
  \bibnamefont {Poulin}}, \ and\ \bibinfo {author} {\bibfnamefont {Th\'eo}\
  \bibnamefont {Simon}},\ }\bibfield  {title} {\enquote {\bibinfo {title}
  {{Assessing the robustness of sound horizon-free determinations of the Hubble
  constant}},}\ }\href@noop {} {\  (\bibinfo {year} {2022})},\ \Eprint
  {http://arxiv.org/abs/2208.12992} {arXiv:2208.12992 [astro-ph.CO]}
  \BibitemShut {NoStop}%
\bibitem [{\citenamefont {Moretti}\ \emph {et~al.}(2023)\citenamefont
  {Moretti}, \citenamefont {Tsedrik}, \citenamefont {Carrilho},\ and\
  \citenamefont {Pourtsidou}}]{Moretti:2023drg}%
  \BibitemOpen
  \bibfield  {author} {\bibinfo {author} {\bibfnamefont {Chiara}\ \bibnamefont
  {Moretti}}, \bibinfo {author} {\bibfnamefont {Maria}\ \bibnamefont
  {Tsedrik}}, \bibinfo {author} {\bibfnamefont {Pedro}\ \bibnamefont
  {Carrilho}}, \ and\ \bibinfo {author} {\bibfnamefont {Alkistis}\ \bibnamefont
  {Pourtsidou}},\ }\bibfield  {title} {\enquote {\bibinfo {title} {{Modified
  gravity and massive neutrinos: constraints from the full shape analysis of
  BOSS galaxies and forecasts for Stage IV surveys}},}\ }\href@noop {} {\
  (\bibinfo {year} {2023})},\ \Eprint {http://arxiv.org/abs/2306.09275}
  {arXiv:2306.09275 [astro-ph.CO]} \BibitemShut {NoStop}%
\bibitem [{\citenamefont {Rubira}\ \emph {et~al.}(2023)\citenamefont {Rubira},
  \citenamefont {Mazoun},\ and\ \citenamefont {Garny}}]{Rubira:2022xhb}%
  \BibitemOpen
  \bibfield  {author} {\bibinfo {author} {\bibfnamefont {Henrique}\
  \bibnamefont {Rubira}}, \bibinfo {author} {\bibfnamefont {Asmaa}\
  \bibnamefont {Mazoun}}, \ and\ \bibinfo {author} {\bibfnamefont {Mathias}\
  \bibnamefont {Garny}},\ }\bibfield  {title} {\enquote {\bibinfo {title}
  {{Full-shape BOSS constraints on dark matter interacting with dark radiation
  and lifting the S8 tension}},}\ }\href {\doibase
  10.1088/1475-7516/2023/01/034} {\bibfield  {journal} {\bibinfo  {journal}
  {JCAP}\ }\textbf {\bibinfo {volume} {01}},\ \bibinfo {pages} {034} (\bibinfo
  {year} {2023})},\ \Eprint {http://arxiv.org/abs/2209.03974} {arXiv:2209.03974
  [astro-ph.CO]} \BibitemShut {NoStop}%
\bibitem [{\citenamefont {Sch\"oneberg}\ \emph {et~al.}(2023)\citenamefont
  {Sch\"oneberg}, \citenamefont {Franco~Abell\'an}, \citenamefont {Simon},
  \citenamefont {Bartlett}, \citenamefont {Patel},\ and\ \citenamefont
  {Smith}}]{Schoneberg:2023rnx}%
  \BibitemOpen
  \bibfield  {author} {\bibinfo {author} {\bibfnamefont {Nils}\ \bibnamefont
  {Sch\"oneberg}}, \bibinfo {author} {\bibfnamefont {Guillermo}\ \bibnamefont
  {Franco~Abell\'an}}, \bibinfo {author} {\bibfnamefont {Th\'eo}\ \bibnamefont
  {Simon}}, \bibinfo {author} {\bibfnamefont {Alexa}\ \bibnamefont {Bartlett}},
  \bibinfo {author} {\bibfnamefont {Yashvi}\ \bibnamefont {Patel}}, \ and\
  \bibinfo {author} {\bibfnamefont {Tristan~L.}\ \bibnamefont {Smith}},\
  }\bibfield  {title} {\enquote {\bibinfo {title} {{The weak, the strong and
  the ugly -- A comparative analysis of interacting stepped dark radiation}},}\
  }\href@noop {} {\  (\bibinfo {year} {2023})},\ \Eprint
  {http://arxiv.org/abs/2306.12469} {arXiv:2306.12469 [astro-ph.CO]}
  \BibitemShut {NoStop}%
\bibitem [{\citenamefont {Holm}\ \emph {et~al.}(2023)\citenamefont {Holm},
  \citenamefont {Herold}, \citenamefont {Simon}, \citenamefont {Ferreira},
  \citenamefont {Hannestad}, \citenamefont {Poulin},\ and\ \citenamefont
  {Tram}}]{Holm:2023laa}%
  \BibitemOpen
  \bibfield  {author} {\bibinfo {author} {\bibfnamefont {Emil~Brinch}\
  \bibnamefont {Holm}}, \bibinfo {author} {\bibfnamefont {Laura}\ \bibnamefont
  {Herold}}, \bibinfo {author} {\bibfnamefont {Th\'eo}\ \bibnamefont {Simon}},
  \bibinfo {author} {\bibfnamefont {Elisa G.~M.}\ \bibnamefont {Ferreira}},
  \bibinfo {author} {\bibfnamefont {Steen}\ \bibnamefont {Hannestad}}, \bibinfo
  {author} {\bibfnamefont {Vivian}\ \bibnamefont {Poulin}}, \ and\ \bibinfo
  {author} {\bibfnamefont {Thomas}\ \bibnamefont {Tram}},\ }\bibfield  {title}
  {\enquote {\bibinfo {title} {{Bayesian and frequentist investigation of prior
  effects in EFTofLSS analyses of full-shape BOSS and eBOSS data}},}\
  }\href@noop {} {\  (\bibinfo {year} {2023})},\ \Eprint
  {http://arxiv.org/abs/2309.04468} {arXiv:2309.04468 [astro-ph.CO]}
  \BibitemShut {NoStop}%
\bibitem [{\citenamefont {Simon}\ \emph
  {et~al.}(2023{\natexlab{c}})\citenamefont {Simon}, \citenamefont {Zhang},
  \citenamefont {Poulin},\ and\ \citenamefont {Smith}}]{Simon:2022adh}%
  \BibitemOpen
  \bibfield  {author} {\bibinfo {author} {\bibfnamefont {Th\'eo}\ \bibnamefont
  {Simon}}, \bibinfo {author} {\bibfnamefont {Pierre}\ \bibnamefont {Zhang}},
  \bibinfo {author} {\bibfnamefont {Vivian}\ \bibnamefont {Poulin}}, \ and\
  \bibinfo {author} {\bibfnamefont {Tristan~L.}\ \bibnamefont {Smith}},\
  }\bibfield  {title} {\enquote {\bibinfo {title} {{Updated constraints from
  the effective field theory analysis of the BOSS power spectrum on early dark
  energy}},}\ }\href {\doibase 10.1103/PhysRevD.107.063505} {\bibfield
  {journal} {\bibinfo  {journal} {Phys. Rev. D}\ }\textbf {\bibinfo {volume}
  {107}},\ \bibinfo {pages} {063505} (\bibinfo {year} {2023}{\natexlab{c}})},\
  \Eprint {http://arxiv.org/abs/2208.05930} {arXiv:2208.05930 [astro-ph.CO]}
  \BibitemShut {NoStop}%
\bibitem [{\citenamefont {D'Amico}\ \emph
  {et~al.}(2020{\natexlab{d}})\citenamefont {D'Amico}, \citenamefont
  {Senatore}, \citenamefont {Zhang},\ and\ \citenamefont
  {Zheng}}]{DAmico:2020ods}%
  \BibitemOpen
  \bibfield  {author} {\bibinfo {author} {\bibfnamefont {Guido}\ \bibnamefont
  {D'Amico}}, \bibinfo {author} {\bibfnamefont {Leonardo}\ \bibnamefont
  {Senatore}}, \bibinfo {author} {\bibfnamefont {Pierre}\ \bibnamefont
  {Zhang}}, \ and\ \bibinfo {author} {\bibfnamefont {Henry}\ \bibnamefont
  {Zheng}},\ }\bibfield  {title} {\enquote {\bibinfo {title} {{The Hubble
  Tension in Light of the Full-Shape Analysis of Large-Scale Structure
  Data}},}\ }\href@noop {} {\  (\bibinfo {year} {2020}{\natexlab{d}})},\
  \Eprint {http://arxiv.org/abs/2006.12420} {arXiv:2006.12420 [astro-ph.CO]}
  \BibitemShut {NoStop}%
\bibitem [{\citenamefont {Smith}\ \emph {et~al.}(2021)\citenamefont {Smith},
  \citenamefont {Poulin}, \citenamefont {Bernal}, \citenamefont {Boddy},
  \citenamefont {Kamionkowski},\ and\ \citenamefont {Murgia}}]{Smith:2020rxx}%
  \BibitemOpen
  \bibfield  {author} {\bibinfo {author} {\bibfnamefont {Tristan~L.}\
  \bibnamefont {Smith}}, \bibinfo {author} {\bibfnamefont {Vivian}\
  \bibnamefont {Poulin}}, \bibinfo {author} {\bibfnamefont {Jos\'e~Luis}\
  \bibnamefont {Bernal}}, \bibinfo {author} {\bibfnamefont {Kimberly~K.}\
  \bibnamefont {Boddy}}, \bibinfo {author} {\bibfnamefont {Marc}\ \bibnamefont
  {Kamionkowski}}, \ and\ \bibinfo {author} {\bibfnamefont {Riccardo}\
  \bibnamefont {Murgia}},\ }\bibfield  {title} {\enquote {\bibinfo {title}
  {{Early dark energy is not excluded by current large-scale structure
  data}},}\ }\href {\doibase 10.1103/PhysRevD.103.123542} {\bibfield  {journal}
  {\bibinfo  {journal} {Phys. Rev. D}\ }\textbf {\bibinfo {volume} {103}},\
  \bibinfo {pages} {123542} (\bibinfo {year} {2021})},\ \Eprint
  {http://arxiv.org/abs/2009.10740} {arXiv:2009.10740 [astro-ph.CO]}
  \BibitemShut {NoStop}%
\bibitem [{\citenamefont {Ivanov}\ \emph
  {et~al.}(2020{\natexlab{b}})\citenamefont {Ivanov}, \citenamefont
  {McDonough}, \citenamefont {Hill}, \citenamefont {Simonovi\'c}, \citenamefont
  {Toomey}, \citenamefont {Alexander},\ and\ \citenamefont
  {Zaldarriaga}}]{Ivanov:2020ril}%
  \BibitemOpen
  \bibfield  {author} {\bibinfo {author} {\bibfnamefont {Mikhail~M.}\
  \bibnamefont {Ivanov}}, \bibinfo {author} {\bibfnamefont {Evan}\ \bibnamefont
  {McDonough}}, \bibinfo {author} {\bibfnamefont {J.~Colin}\ \bibnamefont
  {Hill}}, \bibinfo {author} {\bibfnamefont {Marko}\ \bibnamefont
  {Simonovi\'c}}, \bibinfo {author} {\bibfnamefont {Michael~W.}\ \bibnamefont
  {Toomey}}, \bibinfo {author} {\bibfnamefont {Stephon}\ \bibnamefont
  {Alexander}}, \ and\ \bibinfo {author} {\bibfnamefont {Matias}\ \bibnamefont
  {Zaldarriaga}},\ }\bibfield  {title} {\enquote {\bibinfo {title}
  {{Constraining Early Dark Energy with Large-Scale Structure}},}\ }\href
  {\doibase 10.1103/PhysRevD.102.103502} {\bibfield  {journal} {\bibinfo
  {journal} {Phys. Rev. D}\ }\textbf {\bibinfo {volume} {102}},\ \bibinfo
  {pages} {103502} (\bibinfo {year} {2020}{\natexlab{b}})},\ \Eprint
  {http://arxiv.org/abs/2006.11235} {arXiv:2006.11235 [astro-ph.CO]}
  \BibitemShut {NoStop}%
\bibitem [{\citenamefont {Blanchard}\ \emph {et~al.}(2022)\citenamefont
  {Blanchard}, \citenamefont {H\'eloret}, \citenamefont {Ili\'c}, \citenamefont
  {Lamine},\ and\ \citenamefont {Tutusaus}}]{Blanchard:2022xkk}%
  \BibitemOpen
  \bibfield  {author} {\bibinfo {author} {\bibfnamefont {Alain}\ \bibnamefont
  {Blanchard}}, \bibinfo {author} {\bibfnamefont {Jean-Yves}\ \bibnamefont
  {H\'eloret}}, \bibinfo {author} {\bibfnamefont {St\'ephane}\ \bibnamefont
  {Ili\'c}}, \bibinfo {author} {\bibfnamefont {Brahim}\ \bibnamefont {Lamine}},
  \ and\ \bibinfo {author} {\bibfnamefont {Isaac}\ \bibnamefont {Tutusaus}},\
  }\bibfield  {title} {\enquote {\bibinfo {title} {{$\Lambda$CDM is alive and
  well}},}\ }\href@noop {} {\  (\bibinfo {year} {2022})},\ \Eprint
  {http://arxiv.org/abs/2205.05017} {arXiv:2205.05017 [astro-ph.CO]}
  \BibitemShut {NoStop}%
\bibitem [{\citenamefont {Poulin}\ \emph {et~al.}(2018)\citenamefont {Poulin},
  \citenamefont {Smith}, \citenamefont {Grin}, \citenamefont {Karwal},\ and\
  \citenamefont {Kamionkowski}}]{Poulin:2018dzj}%
  \BibitemOpen
  \bibfield  {author} {\bibinfo {author} {\bibfnamefont {Vivian}\ \bibnamefont
  {Poulin}}, \bibinfo {author} {\bibfnamefont {Tristan~L.}\ \bibnamefont
  {Smith}}, \bibinfo {author} {\bibfnamefont {Daniel}\ \bibnamefont {Grin}},
  \bibinfo {author} {\bibfnamefont {Tanvi}\ \bibnamefont {Karwal}}, \ and\
  \bibinfo {author} {\bibfnamefont {Marc}\ \bibnamefont {Kamionkowski}},\
  }\bibfield  {title} {\enquote {\bibinfo {title} {{Cosmological implications
  of ultralight axionlike fields}},}\ }\href {\doibase
  10.1103/PhysRevD.98.083525} {\bibfield  {journal} {\bibinfo  {journal} {Phys.
  Rev.}\ }\textbf {\bibinfo {volume} {D98}},\ \bibinfo {pages} {083525}
  (\bibinfo {year} {2018})},\ \Eprint {http://arxiv.org/abs/1806.10608}
  {arXiv:1806.10608 [astro-ph.CO]} \BibitemShut {NoStop}%
\bibitem [{\citenamefont {Armendariz-Picon}\ \emph {et~al.}(2001)\citenamefont
  {Armendariz-Picon}, \citenamefont {Mukhanov},\ and\ \citenamefont
  {Steinhardt}}]{Armendariz-Picon:2000ulo}%
  \BibitemOpen
  \bibfield  {author} {\bibinfo {author} {\bibfnamefont {C.}~\bibnamefont
  {Armendariz-Picon}}, \bibinfo {author} {\bibfnamefont {Viatcheslav~F.}\
  \bibnamefont {Mukhanov}}, \ and\ \bibinfo {author} {\bibfnamefont {Paul~J.}\
  \bibnamefont {Steinhardt}},\ }\bibfield  {title} {\enquote {\bibinfo {title}
  {{Essentials of k essence}},}\ }\href {\doibase 10.1103/PhysRevD.63.103510}
  {\bibfield  {journal} {\bibinfo  {journal} {Phys. Rev. D}\ }\textbf {\bibinfo
  {volume} {63}},\ \bibinfo {pages} {103510} (\bibinfo {year} {2001})},\
  \Eprint {http://arxiv.org/abs/astro-ph/0006373} {arXiv:astro-ph/0006373}
  \BibitemShut {NoStop}%
\bibitem [{\citenamefont {Gordon}\ and\ \citenamefont
  {Hu}(2004)}]{Gordon:2004ez}%
  \BibitemOpen
  \bibfield  {author} {\bibinfo {author} {\bibfnamefont {Christopher}\
  \bibnamefont {Gordon}}\ and\ \bibinfo {author} {\bibfnamefont {Wayne}\
  \bibnamefont {Hu}},\ }\bibfield  {title} {\enquote {\bibinfo {title} {{A Low
  CMB quadrupole from dark energy isocurvature perturbations}},}\ }\href
  {\doibase 10.1103/PhysRevD.70.083003} {\bibfield  {journal} {\bibinfo
  {journal} {Phys. Rev. D}\ }\textbf {\bibinfo {volume} {70}},\ \bibinfo
  {pages} {083003} (\bibinfo {year} {2004})},\ \Eprint
  {http://arxiv.org/abs/astro-ph/0406496} {arXiv:astro-ph/0406496} \BibitemShut
  {NoStop}%
\bibitem [{\citenamefont {G\'omez-Valent}\ \emph {et~al.}(2021)\citenamefont
  {G\'omez-Valent}, \citenamefont {Zheng}, \citenamefont {Amendola},
  \citenamefont {Pettorino},\ and\ \citenamefont
  {Wetterich}}]{Gomez-Valent:2021cbe}%
  \BibitemOpen
  \bibfield  {author} {\bibinfo {author} {\bibfnamefont {Adri\`a}\ \bibnamefont
  {G\'omez-Valent}}, \bibinfo {author} {\bibfnamefont {Ziyang}\ \bibnamefont
  {Zheng}}, \bibinfo {author} {\bibfnamefont {Luca}\ \bibnamefont {Amendola}},
  \bibinfo {author} {\bibfnamefont {Valeria}\ \bibnamefont {Pettorino}}, \ and\
  \bibinfo {author} {\bibfnamefont {Christof}\ \bibnamefont {Wetterich}},\
  }\bibfield  {title} {\enquote {\bibinfo {title} {{Early dark energy in the
  pre- and postrecombination epochs}},}\ }\href {\doibase
  10.1103/PhysRevD.104.083536} {\bibfield  {journal} {\bibinfo  {journal}
  {Phys. Rev. D}\ }\textbf {\bibinfo {volume} {104}},\ \bibinfo {pages}
  {083536} (\bibinfo {year} {2021})},\ \Eprint
  {http://arxiv.org/abs/2107.11065} {arXiv:2107.11065 [astro-ph.CO]}
  \BibitemShut {NoStop}%
\bibitem [{\citenamefont {{Turner}}(1983)}]{1983PhRvD..28.1243T}%
  \BibitemOpen
  \bibfield  {author} {\bibinfo {author} {\bibfnamefont {Michael~S.}\
  \bibnamefont {{Turner}}},\ }\bibfield  {title} {\enquote {\bibinfo {title}
  {{Coherent scalar-field oscillations in an expanding universe}},}\ }\href
  {\doibase 10.1103/PhysRevD.28.1243} {\bibfield  {journal} {\bibinfo
  {journal} {\prd}\ }\textbf {\bibinfo {volume} {28}},\ \bibinfo {pages}
  {1243--1247} (\bibinfo {year} {1983})}\BibitemShut {NoStop}%
\bibitem [{\citenamefont {Ma}\ and\ \citenamefont
  {Bertschinger}(1995)}]{Ma:1995ey}%
  \BibitemOpen
  \bibfield  {author} {\bibinfo {author} {\bibfnamefont {Chung-Pei}\
  \bibnamefont {Ma}}\ and\ \bibinfo {author} {\bibfnamefont {Edmund}\
  \bibnamefont {Bertschinger}},\ }\bibfield  {title} {\enquote {\bibinfo
  {title} {{Cosmological perturbation theory in the synchronous and conformal
  Newtonian gauges}},}\ }\href {\doibase 10.1086/176550} {\bibfield  {journal}
  {\bibinfo  {journal} {Astrophys. J.}\ }\textbf {\bibinfo {volume} {455}},\
  \bibinfo {pages} {7--25} (\bibinfo {year} {1995})},\ \Eprint
  {http://arxiv.org/abs/astro-ph/9506072} {arXiv:astro-ph/9506072} \BibitemShut
  {NoStop}%
\bibitem [{\citenamefont {Brinckmann}\ and\ \citenamefont
  {Lesgourgues}(2018)}]{Brinckmann:2018cvx}%
  \BibitemOpen
  \bibfield  {author} {\bibinfo {author} {\bibfnamefont {Thejs}\ \bibnamefont
  {Brinckmann}}\ and\ \bibinfo {author} {\bibfnamefont {Julien}\ \bibnamefont
  {Lesgourgues}},\ }\bibfield  {title} {\enquote {\bibinfo {title}
  {{MontePython 3: boosted MCMC sampler and other features}},}\ }\href@noop {}
  {\  (\bibinfo {year} {2018})},\ \Eprint {http://arxiv.org/abs/1804.07261}
  {arXiv:1804.07261 [astro-ph.CO]} \BibitemShut {NoStop}%
\bibitem [{\citenamefont {Audren}\ \emph {et~al.}(2013)\citenamefont {Audren},
  \citenamefont {Lesgourgues}, \citenamefont {Benabed},\ and\ \citenamefont
  {Prunet}}]{Audren:2012wb}%
  \BibitemOpen
  \bibfield  {author} {\bibinfo {author} {\bibfnamefont {Benjamin}\
  \bibnamefont {Audren}}, \bibinfo {author} {\bibfnamefont {Julien}\
  \bibnamefont {Lesgourgues}}, \bibinfo {author} {\bibfnamefont {Karim}\
  \bibnamefont {Benabed}}, \ and\ \bibinfo {author} {\bibfnamefont {Simon}\
  \bibnamefont {Prunet}},\ }\bibfield  {title} {\enquote {\bibinfo {title}
  {{Conservative Constraints on Early Cosmology: an illustration of the Monte
  Python cosmological parameter inference code}},}\ }\href {\doibase
  10.1088/1475-7516/2013/02/001} {\bibfield  {journal} {\bibinfo  {journal}
  {JCAP}\ }\textbf {\bibinfo {volume} {1302}},\ \bibinfo {pages} {001}
  (\bibinfo {year} {2013})},\ \Eprint {http://arxiv.org/abs/1210.7183}
  {arXiv:1210.7183 [astro-ph.CO]} \BibitemShut {NoStop}%
\bibitem [{\citenamefont {Lesgourgues}(2011)}]{Lesgourgues:2011re}%
  \BibitemOpen
  \bibfield  {author} {\bibinfo {author} {\bibfnamefont {Julien}\ \bibnamefont
  {Lesgourgues}},\ }\bibfield  {title} {\enquote {\bibinfo {title} {{The Cosmic
  Linear Anisotropy Solving System (CLASS) I: Overview}},}\ }\href@noop {} {\
  (\bibinfo {year} {2011})},\ \Eprint {http://arxiv.org/abs/1104.2932}
  {arXiv:1104.2932 [astro-ph.IM]} \BibitemShut {NoStop}%
\bibitem [{\citenamefont {Blas}\ \emph {et~al.}(2011)\citenamefont {Blas},
  \citenamefont {Lesgourgues},\ and\ \citenamefont {Tram}}]{Blas:2011rf}%
  \BibitemOpen
  \bibfield  {author} {\bibinfo {author} {\bibfnamefont {Diego}\ \bibnamefont
  {Blas}}, \bibinfo {author} {\bibfnamefont {Julien}\ \bibnamefont
  {Lesgourgues}}, \ and\ \bibinfo {author} {\bibfnamefont {Thomas}\
  \bibnamefont {Tram}},\ }\bibfield  {title} {\enquote {\bibinfo {title} {{The
  Cosmic Linear Anisotropy Solving System (CLASS) II: Approximation
  schemes}},}\ }\href {\doibase 10.1088/1475-7516/2011/07/034} {\bibfield
  {journal} {\bibinfo  {journal} {JCAP}\ }\textbf {\bibinfo {volume} {1107}},\
  \bibinfo {pages} {034} (\bibinfo {year} {2011})},\ \Eprint
  {http://arxiv.org/abs/1104.2933} {arXiv:1104.2933 [astro-ph.CO]} \BibitemShut
  {NoStop}%
\bibitem [{\citenamefont {Aghanim}\ \emph
  {et~al.}(2020{\natexlab{b}})\citenamefont {Aghanim} \emph
  {et~al.}}]{Planck:2019nip}%
  \BibitemOpen
  \bibfield  {author} {\bibinfo {author} {\bibfnamefont {N.}~\bibnamefont
  {Aghanim}} \emph {et~al.} (\bibinfo {collaboration} {Planck}),\ }\bibfield
  {title} {\enquote {\bibinfo {title} {{Planck 2018 results. V. CMB power
  spectra and likelihoods}},}\ }\href {\doibase 10.1051/0004-6361/201936386}
  {\bibfield  {journal} {\bibinfo  {journal} {Astron. Astrophys.}\ }\textbf
  {\bibinfo {volume} {641}},\ \bibinfo {pages} {A5} (\bibinfo {year}
  {2020}{\natexlab{b}})},\ \Eprint {http://arxiv.org/abs/1907.12875}
  {arXiv:1907.12875 [astro-ph.CO]} \BibitemShut {NoStop}%
\bibitem [{\citenamefont {Aghanim}\ \emph
  {et~al.}(2020{\natexlab{c}})\citenamefont {Aghanim} \emph
  {et~al.}}]{Planck:2018lbu}%
  \BibitemOpen
  \bibfield  {author} {\bibinfo {author} {\bibfnamefont {N.}~\bibnamefont
  {Aghanim}} \emph {et~al.} (\bibinfo {collaboration} {Planck}),\ }\bibfield
  {title} {\enquote {\bibinfo {title} {{Planck 2018 results. VIII.
  Gravitational lensing}},}\ }\href {\doibase 10.1051/0004-6361/201833886}
  {\bibfield  {journal} {\bibinfo  {journal} {Astron. Astrophys.}\ }\textbf
  {\bibinfo {volume} {641}},\ \bibinfo {pages} {A8} (\bibinfo {year}
  {2020}{\natexlab{c}})},\ \Eprint {http://arxiv.org/abs/1807.06210}
  {arXiv:1807.06210 [astro-ph.CO]} \BibitemShut {NoStop}%
\bibitem [{\citenamefont {Beutler}\ \emph {et~al.}(2011)\citenamefont
  {Beutler}, \citenamefont {Blake}, \citenamefont {Colless}, \citenamefont
  {Jones}, \citenamefont {Staveley-Smith}, \citenamefont {Campbell},
  \citenamefont {Parker}, \citenamefont {Saunders},\ and\ \citenamefont
  {Watson}}]{Beutler:2011hx}%
  \BibitemOpen
  \bibfield  {author} {\bibinfo {author} {\bibfnamefont {Florian}\ \bibnamefont
  {Beutler}}, \bibinfo {author} {\bibfnamefont {Chris}\ \bibnamefont {Blake}},
  \bibinfo {author} {\bibfnamefont {Matthew}\ \bibnamefont {Colless}}, \bibinfo
  {author} {\bibfnamefont {D.~Heath}\ \bibnamefont {Jones}}, \bibinfo {author}
  {\bibfnamefont {Lister}\ \bibnamefont {Staveley-Smith}}, \bibinfo {author}
  {\bibfnamefont {Lachlan}\ \bibnamefont {Campbell}}, \bibinfo {author}
  {\bibfnamefont {Quentin}\ \bibnamefont {Parker}}, \bibinfo {author}
  {\bibfnamefont {Will}\ \bibnamefont {Saunders}}, \ and\ \bibinfo {author}
  {\bibfnamefont {Fred}\ \bibnamefont {Watson}},\ }\bibfield  {title} {\enquote
  {\bibinfo {title} {{The 6dF Galaxy Survey: Baryon Acoustic Oscillations and
  the Local Hubble Constant}},}\ }\href {\doibase
  10.1111/j.1365-2966.2011.19250.x} {\bibfield  {journal} {\bibinfo  {journal}
  {Mon. Not. Roy. Astron. Soc.}\ }\textbf {\bibinfo {volume} {416}},\ \bibinfo
  {pages} {3017--3032} (\bibinfo {year} {2011})},\ \Eprint
  {http://arxiv.org/abs/1106.3366} {arXiv:1106.3366 [astro-ph.CO]} \BibitemShut
  {NoStop}%
\bibitem [{\citenamefont {Ross}\ \emph {et~al.}(2015)\citenamefont {Ross},
  \citenamefont {Samushia}, \citenamefont {Howlett}, \citenamefont {Percival},
  \citenamefont {Burden},\ and\ \citenamefont {Manera}}]{Ross:2014qpa}%
  \BibitemOpen
  \bibfield  {author} {\bibinfo {author} {\bibfnamefont {Ashley~J.}\
  \bibnamefont {Ross}}, \bibinfo {author} {\bibfnamefont {Lado}\ \bibnamefont
  {Samushia}}, \bibinfo {author} {\bibfnamefont {Cullan}\ \bibnamefont
  {Howlett}}, \bibinfo {author} {\bibfnamefont {Will~J.}\ \bibnamefont
  {Percival}}, \bibinfo {author} {\bibfnamefont {Angela}\ \bibnamefont
  {Burden}}, \ and\ \bibinfo {author} {\bibfnamefont {Marc}\ \bibnamefont
  {Manera}},\ }\bibfield  {title} {\enquote {\bibinfo {title} {{The clustering
  of the SDSS DR7 main Galaxy sample – I. A 4 per cent distance measure at $z
  = 0.15$}},}\ }\href {\doibase 10.1093/mnras/stv154} {\bibfield  {journal}
  {\bibinfo  {journal} {Mon. Not. Roy. Astron. Soc.}\ }\textbf {\bibinfo
  {volume} {449}},\ \bibinfo {pages} {835--847} (\bibinfo {year} {2015})},\
  \Eprint {http://arxiv.org/abs/1409.3242} {arXiv:1409.3242 [astro-ph.CO]}
  \BibitemShut {NoStop}%
\bibitem [{\citenamefont {Gil-Marín}\ \emph {et~al.}(2016)\citenamefont
  {Gil-Marín} \emph {et~al.}}]{Gil-Marin:2015nqa}%
  \BibitemOpen
  \bibfield  {author} {\bibinfo {author} {\bibfnamefont {Héctor}\ \bibnamefont
  {Gil-Marín}} \emph {et~al.},\ }\bibfield  {title} {\enquote {\bibinfo
  {title} {{The clustering of galaxies in the SDSS-III Baryon Oscillation
  Spectroscopic Survey: BAO measurement from the LOS-dependent power spectrum
  of DR12 BOSS galaxies}},}\ }\href {\doibase 10.1093/mnras/stw1264} {\bibfield
   {journal} {\bibinfo  {journal} {Mon. Not. Roy. Astron. Soc.}\ }\textbf
  {\bibinfo {volume} {460}},\ \bibinfo {pages} {4210--4219} (\bibinfo {year}
  {2016})},\ \Eprint {http://arxiv.org/abs/1509.06373} {arXiv:1509.06373
  [astro-ph.CO]} \BibitemShut {NoStop}%
\bibitem [{\citenamefont {Kitaura}\ \emph {et~al.}(2016)\citenamefont {Kitaura}
  \emph {et~al.}}]{Kitaura:2015uqa}%
  \BibitemOpen
  \bibfield  {author} {\bibinfo {author} {\bibfnamefont {Francisco-Shu}\
  \bibnamefont {Kitaura}} \emph {et~al.},\ }\bibfield  {title} {\enquote
  {\bibinfo {title} {{The clustering of galaxies in the SDSS-III Baryon
  Oscillation Spectroscopic Survey: mock galaxy catalogues for the BOSS Final
  Data Release}},}\ }\href {\doibase 10.1093/mnras/stv2826} {\bibfield
  {journal} {\bibinfo  {journal} {Mon. Not. Roy. Astron. Soc.}\ }\textbf
  {\bibinfo {volume} {456}},\ \bibinfo {pages} {4156--4173} (\bibinfo {year}
  {2016})},\ \Eprint {http://arxiv.org/abs/1509.06400} {arXiv:1509.06400
  [astro-ph.CO]} \BibitemShut {NoStop}%
\bibitem [{\citenamefont {Reid}\ \emph {et~al.}(2016)\citenamefont {Reid} \emph
  {et~al.}}]{Reid:2015gra}%
  \BibitemOpen
  \bibfield  {author} {\bibinfo {author} {\bibfnamefont {Beth}\ \bibnamefont
  {Reid}} \emph {et~al.},\ }\bibfield  {title} {\enquote {\bibinfo {title}
  {{SDSS-III Baryon Oscillation Spectroscopic Survey Data Release 12: galaxy
  target selection and large scale structure catalogues}},}\ }\href {\doibase
  10.1093/mnras/stv2382} {\bibfield  {journal} {\bibinfo  {journal} {Mon. Not.
  Roy. Astron. Soc.}\ }\textbf {\bibinfo {volume} {455}},\ \bibinfo {pages}
  {1553--1573} (\bibinfo {year} {2016})},\ \Eprint
  {http://arxiv.org/abs/1509.06529} {arXiv:1509.06529 [astro-ph.CO]}
  \BibitemShut {NoStop}%
\bibitem [{\citenamefont {Nishimichi}\ \emph {et~al.}(2020)\citenamefont
  {Nishimichi}, \citenamefont {D'Amico}, \citenamefont {Ivanov}, \citenamefont
  {Senatore}, \citenamefont {Simonovi\'c}, \citenamefont {Takada},
  \citenamefont {Zaldarriaga},\ and\ \citenamefont
  {Zhang}}]{Nishimichi:2020tvu}%
  \BibitemOpen
  \bibfield  {author} {\bibinfo {author} {\bibfnamefont {Takahiro}\
  \bibnamefont {Nishimichi}}, \bibinfo {author} {\bibfnamefont {Guido}\
  \bibnamefont {D'Amico}}, \bibinfo {author} {\bibfnamefont {Mikhail~M.}\
  \bibnamefont {Ivanov}}, \bibinfo {author} {\bibfnamefont {Leonardo}\
  \bibnamefont {Senatore}}, \bibinfo {author} {\bibfnamefont {Marko}\
  \bibnamefont {Simonovi\'c}}, \bibinfo {author} {\bibfnamefont {Masahiro}\
  \bibnamefont {Takada}}, \bibinfo {author} {\bibfnamefont {Matias}\
  \bibnamefont {Zaldarriaga}}, \ and\ \bibinfo {author} {\bibfnamefont
  {Pierre}\ \bibnamefont {Zhang}},\ }\bibfield  {title} {\enquote {\bibinfo
  {title} {{Blinded challenge for precision cosmology with large-scale
  structure: results from effective field theory for the redshift-space galaxy
  power spectrum}},}\ }\href {\doibase 10.1103/PhysRevD.102.123541} {\bibfield
  {journal} {\bibinfo  {journal} {Phys. Rev. D}\ }\textbf {\bibinfo {volume}
  {102}},\ \bibinfo {pages} {123541} (\bibinfo {year} {2020})},\ \Eprint
  {http://arxiv.org/abs/2003.08277} {arXiv:2003.08277 [astro-ph.CO]}
  \BibitemShut {NoStop}%
\bibitem [{\citenamefont {Ross}\ \emph {et~al.}(2020)\citenamefont {Ross} \emph
  {et~al.}}]{Ross:2020lqz}%
  \BibitemOpen
  \bibfield  {author} {\bibinfo {author} {\bibfnamefont {Ashley~J.}\
  \bibnamefont {Ross}} \emph {et~al.},\ }\bibfield  {title} {\enquote {\bibinfo
  {title} {{The Completed SDSS-IV extended Baryon Oscillation Spectroscopic
  Survey: Large-scale structure catalogues for cosmological analysis}},}\
  }\href {\doibase 10.1093/mnras/staa2416} {\bibfield  {journal} {\bibinfo
  {journal} {Mon. Not. Roy. Astron. Soc.}\ }\textbf {\bibinfo {volume} {498}},\
  \bibinfo {pages} {2354--2371} (\bibinfo {year} {2020})},\ \Eprint
  {http://arxiv.org/abs/2007.09000} {arXiv:2007.09000 [astro-ph.CO]}
  \BibitemShut {NoStop}%
\bibitem [{\citenamefont {Chuang}\ \emph {et~al.}(2015)\citenamefont {Chuang},
  \citenamefont {Kitaura}, \citenamefont {Prada}, \citenamefont {Zhao},\ and\
  \citenamefont {Yepes}}]{Chuang:2014vfa}%
  \BibitemOpen
  \bibfield  {author} {\bibinfo {author} {\bibfnamefont {Chia-Hsun}\
  \bibnamefont {Chuang}}, \bibinfo {author} {\bibfnamefont {Francisco-Shu}\
  \bibnamefont {Kitaura}}, \bibinfo {author} {\bibfnamefont {Francisco}\
  \bibnamefont {Prada}}, \bibinfo {author} {\bibfnamefont {Cheng}\ \bibnamefont
  {Zhao}}, \ and\ \bibinfo {author} {\bibfnamefont {Gustavo}\ \bibnamefont
  {Yepes}},\ }\bibfield  {title} {\enquote {\bibinfo {title} {{EZmocks:
  extending the Zel'dovich approximation to generate mock galaxy catalogues
  with accurate clustering statistics}},}\ }\href {\doibase
  10.1093/mnras/stu2301} {\bibfield  {journal} {\bibinfo  {journal} {Mon. Not.
  Roy. Astron. Soc.}\ }\textbf {\bibinfo {volume} {446}},\ \bibinfo {pages}
  {2621--2628} (\bibinfo {year} {2015})},\ \Eprint
  {http://arxiv.org/abs/1409.1124} {arXiv:1409.1124 [astro-ph.CO]} \BibitemShut
  {NoStop}%
\bibitem [{\citenamefont {Beutler}\ and\ \citenamefont
  {McDonald}(2021)}]{Beutler:2021eqq}%
  \BibitemOpen
  \bibfield  {author} {\bibinfo {author} {\bibfnamefont {Florian}\ \bibnamefont
  {Beutler}}\ and\ \bibinfo {author} {\bibfnamefont {Patrick}\ \bibnamefont
  {McDonald}},\ }\bibfield  {title} {\enquote {\bibinfo {title} {{Unified
  galaxy power spectrum measurements from 6dFGS, BOSS, and eBOSS}},}\ }\href
  {\doibase 10.1088/1475-7516/2021/11/031} {\bibfield  {journal} {\bibinfo
  {journal} {JCAP}\ }\textbf {\bibinfo {volume} {11}},\ \bibinfo {pages} {031}
  (\bibinfo {year} {2021})},\ \Eprint {http://arxiv.org/abs/2106.06324}
  {arXiv:2106.06324 [astro-ph.CO]} \BibitemShut {NoStop}%
\bibitem [{\citenamefont {Hou}\ \emph {et~al.}(2020)\citenamefont {Hou} \emph
  {et~al.}}]{Hou:2020rse}%
  \BibitemOpen
  \bibfield  {author} {\bibinfo {author} {\bibfnamefont {Jiamin}\ \bibnamefont
  {Hou}} \emph {et~al.},\ }\bibfield  {title} {\enquote {\bibinfo {title} {{The
  Completed SDSS-IV extended Baryon Oscillation Spectroscopic Survey: BAO and
  RSD measurements from anisotropic clustering analysis of the Quasar Sample in
  configuration space between redshift 0.8 and 2.2}},}\ }\href {\doibase
  10.1093/mnras/staa3234} {\bibfield  {journal} {\bibinfo  {journal} {Mon. Not.
  Roy. Astron. Soc.}\ }\textbf {\bibinfo {volume} {500}},\ \bibinfo {pages}
  {1201--1221} (\bibinfo {year} {2020})},\ \Eprint
  {http://arxiv.org/abs/2007.08998} {arXiv:2007.08998 [astro-ph.CO]}
  \BibitemShut {NoStop}%
\bibitem [{\citenamefont {Scolnic}\ \emph {et~al.}(2018)\citenamefont {Scolnic}
  \emph {et~al.}}]{Scolnic:2017caz}%
  \BibitemOpen
  \bibfield  {author} {\bibinfo {author} {\bibfnamefont {D.~M.}\ \bibnamefont
  {Scolnic}} \emph {et~al.},\ }\bibfield  {title} {\enquote {\bibinfo {title}
  {{The Complete Light-curve Sample of Spectroscopically Confirmed SNe Ia from
  Pan-STARRS1 and Cosmological Constraints from the Combined Pantheon
  Sample}},}\ }\href {\doibase 10.3847/1538-4357/aab9bb} {\bibfield  {journal}
  {\bibinfo  {journal} {Astrophys. J.}\ }\textbf {\bibinfo {volume} {859}},\
  \bibinfo {pages} {101} (\bibinfo {year} {2018})},\ \Eprint
  {http://arxiv.org/abs/1710.00845} {arXiv:1710.00845 [astro-ph.CO]}
  \BibitemShut {NoStop}%
\bibitem [{\citenamefont {Mead}\ \emph {et~al.}(2020)\citenamefont {Mead},
  \citenamefont {Brieden}, \citenamefont {Tr\"oster},\ and\ \citenamefont
  {Heymans}}]{Mead:2020vgs}%
  \BibitemOpen
  \bibfield  {author} {\bibinfo {author} {\bibfnamefont {Alexander}\
  \bibnamefont {Mead}}, \bibinfo {author} {\bibfnamefont {Samuel}\ \bibnamefont
  {Brieden}}, \bibinfo {author} {\bibfnamefont {Tilman}\ \bibnamefont
  {Tr\"oster}}, \ and\ \bibinfo {author} {\bibfnamefont {Catherine}\
  \bibnamefont {Heymans}},\ }\bibfield  {title} {\enquote {\bibinfo {title}
  {{HMcode-2020: Improved modelling of non-linear cosmological power spectra
  with baryonic feedback}},}\ }\href {\doibase 10.1093/mnras/stab082} {\
  (\bibinfo {year} {2020}),\ 10.1093/mnras/stab082},\ \Eprint
  {http://arxiv.org/abs/2009.01858} {arXiv:2009.01858 [astro-ph.CO]}
  \BibitemShut {NoStop}%
\bibitem [{\citenamefont {Lewis}(2019)}]{Lewis:2019xzd}%
  \BibitemOpen
  \bibfield  {author} {\bibinfo {author} {\bibfnamefont {Antony}\ \bibnamefont
  {Lewis}},\ }\bibfield  {title} {\enquote {\bibinfo {title} {{GetDist: a
  Python package for analysing Monte Carlo samples}},}\ }\href@noop {} {\
  (\bibinfo {year} {2019})},\ \Eprint {http://arxiv.org/abs/1910.13970}
  {arXiv:1910.13970 [astro-ph.IM]} \BibitemShut {NoStop}%
\bibitem [{\citenamefont {Raveri}\ and\ \citenamefont
  {Hu}(2019)}]{Raveri:2018wln}%
  \BibitemOpen
  \bibfield  {author} {\bibinfo {author} {\bibfnamefont {Marco}\ \bibnamefont
  {Raveri}}\ and\ \bibinfo {author} {\bibfnamefont {Wayne}\ \bibnamefont
  {Hu}},\ }\bibfield  {title} {\enquote {\bibinfo {title} {Concordance and
  discordance in cosmology},}\ }\href {\doibase 10.1103/PhysRevD.99.043506}
  {\bibfield  {journal} {\bibinfo  {journal} {Phys. Rev. D}\ }\textbf {\bibinfo
  {volume} {99}},\ \bibinfo {pages} {043506} (\bibinfo {year} {2019})},\
  \Eprint {http://arxiv.org/abs/1806.04649} {arXiv:1806.04649 [astro-ph.CO]}
  \BibitemShut {NoStop}%
\bibitem [{\citenamefont {Riess}\ \emph {et~al.}(2019)\citenamefont {Riess},
  \citenamefont {Casertano}, \citenamefont {Yuan}, \citenamefont {Macri},\ and\
  \citenamefont {Scolnic}}]{Riess:2019cxk}%
  \BibitemOpen
  \bibfield  {author} {\bibinfo {author} {\bibfnamefont {Adam~G.}\ \bibnamefont
  {Riess}}, \bibinfo {author} {\bibfnamefont {Stefano}\ \bibnamefont
  {Casertano}}, \bibinfo {author} {\bibfnamefont {Wenlong}\ \bibnamefont
  {Yuan}}, \bibinfo {author} {\bibfnamefont {Lucas~M.}\ \bibnamefont {Macri}},
  \ and\ \bibinfo {author} {\bibfnamefont {Dan}\ \bibnamefont {Scolnic}},\
  }\bibfield  {title} {\enquote {\bibinfo {title} {{Large Magellanic Cloud
  Cepheid Standards Provide a 1\% Foundation for the Determination of the
  Hubble Constant and Stronger Evidence for Physics beyond $\Lambda$CDM}},}\
  }\href {\doibase 10.3847/1538-4357/ab1422} {\bibfield  {journal} {\bibinfo
  {journal} {Astrophys. J.}\ }\textbf {\bibinfo {volume} {876}},\ \bibinfo
  {pages} {85} (\bibinfo {year} {2019})},\ \Eprint
  {http://arxiv.org/abs/1903.07603} {arXiv:1903.07603 [astro-ph.CO]}
  \BibitemShut {NoStop}%
\bibitem [{\citenamefont {Abdalla}\ \emph {et~al.}(2022)\citenamefont {Abdalla}
  \emph {et~al.}}]{Abdalla:2022yfr}%
  \BibitemOpen
  \bibfield  {author} {\bibinfo {author} {\bibfnamefont {Elcio}\ \bibnamefont
  {Abdalla}} \emph {et~al.},\ }\bibfield  {title} {\enquote {\bibinfo {title}
  {{Cosmology intertwined: A review of the particle physics, astrophysics, and
  cosmology associated with the cosmological tensions and anomalies}},}\ }\href
  {\doibase 10.1016/j.jheap.2022.04.002} {\bibfield  {journal} {\bibinfo
  {journal} {JHEAp}\ }\textbf {\bibinfo {volume} {34}},\ \bibinfo {pages}
  {49--211} (\bibinfo {year} {2022})},\ \Eprint
  {http://arxiv.org/abs/2203.06142} {arXiv:2203.06142 [astro-ph.CO]}
  \BibitemShut {NoStop}%
\bibitem [{\citenamefont {Vagnozzi}(2021)}]{Vagnozzi:2021gjh}%
  \BibitemOpen
  \bibfield  {author} {\bibinfo {author} {\bibfnamefont {Sunny}\ \bibnamefont
  {Vagnozzi}},\ }\bibfield  {title} {\enquote {\bibinfo {title} {{Consistency
  tests of \ensuremath{\Lambda}CDM from the early integrated Sachs-Wolfe
  effect: Implications for early-time new physics and the Hubble tension}},}\
  }\href {\doibase 10.1103/PhysRevD.104.063524} {\bibfield  {journal} {\bibinfo
   {journal} {Phys. Rev. D}\ }\textbf {\bibinfo {volume} {104}},\ \bibinfo
  {pages} {063524} (\bibinfo {year} {2021})},\ \Eprint
  {http://arxiv.org/abs/2105.10425} {arXiv:2105.10425 [astro-ph.CO]}
  \BibitemShut {NoStop}%
\bibitem [{\citenamefont {Heymans}\ \emph {et~al.}(2021)\citenamefont {Heymans}
  \emph {et~al.}}]{Heymans:2020gsg}%
  \BibitemOpen
  \bibfield  {author} {\bibinfo {author} {\bibfnamefont {Catherine}\
  \bibnamefont {Heymans}} \emph {et~al.},\ }\bibfield  {title} {\enquote
  {\bibinfo {title} {{KiDS-1000 Cosmology: Multi-probe weak gravitational
  lensing and spectroscopic galaxy clustering constraints}},}\ }\href {\doibase
  10.1051/0004-6361/202039063} {\bibfield  {journal} {\bibinfo  {journal}
  {Astron. Astrophys.}\ }\textbf {\bibinfo {volume} {646}},\ \bibinfo {pages}
  {A140} (\bibinfo {year} {2021})},\ \Eprint {http://arxiv.org/abs/2007.15632}
  {arXiv:2007.15632 [astro-ph.CO]} \BibitemShut {NoStop}%
\bibitem [{\citenamefont {Abbott}\ \emph {et~al.}(2021)\citenamefont {Abbott}
  \emph {et~al.}}]{DES:2021wwk}%
  \BibitemOpen
  \bibfield  {author} {\bibinfo {author} {\bibfnamefont {T.~M.~C.}\
  \bibnamefont {Abbott}} \emph {et~al.} (\bibinfo {collaboration} {DES}),\
  }\bibfield  {title} {\enquote {\bibinfo {title} {{Dark Energy Survey Year 3
  Results: Cosmological Constraints from Galaxy Clustering and Weak
  Lensing}},}\ }\href@noop {} {\  (\bibinfo {year} {2021})},\ \Eprint
  {http://arxiv.org/abs/2105.13549} {arXiv:2105.13549 [astro-ph.CO]}
  \BibitemShut {NoStop}%
\bibitem [{\citenamefont {Liu}\ \emph {et~al.}(2023)\citenamefont {Liu},
  \citenamefont {Gao}, \citenamefont {Han}, \citenamefont {Mu},\ and\
  \citenamefont {Xu}}]{Liu:2023haw}%
  \BibitemOpen
  \bibfield  {author} {\bibinfo {author} {\bibfnamefont {Gang}\ \bibnamefont
  {Liu}}, \bibinfo {author} {\bibfnamefont {Jiaze}\ \bibnamefont {Gao}},
  \bibinfo {author} {\bibfnamefont {Yufen}\ \bibnamefont {Han}}, \bibinfo
  {author} {\bibfnamefont {Yuhao}\ \bibnamefont {Mu}}, \ and\ \bibinfo {author}
  {\bibfnamefont {Lixin}\ \bibnamefont {Xu}},\ }\bibfield  {title} {\enquote
  {\bibinfo {title} {{Mitigating Cosmological Tensions via Momentum-Coupled
  Dark Sector Model}},}\ }\href@noop {} {\  (\bibinfo {year} {2023})},\ \Eprint
  {http://arxiv.org/abs/2310.09798} {arXiv:2310.09798 [astro-ph.CO]}
  \BibitemShut {NoStop}%
\bibitem [{\citenamefont {Franco~Abell\'an}\ \emph {et~al.}(2023)\citenamefont
  {Franco~Abell\'an}, \citenamefont {Braglia}, \citenamefont {Ballardini},
  \citenamefont {Finelli},\ and\ \citenamefont
  {Poulin}}]{FrancoAbellan:2023gec}%
  \BibitemOpen
  \bibfield  {author} {\bibinfo {author} {\bibfnamefont {Guillermo}\
  \bibnamefont {Franco~Abell\'an}}, \bibinfo {author} {\bibfnamefont {Matteo}\
  \bibnamefont {Braglia}}, \bibinfo {author} {\bibfnamefont {Mario}\
  \bibnamefont {Ballardini}}, \bibinfo {author} {\bibfnamefont {Fabio}\
  \bibnamefont {Finelli}}, \ and\ \bibinfo {author} {\bibfnamefont {Vivian}\
  \bibnamefont {Poulin}},\ }\bibfield  {title} {\enquote {\bibinfo {title}
  {{Probing Early Modification of Gravity with Planck, ACT and SPT}},}\
  }\href@noop {} {\  (\bibinfo {year} {2023})},\ \Eprint
  {http://arxiv.org/abs/2308.12345} {arXiv:2308.12345 [astro-ph.CO]}
  \BibitemShut {NoStop}%
\end{thebibliography}%

\end{document}